\documentclass[a4paper,11pt]{article}
\usepackage{jheppub} 
\usepackage{lineno}
\nolinenumbers
\usepackage{mystuff}
\pgfplotsset{compat=1.18}
\usepackage{graphicx} 
\usepackage{xcolor}

\definecolor{amethyst}{rgb}{0.54, 0.17, 0.89}
\definecolor{coral}{rgb}{1.0, 0.3, 0.4}

\newcommand{\dket}[1]{\lvert #1 \rangle\!\rangle}
\newcommand{\dbra}[1]{\langle\!\langle #1 \rvert}
\newcommand{\dbraket}[2]{\langle\!\langle #1 \vert #2 \rangle\!\rangle}
\newcommand{\dnorm}[1]{\langle\!\langle #1 \vert #1 \rangle\!\rangle}

\newcommand{\Hext}[1]{(\Ha_{\mathrm{ext}})_{#1}}
\newcommand{\Hmatt}[1]{\Ha_{\mathrm{matt}}^{#1}}

\title{
Entanglement entropy in topological tensor networks
}

\author[a,b,c]{Vijay Balasubramanian,}
\author[a,d]{and Charlie Cummings}
\affiliation[a]{David Rittenhouse Laboratory, University of Pennsylvania,  209 S. 33rd Street, Philadelphia, Pennsylvania 19104, USA}
\affiliation[b]{Santa Fe Institute, 1399 Hyde Park Road, Santa Fe, NM 87501, USA}
\affiliation[c]{Theoretische Natuurkunde, Vrije Universiteit Brussel, Pleinlaan 2, B-1050 Brussels, Belgium}
\affiliation[d]{Kavli Institute for Theoretical Physics,
University of California, Santa Barbara, Kohn Hall, Lagoon Rd, Santa Barbara, CA 93106, USA}

\emailAdd{charlie5@sas.upenn.edu}

\abstract{
We derive an entropy formula for recently proposed tensor network models which prepare diffeomorphism invariant states of topological quantum field theories with non-compact and/or continuous gauge groups.  
We  show that our entropy formula generalizes the notion of ``topological entanglement entropy'' to  incorporate the infinite number of particle-like excitations in such theories. When our  networks are endowed with gauge group $\SL(2,\R)$, we can interpret them as  models of three-dimensional gravity with small Newton's constant and possibly non-invertible metrics.}

\begin{document}
\maketitle
\flushbottom

\section{Introduction}

Quantum information theory has played a crucial role in our understanding of the structure of holographic theories of quantum gravity \cite{Ryu_2006,Hubeny:2007xt,Faulkner:2013ana,Engelhardt:2014gca,Hayden:2011ag,Rangamani:2016dms}, in particular through tensor network constructions of quantum states \cite{Balasubramanian:2025rcr,Akers:2024wab,Dong2024,Akers:2024ixq,Swingle_2012,Pastawski_2015,Hayden_2016,Dong:2018seb,Donnelly:2016qqt,Qi:2022lbd,Singh_2010,Colafranceschi:2020ern,Basteiro:2022xvu,Basteiro:2024cuh,Basteiro:2024crz,Basteiro:2022zur,Caputa:2020fbc,Frenkel:2024smt,Singh:2017tet,Colafranceschi:2022dig,Chirco:2017wgl,Chirco:2021chk,Yang:2015uoa,Sahu:2025upe}. A tensor network is a toy model of the holographic map from the bulk to the boundary.  Specifically, given  bulk and boundary Hilbert spaces $\Ha_{\mathrm{bulk}}$ and $\Ha_{\mathrm{bdry}}$,  a tensor network provides an isometry $\Ha_{\mathrm{bulk}} \to \Ha_{\mathrm{bdry}}$ constructed by  contracting  legs of local tensors according to a graph.  We think of the graph  as tessellating a Cauchy slice of a holographic spacetime. The quantum state on $\Ha_{\mathrm{bdry}}$ is  defined on the  uncontracted legs of the tensor network (Fig.~\ref{fig:tensornetwork}) which are understood as anchored to  the spacetime boundary.

This picture has some awkward features.  For example, Fig.~\ref{fig:tensornetwork} makes it clear that a naive implementation of a tensor network involves a discretization of  spacetime, which breaks diffeomorphism invariance, the gauge symmetry of gravity. This is in  contrast with lattice gauge theory, where the lattice does not break gauge invariance. Furthermore,  quantum states prepared by traditional tensor networks have a flat R\'enyi spectrum, in contrast to semiclassical states of quantum gravity \cite{Dong:2018seb}.\footnote{This need not be a problem, because traditional tensor networks can serve as a basis for expanding  semiclassical states \cite{Dong:2018seb}. Indeed, this is essentially the structure in the model we develop below.} Thus, while  conventional tensor networks prepare states that share many features of  quantum gravity, open questions remain.

Recently, we proposed a refined notion of tensor network which addresses some of these difficulties \cite{Balasubramanian:2025rcr} (see also  \cite{Akers:2024wab,Dong2024,Delcamp:2016eya,Chandra:2023dgq,Bonderson:2017osr,Fliss:2023dze,Buerschaper:2008eyf,Gu:2009xyz,McGough:2013gka,Carlip:1994gy} for related models/computations). Our tensor networks are based on {\it string nets} \cite{kirillov2011stringnet,Kitaev:1997wr,Levin_2005}, a tool for constructing quantum states of three-dimensional topological field theories (TQFTs) in terms of discrete two dimensional graphs.  Our method can prepare states of TQFTs with non-compact and continuous gauge groups, including, for example, the large level limit of $\SL(2,\R) \times \SL(2,\R)$ Chern--Simons theory. Following \cite{Witten:1988hc}, states of the latter theory can be interpreted as gauge invariant states of three-dimensional gravity with a small Newton constant ($G_N \to 0$), albeit including non-invertible metrics,\footnote{See Sec.~\ref{sec:CTV} for a discussion about how to account for differences between this Chern--Simons theory and three-dimensional gravity. We will discuss these points in more detail in \cite{Balasubramanian:2026xyz}.} even though they are defined by  discrete graphs $\Lambda$. This is because after imposing the constraints of gauge invariance, the Hilbert space $\Ha_{\mathrm{phys}}(\Lambda)$ of our {\it topological tensor networks} only depends on the spatial surface $\Sigma$ that $\Lambda$ tessellates. So we should actually denote the Hilbert space as $\Ha_{\mathrm{phys}}(\Sigma)$, with $\Sigma$ understood as a Cauchy slice of a spacetime.  The graph $\Lambda$ is simply a tool for constructing  $\Ha_{\mathrm{phys}}(\Sigma)$. 

In this paper, we will derive a universal expression for the entropy of boundary subregions in this model.  In a companion paper \cite{Balasubramanian:2026xyz} we will study the entropy of topological tensor networks with added matter, and compare the resulting entropy formula to the holographic entropy formula for three-dimensional gravity.

\subsection{Synopsis of the entropy formula derivation} \label{sec:synopsis}

Let $G$ be a transformable Lie group,\footnote{See \cite{Balasubramanian:2025rcr} for the definition of a transformable Lie group (unimodular and type I). For example, every semisimple Lie group (compact or non-compact) is a transformable group.} and $\Sigma$ be a spatial surface (possibly with boundary). 
Then, as discussed above  (details in Sec.~\ref{sec:themodel}), a Hilbert space of a  gauge theory with group $G$\footnote{More precisely, the theory we consider has a $D[G]$ gauge symmetry, where $D[G]$ is the Drinfeld double of $\mathrm{Rep}(G)$. See Secs.~\ref{sec:themodel} and \ref{sec:doublemodel} for more details about this ``doubling'' of the gauge symmetry.} defined on $\Sigma$, which we denote as $\Ha_{\mathrm{phys}}(\Sigma)$, can be constructed by tensor network methods \cite{Balasubramanian:2025rcr}.
Here, we study the entropy of such states in $\Ha_{\mathrm{phys}}(\Sigma)$, bipartitioned into subsystems. 

To define an entropy (or any other information-theoretic quantity), we must typically first define a notion of subregion across which the Hilbert space $\Ha_{\mathrm{phys}}(\Sigma)$ factorizes. Indeed, given a quantum state $\ket{\psi}$ and a subregion $R$, the reduced density matrix of $R$, which is  used to define  information theoretic quantities, is defined as $\rho_R = \tr_{\overline{R}}[\ketbra{\psi}]$. The partial trace over the complementary region $\overline{R}$ requires the total Hilbert space to factorize as $\Ha(\Sigma) = \Ha_R \otimes \Ha_{\overline{R}}$.

However, $\Ha_{\mathrm{phys}}(\Sigma) $ does not factorize in this way because of the gauge constraints. Thus, our first step will be to define an auxiliary map $V$ which embeds the physical Hilbert space $\Ha_{\mathrm{phys}}(\Sigma)$ into a factorizing Hilbert space associated with each subregion, i.e.,
\begin{align}
    V: \Ha_{\mathrm{phys}}(\Sigma) \to \Hext{R} \otimes \Hext{\overline{R}}\,.
\end{align}
The map $V$ will introduce additional degrees of freedom, called \emph{edge modes}, in order to factorize the Hilbert space.
Such a map is not unique, but we will be careful to identify the ambiguity, explain its physical origin, and determine a preferred choice among such factorization maps.

There is an alternative approach to computing the entropy shared between two boundary subregions using operator algebras which does not require the auxiliary map $V$. This relies on noting that (see Sec.~\ref{sec:themodel} around \eqref{eq:HphysAlgebras}) when $\Sigma$ is a disk, the Hilbert space of the disk decomposes as
\begin{align}
    \Ha_{\mathrm{phys}}(\Sigma) = \int_{\widehat{G}}^{\oplus} d\mu(\pi)\, \Pi_A[\Ha_R(\pi) \otimes \Ha_{\overline{R}}(\overline{\pi})] \,. \label{eq:fixedareadecompositionintro}
\end{align}
Here, $\pi$ is a unitary irreducible representation of the gauge group $G$, and $d\mu(\pi)$ is the Plancherel measure of $G$. 
Furthermore, $\Ha_R(\pi)$ is an auxiliary Hilbert space associated with the boundary subregion $R$, and similarly for $\Ha_{\overline{R}}(\overline{\pi})$, and the $\Pi_A$ indicates we should restrict to the gauge invariant subspace of the tensor product in each sector.
This direct integral is a decomposition of the disk Hilbert space into a basis of ``fixed area states'' \cite{Dong:2018seb,Dong:2022ilf}. 
We will call  degrees of freedom within $\Ha_R(\pi)$ which restrict the subregion Hilbert space to the $\pi$ sector the \emph{gauge fixing edge modes}. Thus, the gauge fixing edge modes are inevitably introduced in the computation of the entropy of a subregion.
We will explain in detail how the factorization map approach and the algebraic approach are related, and will argue that they are equivalent.

In particular, we will see that the factorized Hilbert spaces $\Hext{R}$ and $\Hext{\overline{R}} $, called the extended Hilbert spaces of $R$ and  $\overline{R}$, respectively, are related to the ``fixed area'' Hilbert spaces $\Ha_R(\pi)$ as
\begin{align}
    \Hext{R} = \int_{\widehat{G}}^{\oplus} d\mu(\pi) \, \Ha_R(\pi) \otimes V_\pi \,, \label{eq:Hextdecomp}
\end{align}
where the extended Hilbert space introduces additional degrees of freedom, the \emph{observer edge modes} \cite{Donnelly_2016}, captured by the auxiliary Hilbert space $V_\pi$ in \eqref{eq:Hextdecomp}. The representations $\pi$ which label the observer edge modes are the same representations that appear in the decomposition of the Hilbert space \eqref{eq:fixedareadecompositionintro} in the algebraic approach. The observer edge modes are entangled with the gauge fixing edge modes in a precise way, which allows us to ``measure'' properties of the bulk subregion, despite the diffeomorphism invariance which the gauge fixing edge modes enforce. The observer edge modes only arise when we use the Hilbert space factorization map $V$ approach instead of the algebraic approach. As suggested by their name, we will comment on their relationship to recent investigations of observers in quantum gravity. 

We use the factorization map $V$ to define the reduced state $\rho_R$ of a boundary subregion. This state must be  normalizable and we will explain how to ensure this. We then compute the entropy of $\rho_R$, and find that (see \eqref{eq:resultsummary})
\begin{align}
    S(\rho_R) = H[\,p_\pi] + \langle \hat{A} \rangle_\rho + S_{\mathrm{bulk}}(\rho_R) \,,
\end{align}
where $p_\pi$ is the probability that the subregion is found in the $\pi$ sector of \eqref{eq:fixedareadecompositionintro} and $H[\,p_\pi]$ is its differential entropy, $S_{\mathrm{bulk}}(\rho_R)$ is the average over sectors of the entropy of the state within $\Ha_R(\pi)$, and $\hat{A}$ is a state independent operator in the center of the subregion algebra whose eigenvalue on the $\pi$ sector is $\ln(\mu(\pi))$, the ratio between the Plancherel measure and a second, ``microcanonical'' measure on $\widehat{G}$ which we will define. This last term is universal: it depends only on $G$ and on the boundary condition at the cut, and not on the tessellation or on the number of marked points, and it is what plays the role of the area term in the holographic entropy formula. We will state each of these pieces precisely once the factorization map and the renormalized traces have been set up, in Sec.~\ref{sec:entropy_nomatter}.

\subsection{Outline}

The rest of the paper is organized as follows. 
In Sec.~\ref{sec:themodel}, we review the construction of topological tensor networks. 
In Sec.~\ref{sec:factorization}, we explain how to factorize the Hilbert space. 
In Sec.~\ref{sec:algebras}, we compare the Hilbert space factorization approach to the operator algebra approach, and conclude they are equivalent.
In Sec.~\ref{sec:entropy_nomatter}, we compute the entropy of a state, and find that it matches an analogue of the holographic entropy formula in gravity for boundary anchored subregions without matter.
In Sec.~\ref{sec:topEE}, we explain how this entropy formula is related to ``topological entanglement entropy'' in condensed matter physics, and give a brief account of how our results may generalize to three-dimensional gravity. 
We conclude in Sec.~\ref{sec:discussion} with a discussion of our results.
Additionally, in Appendix~\ref{apx:topvscon}, we explain the boundary conditions that we impose to define the Hilbert space on a spatial surface-with-boundary, and in Appendix~\ref{app:microcanonical}, we include some details about how to define a preferred measure over the edge modes that appear in the factorization map.

\section{Topological tensor networks}\label{sec:themodel}

Below we review the construction of the Hilbert space of a topological tensor network for arbitrary transformable Lie groups $G$ \cite{Balasubramanian:2025rcr}. When $G$ is a finite group these structures are sometimes called Kitaev's double models; also see \cite{Akers:2024wab,Kitaev:1997wr,Levin_2005}.  The case  $G = \SL(2,\R)$ has a close relation with AdS gravity in 2+1 dimensions. 

\subsection{The physical Hilbert space}

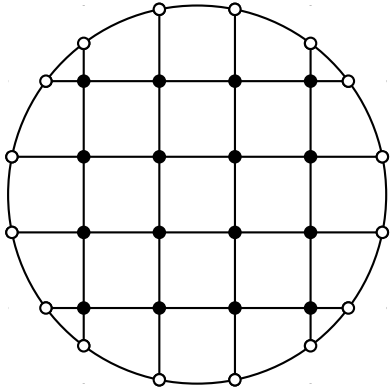
\begin{figure}
    \centering
    \begin{tikzpicture}[thick,scale=5]
        \def\nn{4}
        \foreach\xx in {1,...,\nn}{
            \draw ({\xx/(\nn+1)},0) -- ({\xx/(\nn+1)},1);
            \draw (0,{\xx/(\nn+1)}) -- (1,{\xx/(\nn+1)});

            \draw ({\xx/(\nn+1)},{0.5+sqrt(\xx/(\nn+1))*sqrt(1-\xx/(\nn+1) )}) circle (0.015);
            \draw ({\xx/(\nn+1)},{0.5-sqrt(\xx/(\nn+1))*sqrt(1-\xx/(\nn+1) )}) circle (0.015);
            \draw({0.5+sqrt(\xx/(\nn+1))*sqrt(1-\xx/(\nn+1) )},{\xx/(\nn+1)}) circle (0.015);
            \draw ({0.5-sqrt(\xx/(\nn+1))*sqrt(1-\xx/(\nn+1) )},{\xx/(\nn+1)}) circle (0.015);
            \foreach\yy in {1,...,\nn}{
                \filldraw ({\xx/(\nn+1)},{\yy/(\nn+1)}) circle (0.015);
        }}
        \fill[white,even odd rule]  (0.5,0.5) circle (0.5)
        (0,0)--(1,0)--(1,1)--(0,1)--cycle;
        \draw (0.5,0.5) circle (0.5);
        \foreach\xx in {1,...,\nn}{
            \filldraw[fill=white] ({\xx/(\nn+1)},{0.5+sqrt(\xx/(\nn+1))*sqrt(1-\xx/(\nn+1) )}) circle (0.015);
            \filldraw[fill=white] ({\xx/(\nn+1)},{0.5-sqrt(\xx/(\nn+1))*sqrt(1-\xx/(\nn+1) )}) circle (0.015);
            \filldraw[fill=white] ({0.5+sqrt(\xx/(\nn+1))*sqrt(1-\xx/(\nn+1) )},{\xx/(\nn+1)}) circle (0.015);
            \filldraw[fill=white] ({0.5-sqrt(\xx/(\nn+1))*sqrt(1-\xx/(\nn+1) )},{\xx/(\nn+1)}) circle (0.015);
        }
    \end{tikzpicture}
    \caption{A lattice $\Lambda$ tessellating the disk $\Sigma$. The bulk vertices are in black, and the boundary vertices are in white. For a topological tensor network, we should take all the edges to be oriented.
    }
    \label{fig:tensornetwork}
\end{figure}

Topological tensor networks are so named because they satisfy  constraints defining gauge invariant states of an associated topological field theory. The physical Hilbert space of such networks is constructed as follows. First, we define a kinematic Hilbert space $\Ha(\Lambda)$ associated with a tessellation $\Lambda$ of space.   $\Ha(\Lambda)$ labels a basis of states before the constraints are imposed. Then, we define an algebra of operators on the kinematic Hilbert space and use these operators to construct the constraint equations.  Finally, we impose the constraints, thus constructing the physical Hilbert space.  We will describe these steps below.

First, consider an orientable spatial surface $\Sigma_{g,n}$ with genus $g$ and $n$ boundary components. We will mostly consider the disk $\Sigma_{0,1}$ but generalizing  to other surfaces is straightforward. We  approximate this surface by a {\it tessellation} $\Lambda = (V,E,P)$, with vertices $V$, oriented edges $E$, and plaquettes $P$. {\it Boundary vertices} are vertices embedded on boundary components, and {\it boundary legs} are edges that intersect a boundary vertex. See Fig.~\ref{fig:tensornetwork} for an illustration. 

\paragraph{Boundary conditions.} \label{sec:bcs}

As with any quantum system, the physical Hilbert space $\Ha_{\mathrm{phys}}(\Sigma)$ depends on the conditions we impose at the boundary components of $\Sigma$. In this paper we will consider states prepared with what we will call \emph{open} boundary conditions \cite{Cirac:2011oss,Cong:2017ffh,Cheipesh:2018imk}.\footnote{In the language of Appendix~\ref{apx:topvscon}, open boundary conditions with $n$ marked points are a boundary on which $n$ segments of rough boundary condition alternate with $n$ segments of smooth boundary condition, with the rough segments shrunk to the lattice scale. Line operators end on the rough segments, and the smooth arcs between them carry no degrees of freedom of their own.} In the tensor network, they are specified by choosing $n$ marked points on each boundary component of $\Sigma$, taking a tessellation $\Lambda$ whose boundary vertices sit at the marked points, and imposing the constraints of the theory (defined below) everywhere except at those vertices: the electric constraint is dropped at the $n$ boundary vertices and nowhere else, while the magnetic constraint is imposed at every plaquette of $\Lambda$. We denote the resulting Hilbert space by $\Ha_{\mathrm{phys}}(\Sigma^{(n)})$ when we wish to emphasize the dependence on $n$. In the continuum, open boundary conditions are the statement that there are exactly $n$ points on the boundary at which a line operator of the theory may end.\footnote{We thank Chris Akers and Juan Maldacena for helpful discussions about this point.}

A boundary component of $\Sigma$ could instead be given a \emph{conformal} boundary condition, in which the boundary hosts a full two dimensional CFT \cite{Kong:2019byq,Kong:2019cuu}. This is the boundary condition relevant to three-dimensional gravity with asymptotically AdS boundary conditions \cite{Brown1986}. The two are related: if we take open boundary conditions, tune a boundary Hamiltonian at the marked points to a critical point, and send $n \to \infty$, the boundary becomes gapless and line operators can end at any point of the boundary. At the same time, because the boundary Hamiltonian has been tuned to criticality, a gapless CFT will emerge there, which is a conformal boundary condition. 

The reason we can nonetheless work at finite $n$ is the following. The entropy of a boundary subregion will turn out to contain terms which are extensive in the number of marked points that the subregion contains, and these terms depend  on the boundary conditions, so they are not universal. But it will also contain a term which is independent of $n$ altogether. This term is fixed by the gauge group and by the boundary conditions we impose near the cut which factorizes the Hilbert space, and it is unaffected by the $n\to\infty$ limit. This is the contribution to the entropy we will be interested in, and its independence of $n$ is what allows us to compute it at finite $n$. We review the two families of boundary conditions in more detail in Appendix~\ref{apx:topvscon}.

\paragraph{Kinematic Hilbert space.} To define the {\it kinematic Hilbert space} of a  network defined by a tessellation $\Lambda$  we associate a Hilbert space of square integrable wave functions on a gauge group, i.e., $L^2(G)$, to each edge $\ell \in E$.
Then the total kinematic Hilbert space is  
\begin{align}
    \Ha(\Lambda) = \bigotimes_{\ell \in E} L^2(G) \,.
\end{align}
Letting $\widehat{G}$ denote the space of  unitary irreducible representations $V_\pi$ of $G$,  also called the unitary dual of $G$, we can expand $L^2(G)$ as
\begin{align}
    L^2(G) = \int^{\oplus}_{\widehat{G}} d\mu(\pi) \, V_\pi \otimes V_\pi^* \,.
    \label{eq:peterweylnoncompact}
\end{align}
Here, $d\mu(\pi)$ is the Plancherel measure and $V_\pi^*$ is the dual space of $V_\pi$. When $G$ is compact, this is called the Peter--Weyl theorem and the integral \eqref{eq:peterweylnoncompact} reduces to a discrete sum.   Geometrically,  for compact $G$ the unitary dual $\widehat{G}$ is a collection of points, so the integral in \eqref{eq:peterweylnoncompact} becomes a sum over those points with the Plancherel measure giving weights proportional to the dimension of $V_\pi$.  For example, for $\SU(2)$ $\pi = j$  labels  half integer spin quantum numbers,  $V_\pi = \C^{2j+1}$ are the associated vector spaces on which the spin $j$ representation acts linearly, and the weight induced by the Plancherel measure is $2j + 1$. When $G$ is non-compact, $\widehat{G}$ generally contains subspaces of various dimensions including points and lines. 

One basis of states for $L^2(G)$ is $\ket{g}$, i.e., delta functions of group elements which satisfy $\braket{g}{h} = \delta(g^{-1}h)$. Using \eqref{eq:peterweylnoncompact}, there is another basis for $L^2(G)$ written as $\ket{\pi,ab}$, labeled by the unitary irreducible representation $\pi$ and  indices $a$ and $b$ for $V_\pi, V_\pi^*$, respectively. These states have overlap
\begin{align}
    \braket{\pi,ab}{\omega,mn} = \delta(\pi,\omega) \delta_{am}\delta_{bn}\,, \label{eq:repbasisoverlap}
\end{align}
where $\delta(\pi,\omega)$ is the delta function with respect to the Plancherel measure. The overlap between these bases is 
\begin{align}
    \braket{g}{\pi,ab} = \pi(g)_{ab}
\end{align}
where $\pi(g)_{ab}$ is the $(a,b)$ matrix element of $\pi(g)$, the matrix representation of $g$ in the $\pi$ representation. This expression can be thought of as a generalization of the Fourier transform to arbitrary semisimple Lie groups, and the Plancherel measure uniquely ensures that the transform between the group basis and the representation basis is unitary.

Note that  $d\mu(\pi)$ is \emph{not} uniform over $\widehat{G}$ when $G$ is non-abelian. For example, when $G$ is compact, the Plancherel measure assigns a weight $\mu(\pi) = d_\pi / \mathrm{Vol}(G)$ to the $\pi$ representation, i.e., the dimension of the representation divided by the volume of the gauge group computed in the Haar measure. When $G$ is non-compact, the Plancherel measure assigns a finite measure to any compact subset of $\widehat{G}$, but is not simply related to the dimension of these representations. Indeed, any non-trivial unitary representation of a non-compact, semisimple Lie group is infinite dimensional. Nevertheless, this weight still exists, is finite, and is intrinsic to the group $G$.   It will play a crucial role in what follows. The Plancherel measure is much more complex to calculate explicitly for non-compact groups than for compact ones, but is in principle determined; see for example the famous work of Harish-Chandra on semisimple groups \cite{HarishChandra1952,HarishChandra1954complex,HarishChandra1976}. Fortunately, for most of what follows  we  only need general properties of the measure, and not explicit formulae.

Now that we have described the kinematic Hilbert space, we will construct operators that act on it, and then write the physical constraints in terms of these operators.

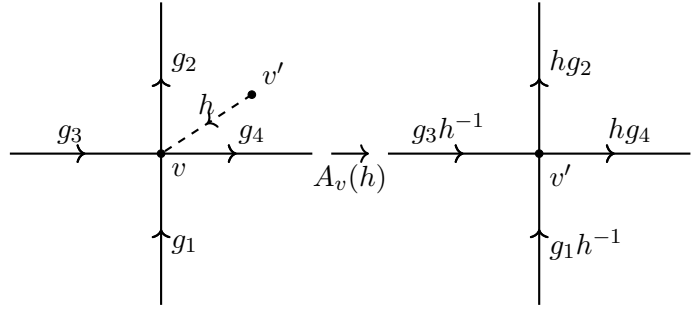
\begin{figure}
    \centering
    \begin{tikzpicture}
        \def\sep{0.5}
        \def\xx{0.6}
        \node at (-2-\sep,0) {\begin{tikzpicture}[scale=2]
                \draw[thick,->-=0.25,->-=0.75] (0,-1) -- (0,1);
                \draw[thick,->-=0.25,->-=0.75]  (-1,0) -- (1,0);
                \filldraw (0,0) circle (0.025);
                \node[anchor = north west] at (0,0) {$v$};
                \node[anchor=west] at (0,-\xx) {$g_1$};
                \node[anchor=west] at (0,\xx) {$g_2$};
                \node[anchor=south] at (-\xx,0) {$g_3$};
                \node[anchor=south] at (\xx,0) {$g_4$};
                \draw[thick,->-=0.5,dashed]  (\xx,\xx*0.65) -- (0,0);
                \filldraw (\xx,\xx*0.65) circle (0.025) node[anchor=south west] {$v'$};
                \node[anchor=south] at (\xx/2,\xx*0.65/2) {$h$};
        \end{tikzpicture}};
        \draw[thick,->] (-\sep/2,0) -- (\sep/2,0);
        \node[anchor=north] at (0,0) {$A_v(h)$};
        \node at (2+\sep,0) {\begin{tikzpicture}[scale=2]
                \draw[thick,->-=0.25,->-=0.75] (0,-1) -- (0,1);
                \draw[thick,->-=0.25,->-=0.75]  (-1,0) -- (1,0);
                \filldraw (0,0) circle (0.025);
                \node[anchor = north west] at (0,0) {$v'$};
                \node[anchor=west] at (0,-\xx) {$g_1 h^{-1}$};
                \node[anchor=west] at (0,\xx) {$hg_2$};
                \node[anchor=south] at (-\xx,0) {$g_3 h^{-1}$};
                \node[anchor=south] at (\xx,0) {$hg_4$};
        \end{tikzpicture}};
    \end{tikzpicture}
    \caption{The action of $A_v(h)$ on a state $\ket{g_1, g_2, g_3, g_4}$. $A_v(h)$ acts trivially on any edge which does not end at the vertex $v$. }
    \label{fig:Adef}
\end{figure}

\paragraph{Electric and magnetic operators.}
{\it Electric operators} act at  vertices of $\Lambda$.
For every group element $h \in G$ and $n$-valent vertex $v \in V$, there is an electric operator $A_v(h)$ which acts on an element of the group basis for tensor network states $\ket{g_1,\cdots, g_n}$ as in Fig.~\ref{fig:Adef}. These operators enact gauge transformations locally at the vertex $v$. To understand why this is a gauge transformation, let $v'$ be another vertex such that the holonomy from $v$ to $v'$ is given by $h$. If we had chosen to measure the holonomies of the gauge field using Wilson lines anchored at $v'$ instead of $v$, then we would change our coordinates on the space of holonomies precisely as in Fig.~\ref{fig:Adef}. So we can interpret gauge transformations as translations of the vertices $v$ on the surface $\Sigma$. Gauge invariance, then, imposes that the locations of the vertices defining the tensor network are arbitrary. 

\begin{figure}
    \centering
    \begin{tikzpicture}[scale=4]
        \def\rr{0.15}
        \def\vv{0.05}
        \def\eps{0.2}

        \draw[thick,->-=0.5] (-\eps,0) -- (1+\eps,0);
        \draw[thick,->-=0.5] (1,-\eps) -- (1,1+\eps);
        \draw[thick,->-=0.5] (-\eps,1) -- (1+\eps,1);
        \draw[thick,->-=0.5] (0,-\eps) -- (0,1+\eps);

        \filldraw (0,0) circle (0.01);
        \filldraw (1,0) circle (0.01);
        \filldraw (0,1) circle (0.01);
        \filldraw (1,1) circle (0.01);
        \node[anchor=north west] at (0,0) {$v$};
        \node[anchor=north] at (0.5,0) {$h_1$};
        \node[anchor=west] at (1,0.5) {$h_2$};
        \node[anchor=south] at (0.5,1) {$h_3$};
        \node[anchor=east] at (0,0.5) {$h_4$};

        \draw[thick] (0.5,{0.5-\rr}) arc (-90:180:\rr)
        ({0.5-\rr - \vv},{0.5 + \vv}) --  ({0.5-\rr},{0.5}) --  ({0.5-\rr + \vv},{0.5 + \vv}); 
        \node at (0.5,0.5) {$p$};
        \node[anchor=north] at (0.5,{0.5-\rr}) {$g_{(v,p)} = h_1 h_2 h_3^{-1} h_4^{-1}$};
    \end{tikzpicture}
    \caption{An example of a plaquette that $B_{(v,p)}(h)$ acts on. }
    \label{fig:Bdef}
\end{figure}
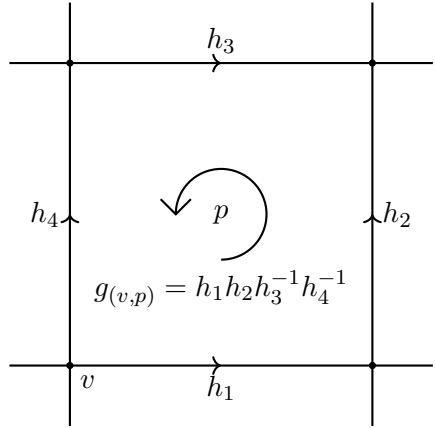

{\it Magnetic operators} act on the plaquettes of $\Lambda$. For every group element $h \in G$ and plaquette $p$ (and a vertex $v \in \partial p$), there is a magnetic operator $B_{(v,p)}(h)$ which acts as 
\begin{equation}
    B_{(v,p)}(h)\ket{g_1, \cdots, g_n} = \delta(h^{-1} g_{(v,p)}) \ket{g_1, \cdots, g_n}. 
\end{equation}
Here, $g_{(v,p)}$ is defined as in Fig.~\ref{fig:Bdef} and $\ket{g_1,\cdots, g_n}$ is an element of the group basis for tensor network states. These operators measure the flux through the plaquette $p$, defined by a line integral starting at the vertex $v$, and annihilate any state which does not have flux $h$. The operator $B_{(v,p)}(h)$ measures the difference in holonomy between two paths $\gamma, \gamma'$ which start at the vertex $v \in \partial p$ and end at a different vertex $v' \in \partial p$, such that $\gamma \sqcup \gamma' = \partial p$. This is the usual story in gauge theories: non-trivial curvature in the connection leads to path dependence for Wilson lines. Later, we will impose a flatness constraint on our tensor networks, which amounts to taking tensor network states in the image of $B_{(v,p)}(e)$ (where $e$ is the identity) for each plaquette $p$. For such states, the particular choice of edges embedded within $\Sigma$ are physically irrelevant. In other words, if we act on $\Sigma$ by a diffeomorphism which moves the edges of the lattice on the surface, a state which satisfies the flatness constraint will remain invariant. Together with the electric constraint, this implies that tensor network states that satisfy both the constraint equations are invariant under small diffeomorphisms of the surface $\Sigma$, i.e., diffeomorphisms which are continuously connected to the identity map of the surface to itself.

\paragraph{Constraint operators.}  We now have to impose two kinds of physical constraints on the kinematic Hilbert space.  The first is the {\it electric constraint} that says that the tensor network must represent a gauge invariant state.  The second is the {\it magnetic constraint} that requires that the holonomy around each plaquette $p$ be trivial. The electric constraint is required because a diffeomorphism invariant tensor network should be independent of the coordinate values of the vertices $v$ which define the lattice. The magnetic constraint is required because a diffeomorphism can deform an edge $\gamma$ to another homotopic edge $\gamma'$, and the difference between states defined on the lattice with the $\gamma$ edge and the $\gamma'$ edge is precisely the flux between the plaquette bounded by $\gamma \sqcup\gamma'$. So states satisfying the flatness constraint are invariant under such deformations. We will use the electric and magnetic operators defined above to impose these constraints.

To impose the {\it electric constraint} saying that the tensor network is invariant under gauge transformations, at each vertex
$v$ we  define the ``projector''
\begin{align}
    \Pi_A^{(v)} = \int dg \, A_v(g) \,. \label{eq:PiAdef}
\end{align}
This ``projector'' is Hermitian for any semisimple Lie group,\footnote{More generally, it is Hermitian for any unimodular group \cite{Haar,Dixmier1977}.} and because of the left invariance of the Haar measure, 
\begin{align}
    A_v(g) \Pi_A^{(v)} = \Pi_A^{(v)} \,. \label{eq:gaugeinv}
\end{align}
So states in the image of $\Pi_A^{(v)}$ are invariant under gauge transformations at the vertex $v$. We are placing ``projector'' in quotes  because $\Pi_A^{(v)}$ does not square to itself. Indeed, integrating both sides of \eqref{eq:gaugeinv} with respect to $g$, we can use \eqref{eq:PiAdef} to see that $(\Pi_A^{(v)})^2$ diverges as the volume of the gauge group. However, matrix elements of $\Pi_A^{(v)}$ between compactly supported test functions remain finite, so it is well-defined on this dense subspace of the kinematic Hilbert space \cite{Balasubramanian:2025rcr}.

To impose the  {\it magnetic constraint} saying that  the holonomy around a particular plaquette $p$ should vanish, we define a  projector for a plaquette $p$ as 
\begin{align}
    \Pi_B^{(p)} = B_{(v,p)}(e)\,, \label{eq:PiBdef}
\end{align}
where $e$ is the identity element of the gauge group. This operator has the property that
\begin{align}
    B_{(v,p)}(h) \Pi^{(p)}_B = \delta(h) \Pi^{(p)}_B\,, \label{eq:flateq}
\end{align}
where $\delta(h)$ is the delta function for the group.
So states in the image of $\Pi_B^{(p)}$ have vanishing holonomy around plaquette $p$ independently of the associated vertex $v$. Like the electric operator, $\Pi_B^{(p)}$ is Hermitian. But setting $h=e$ within \eqref{eq:flateq}, and using the definition \eqref{eq:PiBdef}, we see that $(\Pi_B^{(p)})^2$ diverges as $\delta(e)$. So $\Pi_B^{(p)}$ is not a conventional projector for continuous groups. However, bounded, compactly supported functions have finite matrix elements with $\Pi_B^{(p)}$, so it is well-defined on a dense subspace of the kinematic Hilbert space \cite{Balasubramanian:2025rcr}. 

We would like to impose the electric constraint at every vertex and the magnetic constraint at every plaquette.  Consistency requires that $[\Pi_A^{(v)},\Pi_A^{(v')}] = [\Pi_B^{(p)},\Pi_B^{(p')}] = [\Pi_A^{(v)},\Pi_B^{(p)}] = 0$ for any choice of vertices and plaquettes. These conditions follow  immediately from the definitions. Thus, the complete constraint operator
\begin{align}
    \Pi = \prod_{v \in V_{\mathrm{bulk}}} \Pi_A^{(v)} \prod_{p \in P} \Pi_B^{(p)} \, ,
\end{align}
which annihilates unphysical states in the kinematic Hilbert space, is well-defined.   Crucially, we do \emph{not} impose the electric constraint at the vertices on the boundary of the tensor network, or the magnetic constraint outside of the disk. This is the open boundary condition defined in Sec.~\ref{sec:bcs}. Again, $\Pi$ diverges if we try to square it for the same reasons as explained above for the individual electric and magnetic operators, so it is not a strictly speaking projector. But given two bounded, compactly supported wave functions in the kinematic Hilbert space, $\Pi$ has finite matrix elements \cite{Balasubramanian:2025rcr}.

\paragraph{Physical Hilbert space and improved (co-invariant) inner product.} We will define the physical Hilbert space by using the constraint operator $\Pi$ to remove unphysical parts of the kinematic Hilbert space. To this end, let $\Ha_{\mathrm{null}}$ be the vector subspace of $\Ha(\Lambda)$ in the kernel of $\Pi$.   Then we can define the pre-Hilbert space of equivalence classes
\begin{align}
    \dket{\psi} \equiv [\ket{\psi} \sim \ket{\psi} + \ket{\chi}] & \, \text{ for all }  \ket{\psi} \in \Ha(\Lambda) \,\text{ and } \, \ket{\chi} \in \Ha_{\mathrm{null}}\,. \label{eq:electricnullstates}
\end{align}
The idea is that  each physical state is defined by such an  equivalence class $\dket{\psi}$ in the kinematic Hilbert space whose elements differ by elements of the kernel of $\Pi$.
We next define an improved inner product, often called the {\it co-invariant inner product},
\begin{equation}
    \dbraket{\psi}{\sigma} := \bra{\psi} \Pi \ket{\sigma}
    \label{eq:ImpInnerProduct}
\end{equation}
under which the equivalence classes in \eqref{eq:electricnullstates} become an inner product space without any null states. This construction is well-defined for  bounded, compactly supported wave functions in the kinematic Hilbert space, between which $\Pi$ has finite matrix elements. Taking limits, we can complete this space of functions with respect to the inner product \eqref{eq:ImpInnerProduct} to get a Hilbert space. This is the physical Hilbert space $\Ha_{\mathrm{phys}}(\Sigma)$.  We write the physical Hilbert space as a function of the spatial surface $\Sigma$, not of the particular tessellation $\Lambda$, because, as we will discuss in Sec.~\ref{sec:LatticeIndep}, it is an invariant of the surface itself.\footnote{More precisely, $\Ha_{\mathrm{phys}}(\Sigma)$ depends on $\Sigma$ together with its boundary conditions, which for the open boundary conditions of Sec.~\ref{sec:bcs} includes the number $n$ of marked points. Tessellations with different numbers of boundary vertices therefore construct different Hilbert spaces, $\Ha_{\mathrm{phys}}(\Sigma^{(n)})$ for different $n$, in the same way that a field theory on a manifold with boundary has a different Hilbert space for each boundary condition. What is independent of the tessellation is the Hilbert space at fixed $n$. We will see in Sec.~\ref{sec:entropy_nomatter} that $n$ enters the entropy of a boundary subregion only through its non-universal part.}   This is a form of background independence.

Notice that we did {\it not} define the inner product by first projecting states in the kinematic Hilbert space as $\ket{\psi} \to \Pi \ket{\psi}$  and then applying the standard inner product to define $\bra{\psi}\Pi^\dagger \Pi \ket{\psi}$.  The latter expression is not well-defined because, as we explained above, the square of $\Pi$ diverges for groups which are continuous and/or non-compact. The equivalence class construction \eqref{eq:electricnullstates} equipped with the improved inner product \eqref{eq:ImpInnerProduct} circumvents this issue while nevertheless projecting out components of the kinematic Hilbert space that do not satisfy the physical constraints.  For finite groups, the construction described here has the same effect as simply acting with the physical projector on kinematic states and then employing the standard inner product.  

\subsection{Lattice independence}
\label{sec:LatticeIndep}
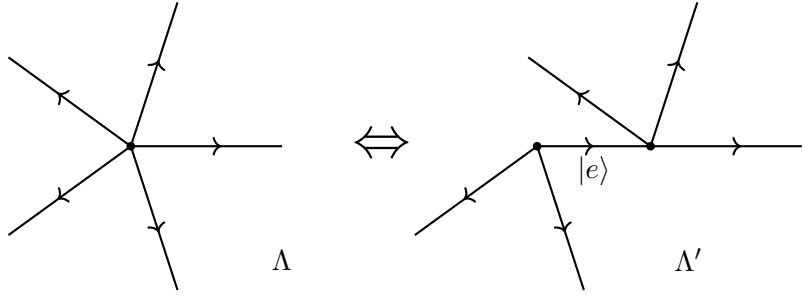
\begin{figure}
    \centering
    \begin{tikzpicture}
        \def\sep{3}
        \def\nn{5}
        \def\scale{2}
        \def\hh{0.75}
        \node at (-\sep,0) {\begin{tikzpicture}[scale=\scale]

                \filldraw (0,0) circle (0.025);
                \foreach \xx in {1,...,\nn}{
                    \draw[thick,->-=0.6] (0,0) -- ({cos(360*\xx/\nn)},{sin(360*\xx/\nn)});
                }
                \node at (1,-0.75) {$\Lambda$};
        \end{tikzpicture}};
        \node[scale=2] at (0,0) {$\Leftrightarrow $};
        \node at (\sep,0) {\begin{tikzpicture}[scale=\scale]
                \filldraw (0,0) circle (0.025);
                \filldraw (\hh,0) circle (0.025);
                \foreach \xx in {1,...,3}{
                    \draw[thick,->-=0.6] (\hh,0) -- ({\hh+cos(360*(\xx-1)/\nn)},{sin(360*(\xx-1)/\nn)});
                }
                \foreach \xx in {4,...,\nn}{
                    \draw[thick,->-=0.6] (0,0) -- ({cos(360*(\xx-1)/\nn)},{sin(360*(\xx-1)/\nn)});
                }
                \draw[thick,->-=0.5] (0,0) -- (\hh,0);
                \node[anchor=north] at (\hh/2,0) {$\ket{e}$};
                \node at (1,-0.75) {$\Lambda'$};
        \end{tikzpicture}};

    \end{tikzpicture}
    \caption{An example of move 1, which can be performed to add or remove a bulk vertex/leg from the graph $\Lambda$. In other words, we can split or combine bulk vertices. 
    }
    \label{fig:move1}
\end{figure}

\begin{figure}
    \centering
    \begin{tikzpicture}
        \def\sep{4}
        \def\nn{5}
        \def\rr{1.3}
        \def\scale{2}
        \def\hh{0.75}
        \node at (-\sep,0) {\begin{tikzpicture}[scale=\scale]

                \foreach \xx in {1,...,\nn}{
                    \filldraw ({cos(360*(\xx-1)/\nn)},{sin(360*(\xx-1)/\nn)}) circle (0.025);
                    \draw[thick,->-=0.6] ({cos(360*(\xx-1)/\nn)},{sin(360*(\xx-1)/\nn)}) -- ({\rr*cos(360*(\xx-1)/\nn)},{\rr*sin(360*(\xx-1)/\nn)});
                    \draw[thick,->-=0.6] ({cos(360*(\xx-1)/\nn)},{sin(360*(\xx-1)/\nn)}) -- ({cos(360*\xx/\nn)},{sin(360*\xx/\nn)});
                }
                \node at (1,-1) {$\Lambda$};
        \end{tikzpicture}};
        \node[scale=2] at (0,0) {$\Leftrightarrow $};
        \node at (\sep,0)  {\begin{tikzpicture}[scale=\scale]

                \draw[thick,->-=0.5] (1,0) -- ({cos(360*(2)/\nn)},{sin(360*(2)/\nn)});

                \foreach \xx in {1,...,\nn}{
                    \filldraw ({cos(360*(\xx-1)/\nn)},{sin(360*(\xx-1)/\nn)}) circle (0.025);
                    \draw[thick,->-=0.6] ({cos(360*(\xx-1)/\nn)},{sin(360*(\xx-1)/\nn)}) -- ({\rr*cos(360*(\xx-1)/\nn)},{\rr*sin(360*(\xx-1)/\nn)});
                    \draw[thick,->-=0.6] ({cos(360*(\xx-1)/\nn)},{sin(360*(\xx-1)/\nn)}) -- ({cos(360*\xx/\nn)},{sin(360*\xx/\nn)});
                }
                \node[anchor=north] at ({0.5+cos(360*(2)/\nn)/2},{sin(360*(2)/\nn)/2}) {$\ket{1}$};
                \node at (1,-1) {$\Lambda'$};
        \end{tikzpicture}};

    \end{tikzpicture}
    \caption{An example of move 2, which can add or remove a leg/plaquette from the graph $\Lambda$. In other words, we can split or fuse two bulk plaquettes together. 
    }
    \label{fig:move2}
\end{figure}
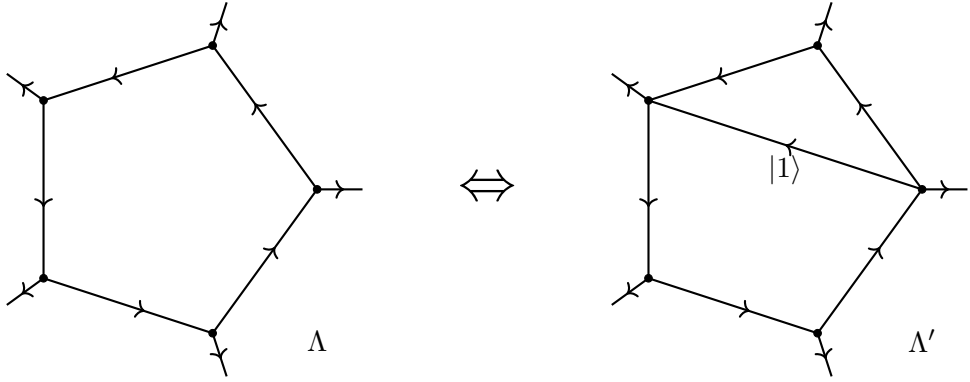

We want to show that the  physical Hilbert space as defined above is a topological invariant of the spatial surface $\Sigma$, i.e., it does not depend on the  tessellation $\Lambda$ (up to a choice of boundary vertices). We refer the reader to \cite{Balasubramanian:2025rcr,Akers:2024wab} for more details, but briefly review the proof in this subsection. We first define two families of graphical moves that transform between any two tessellations $\Lambda $ and $\Lambda'$ of a spatial surface of  $\Sigma$. These are shown in Fig.~\ref{fig:move1} and Fig.~\ref{fig:move2}.  We will refer to the physical Hilbert space constructed with the tessellation $\Lambda$ as $\Ha_{\mathrm{phys}}(\Lambda)$.

Associated with these graphical moves are unitary maps $ \Delta_{1,2}: \Ha_{\mathrm{phys}}(\Lambda) \to \Ha_{\mathrm{phys}}(\Lambda')$ which transform the physical Hilbert space from one presentation to another:
\begin{align}
    \Delta_1\dket{\Psi} & = [\ket{e} \ket{\Psi} \sim \ket{e} \ket{\Psi} + \ket{\chi}] \equiv \dket{e,\Psi} \,,\\
    \Delta_2 \dket{\Psi} & = \left[\left(\int dg \ket{g}\right) \ket{\Psi} \sim \left(\int dg \ket{g}\right) \ket{\Psi} + \ket{\chi}\right] \equiv \dket{1,\Psi}\,.
\end{align}
In these expressions, the states $\ket{e}$ and $\ket{1} = \int dg \ket{g}$ are associated with the new legs introduced by moves 1 and 2, respectively. The states $\ket{\chi}$ in the middle expressions are  null with respect to a constraint operator $\Pi'$ associated with the transformed tessellation $\Lambda'$. Although the original state $\dket{\Psi}$ and the auxiliary states $\ket{e}, \ket{1}$ introduced by the unitary maps $\Delta_1, \Delta_2$ may seem disentangled, they are actually inherently entangled in the physical Hilbert space because of the quotient by null states. 

One can show explicitly that $\Delta_1$ and $\Delta_2$ are unitary \cite{Balasubramanian:2025rcr} by computing the inner products of $\dbra{\Psi} \Delta_i^\dagger \Delta_i \dket{\Psi}$ and $\dnorm{\Psi}$, and seeing that they explicitly agree for both $i=1,2$. This proves that $\Delta_{1,2}$ are isometries from $\Ha_{\mathrm{phys}}(\Lambda)$ into $\Ha_{\mathrm{phys}}(\Lambda')$. To complete the proof that these maps are unitaries, we need to also show that $ \dbra{\Psi'} \Delta_i \Delta_i^\dagger \dket{\Psi'} = \dnorm{\Psi'}$. To see this, let $\ell$ be the leg that $\Delta^\dagger$ is poised to remove when passing from $\Lambda' \to \Lambda$. Consider a particular basis element $\dket{\Psi'} =\dket{g_\ell , \Psi}$ for $\Ha_{\mathrm{phys}}(\Lambda')$. Through a gauge transformation at either vertex connected to the edge $\ell$, we can rotate $g_\ell$ to become the identity element. At the same time, this gauge transformation will unitarily rotate $\Psi$ to some other state $\Psi_{g}$ because the gauge transformation also acts on the other legs attached to the same vertex. In other words, as vectors in $\Ha_{\mathrm{phys}}(\Lambda')$, $\dket{g_\ell , \Psi} = \dket{e , \Psi_{g}}$. We can think of $\dket{e , \Psi_{g}}$ as a partially gauge fixed version of the state $\dket{g_\ell , \Psi}$. Then it is clear that the map $\Delta_1^\dagger\dket{e , \Psi_{g}} = \dket{ \Psi_{g}}$ is also an isometry, as the vector on the right hand side is by definition a normalized state in $\Ha_{\mathrm{phys}}(\Lambda)$. A similar argument also holds for $\Delta_2$, where we instead invoke the flatness condition to ``gauge fix'' the support of the state on the $\ell$ leg to be $\ket{1}$, as defined above. Thus, the maps $\Delta_{1,2}$ are unitaries.

Finally, noting that the two graphical moves can be composed to map between any two tessellations of the spatial surface $\Sigma$,\footnote{One way to see this is that we can always use a sequence of these moves to go from $\Lambda$ to the reduced lattice (Fig.~\ref{fig:reduced}) of the spatial surface that $\Lambda$ tessellates. Composing with the inverse of a similar reduction of a different lattice $\Lambda'$ defines a move sequence from $\Lambda \to \Lambda'$.} and the associated composition of unitary maps $\Delta_1, \Delta_2$ is unitary, the physical Hilbert spaces associated with any two tessellations $\Lambda , \Lambda'$ are unitarily equivalent. Thus, the physical Hilbert space is independent of the tessellation, and is an invariant of the surface $\Sigma$. This justifies the notation $\Ha_{\mathrm{phys}}(\Sigma)$ we will continue to use for the rest of this paper.

\paragraph{Relation to Topological Field Theory.}
Because $\Ha_{\mathrm{phys}}(\Sigma)$ is a topological invariant of the spatial surface $\Sigma$, we can think of it as the Hilbert space of a topological quantum field theory (TQFT).  If we reverse the orientation of a leg $\ell \in E$ while simultaneously sending $g_\ell \to g_\ell^{-1}$, then the states in the physical Hilbert space are unchanged. Thus, if we think of this operation as a kind of parity transformation, $\Ha_{\mathrm{phys}}(\Sigma)$ must describe a parity invariant theory \cite{Levin_2005}. 

When $G$ is a finite group, these TQFTs are known: they are the  Dijkgraaf--Witten theories \cite{Dijkgraaf:1989pz,Freed:1991bn} associated with $G$ \cite{Hu:2012wx}. 
In this setting, topological tensor networks are also known as Kitaev's double model \cite{Kitaev:1997wr}. 
The gravity-like properties of the double model were studied in the interesting paper \cite{Akers:2024wab}.

When the gauge group on each leg is instead taken to be the compact quantum group $G_k$,\footnote{More precisely, we label each leg by an integrable representation of $G_k$, i.e., the representations of Wilson lines in the spectrum of $G$ Chern--Simons theory at level $k$.} then the corresponding TQFT is the doubled $G_k \times G_{-k}$ Chern--Simons theory \cite{Levin_2005,kirillov2011stringnet}. Indeed, $G_k$ Chern--Simons theory is not parity invariant (it maps to $G_{-k}$ Chern--Simons theory), so this is the minimal parity invariant TQFT with a $G_k$ gauge symmetry. In this context, topological tensor networks are also known as a Levin--Wen model, or a ``string net'', in the condensed matter community. 

In this paper we are working with general transformable groups, rather than finite groups or compact quantum groups.  To the best of our knowledge, the explicit TQFT associated with $\Ha_{\mathrm{phys}}(\Sigma)$ is not generally known for such groups. Note that in the limit of large level $k$, the representation theory of a quantum group $\mathrm{Rep}(G_k)$ limits to $\mathrm{Rep}(G)$, the category we are using to construct our model. 
Because the string net construction using $\mathrm{Rep}(G_k)$ is equivalent to $G_k \times G_{-k}$ Chern--Simons theory, it is natural to conjecture that the TQFTs described by our networks can be obtained as a large level limit of doubled $G_k \times G_{-k}$ Chern--Simons theories. To see some of the properties the TQFT associated with arbitrary transformable gauge groups may have, we will now explain the finite $G$ case in more detail, for the sake of comparison.

Dijkgraaf--Witten theory and compact $G_k$ Chern--Simons theory associate finite dimensional Hilbert spaces to any closed surface $\Sigma$, which means that the path integrals for these theories can often be reduced to finite sums over simpler building blocks. Physically, the Hilbert space of a closed surface is spanned by the action of closed line operators which wrap the non-contractible cycles of $\Sigma$ on a choice of vacuum state.\footnote{More precisely, this is true for any choice of cyclic and separating state \cite{Dixmier1977} for this algebra of line operators.} If we instead let these line operators ``pierce'' the surface $\Sigma$ in the path integral of the TQFT, then they represent trajectories of  fractionally charged particles called {\it anyons}. The finite dimensionality of the closed-surface Hilbert spaces (in particular, the torus Hilbert space \cite{Witten1989Jones}) of Dijkgraaf--Witten theory and compact $G_k$ Chern--Simons theory then implies that these theories have a finite number of anyon types. 

Because the path integral and canonical quantization (i.e., the collection of Hilbert spaces associated to $\Sigma$ and the possible anyonic line operators acting on these Hilbert spaces) are equivalent,
the TQFT itself can actually be uniquely identified with the allowed anyons of the theory. In modern language, the mathematical structure describing a consistent theory of finitely many anyons and their braiding/fusion rules is called a ``modular tensor category'' (MTC) \cite{Turaev:1994xb}, and conversely, one can explicitly construct a three-dimensional TQFT with finitely many anyon types from any MTC \cite{TuraevViro1992,ReshetikhinTuraev1991}. 

However, as we show in Sec.~\ref{sec:torus}, the Hilbert space of the torus constructed by our tensor networks is infinite dimensional, which implies that the TQFTs associated to our networks will contain infinitely many anyon types, which implies they cannot be constructed from any MTC. Physically, the conformal field theory (CFT) dual of a TQFT with infinitely many anyons will have infinitely many (extended) primaries, and therefore will not be a rational CFT \cite{Moore:1988qv}. Irrational CFTs are less well-understood than their rational counterparts, and so explicitly constructing their TQFT duals is not as straightforward as in the rational case.
Nevertheless, a key example of a TQFT which is dual to an irrational CFT (Liouville CFT) is Virasoro TQFT \cite{Collier_2023,Collier:2024mgv}, and our construction of $\Ha_{\mathrm{phys}}(\Sigma)$ suggests the existence of a wide new class of TQFTs which, similar to Virasoro TQFT, have infinitely many anyons. It would be interesting to construct these TQFTs more explicitly from first principles, perhaps along the lines of \cite{Hartman:2025cyj,Hartman:2025ula}, but we leave this for future work. 

\subsection{Adding matter}

For completeness, we now describe how to incorporate matter into a topological tensor network. We do so by adding new degrees of freedom to the tessellation $\Lambda$ in the form of additional {\it matter legs}. Operationally, in the kinematic Hilbert space a matter leg is an auxiliary Hilbert space $\Hmatt{}$ which supports the matter degrees of freedom. Having no matter leg is equivalent to setting $\Hmatt{} = \C$, so the matter-free case described above can be thought of as a special case. To dress the matter so that the constraint equations remain satisfied, each matter leg must support an action of the electric and magnetic operators. Therefore, rather than simply attaching each matter leg to a bulk vertex $v$ of our tensor network, we must also associate each matter leg with a network plaquette $p$, so that both the electric and  magnetic constraints can be imposed on the matter as well. Using the graphical moves 1 and 2 as needed, we can always associate each matter leg to distinct vertices and plaquettes, a convention we adopt from now on. So, let $A_v^{\mathrm{matt}}(h),B^{\mathrm{matt}}_{(v,p)}(g) $ describe the action of the electric and magnetic operators on the matter Hilbert space. These operators are part of the definition of the matter leg, similar to how the matrices $\pi(g)$ of the unitary $\pi$ representation of $G$ are the part of the definition of the group representation.
To incorporate matter legs into our theory we simply extend the formalism from above as follows:
\begin{align}
    \Ha(\Lambda) &= \bigotimes_{\ell \in E} L^2(G)_\ell \bigotimes_{(v,p) \in V \times P} \Hmatt{(v,p)} \,, \label{eq:HLambdamatter}\\
    A^{\mathrm{tot}}_v(h) &= A_v(h) A_v^{\mathrm{matt}}(h) \,,\\
    B^{\mathrm{tot}}_{(v,p)}(h) &= \int dg\, B_{(v,p)}(hg^{-1})B^{\mathrm{matt}}_{(v,p)}(g) \,.
\end{align}
The definitions of the bulk electric and magnetic operators $A_v(g)$ and $B_{(v,p)}(h)$ are the same as in the previous subsection. The multiplication of the electric operators and convolution of the magnetic operators is the correct combination to ensure that the resulting Hilbert space continues to describe a topological theory \cite{Akers:2024wab}. The constraint equations are then defined using the total electric and magnetic operators in the same formulas as \eqref{eq:PiAdef} and \eqref{eq:PiBdef}. Explicitly, the constraints  now take the form
\begin{align}
    \Pi_A^{(v)} &= \int dg \, A_v^{\mathrm{tot}}(g)\,, \\
    \Pi_B^{(p)} &= B^{\mathrm{tot}}_{(v,p)}(e)\,,\\
    \Pi &= \prod_{v \in V_{bulk}} \Pi_A^{(v)} \prod_{p \in P} \Pi_B^{(p)} \,.
\end{align}
We then quotient \eqref{eq:HLambdamatter} by the null states defined by $\Pi$ to obtain the physical Hilbert space.
Again, if there is no matter leg at a particular plaquette, then we can just set $\Hmatt{(v,p)} = \C$, $A_v^{\mathrm{matt}}(h)=1$, $B_{(v,p)}^{\mathrm{matt}}(h) = \delta(h)$, and then we recover the matter-free formalism above.  

For the rest of this paper, we will consider tensor networks without matter legs. We will consider the role of the matter legs in detail in \cite{Balasubramanian:2026xyz}. 

\subsection{Example: the torus} \label{sec:torus}

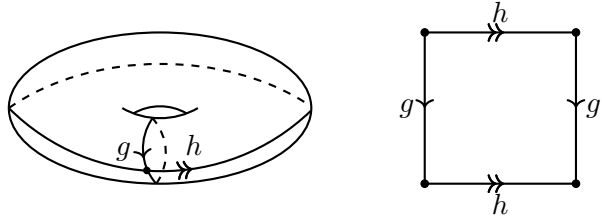
\begin{figure}
    \centering
    \begin{tikzpicture}

        \draw[thick] (.5,0) ellipse (2cm and 1cm);
        \draw[thick] (0.85,-.07) arc[start angle=60, end angle=120, radius=.7cm];
        \draw[thick] (0,0) arc[start angle=240, end angle=300, radius=1cm];

        \draw[thick,->-=0.6] (.4,-.135) arc[start angle=90+55, end angle=270-49, radius=.7cm];
        \draw[scale=-1,dashed,thick] (-.45,1) arc[start angle=90+55, end angle=270-49, radius=.7cm];

        \draw[thick,->-=0.6,->-=0.58] (-1.5,0) arc[start angle=180+45, end angle=360-46, radius=2.85cm];
        \draw[thick,dashed,yscale=0.7] (-1.5,0) arc[start angle=180-45, end angle=46, radius=2.85cm];
        \fill (.325,-0.825) circle (0.05);
        \node[anchor=south east] at (.25,-0.825)  {$g$};
        \node[anchor=south west] at (.7,-0.75)  {$h$};

        \node at (5,0) {\begin{tikzpicture}[scale=2]
                \draw[thick,->-=0.5] (0,1) -- (0,0) ;
                \draw[thick,->-=0.5] (1,1) -- (1,0);
                \draw[thick,->-=0.475,->-=0.525] (0,0) -- (1,0);
                \draw[thick,->-=0.475,->-=0.525] (0,1) -- (1,1);
                \filldraw (0,0) circle (0.025);
                \filldraw (1,0) circle (0.025);
                \filldraw (0,1) circle (0.025);
                \filldraw (1,1) circle (0.025);
                \node[anchor=east] at (0,0.5) {$g$};
                \node[anchor=west] at (1,0.5) {$g$};
                \node[anchor=north] at (0.5,0) {$h$};
                \node[anchor=south] at (0.5,1) {$h$};

        \end{tikzpicture}};

    \end{tikzpicture}
    \caption{The torus $T^2$, and its tessellation with one vertex, two edges, and one plaquette.}
    \label{fig:abcycle}
\end{figure}

As an example of how this procedure works, consider the torus $T^2$. To construct the physical Hilbert space $\Ha_{\mathrm{phys}}(T^2)$  we pick a   tessellation $\Lambda$ with one vertex, two edges, and one plaquette, as in Fig.~\ref{fig:abcycle}.  We can draw this tessellation on the torus represented as a square with opposite edges identified. The kinematic Hilbert space is $\Ha(\Lambda) = L^2(G) \otimes L^2(G)$, one group element for each non-contractible cycle of the torus. Because there is only one vertex and one plaquette, the electric and magnetic operators are  simple to write down:
\begin{align}
    A_v(k) \ket{g,h} & = \ket{kgk^{-1} , khk^{-1}}\,, \label{eq:torusAconstraint} \\
    B_{(v,p)}(k) \ket{g,h} & = \delta(k^{-1} ghg^{-1} h^{-1}) \ket{g,h} \,. \label{eq:torusBconstraint}
\end{align}
These are the electric and magnetic operators that we use to define the constraints in \eqref{eq:PiAdef} and \eqref{eq:PiBdef}.

From \eqref{eq:torusAconstraint}, the electric constraint identifies $\dket{g,h} \sim  \dket{kgk^{-1} , khk^{-1}}$ for every $k$. Thus, the physical Hilbert space only depends on the simultaneous conjugacy class of $g,h$. This ensures that the physical Hilbert space is independent of our choice of reference vertex $v$ to anchor the holonomies. Taken together, these constraints mean we can identify a particular state $\dket{g,h}$ with a homomorphism from the fundamental group $\pi_1(T^2) = \Z^2$ into the gauge group $G$,\footnote{In other words, for a choice of $a$,$b$ cycles of the torus, we have a pair of group elements $(g,h)$ which satisfy the same algebraic relations under composition as the $a,b$ cycles of the torus.} up to conjugation in $G$. In fact, this is precisely the same as the moduli space of flat $G$ connections on the torus, which is the phase space of  Chern--Simons theory of the group $G$ \cite{Witten1989Jones, Goldman1988Components}.
However, recall that a Hilbert space has support on \emph{half} the phase space of the associated classical theory. So the theory we have quantized to obtain the Hilbert space $\Ha(T^2)$ is actually not just $G$ Chern--Simons theory,\footnote{Recall that we have placed a copy of $L^2(G)$ at each leg, which by the Peter--Weyl theorem means that any irreducible representation in the support of the Plancherel measure is allowed. If we restricted the allowed representations to those which are integrable at level $k$, the Hilbert space we obtain would be that of a $G_k \times G_{-k}$ Chern--Simons theory \cite{Levin_2005}. Thus, because we are allowing arbitrary representations, our construction corresponds to the $k\to\infty$ limit of this doubled Chern--Simons theory. 
} but a theory whose phase space has twice the dimension of the Chern--Simons theory of group $G$. This is one way to think about the doubling of the gauge group of the theory the topological tensor networks describe.

Indeed, recall that the phase space of a particle propagating on a manifold $\mathcal{M}$ is the cotangent space $T^*\mathcal{M}$. Thus, the phase space of the theory we quantized to obtain $\Ha_{\mathrm{phys}}(\Sigma)$ was not the phase space $\mathcal{M}$ of $G$ Chern--Simons theory: it is the symplectic manifold $T^*\mathcal{M}$, because we are considering the Hilbert space of wave functions on $\mathcal{M}$. This is a subtle point, because the phase space of Chern--Simons theory is a symplectic manifold in its own right. We  are not quantizing this single-copy Chern--Simons phase space directly. Rather, the single-copy Chern--Simons phase space appears here as the configuration space for the theory we are actually interested in, and we are treating the flat $G$ connections as the position space $\mathcal{M}$. We will comment more about the continuum interpretation of this phase space in \cite{Balasubramanian:2026xyz}.

\subsection{Example: the disk} \label{sec:disk}

Next, we investigate the Hilbert space of the disk with open boundary conditions of Sec.~\ref{sec:bcs} (see Appendix~\ref{apx:topvscon} for more details about boundary conditions in this model). Specifically, we consider $n$ marked points on the boundary of the disk at which we do not impose the electric constraint. When constructing the physical Hilbert space of the disk, we can pick any tessellation that is convenient for our calculations, because as we discussed, the physical Hilbert space is independent of this choice even though the presentation of the state will depend on it, like any gauge choice. 

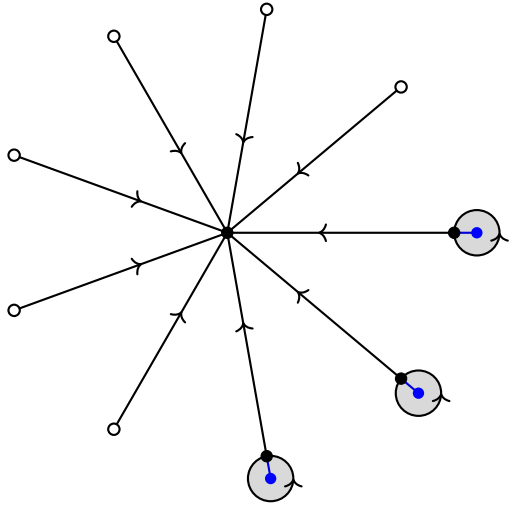
\begin{figure}
    \centering
    \begin{tikzpicture}[scale=3]
        \def\nn{6}
        \def\mm{3}
        \def\rr{1.1}

        \foreach\xx in {1,...,\nn}{
            \draw[thick,->-=0.6]  ({cos(360*(\xx)/(\nn+\mm))},{sin(360*(\xx)/(\nn+\mm))}) -- (0,0);
            \filldraw[thick,fill=white] ({cos(360*(\xx)/(\nn+\mm))},{sin(360*(\xx)/(\nn+\mm))}) circle (0.025);
        }
        \foreach\xx in {1,...,\mm}{
            \draw[thick,->-=0.6] ({cos(360*(\xx+\nn)/(\nn+\mm))},{sin(360*(\xx+\nn)/(\nn+\mm))}) -- (0,0);

            \draw[blue,thick] ({cos(360*(\xx+\nn)/(\nn+\mm))},{sin(360*(\xx+\nn)/(\nn+\mm))}) -- ({\rr*cos(360*(\xx+\nn)/(\nn+\mm))},{\rr*sin(360*(\xx+\nn)/(\nn+\mm))});

            \filldraw ({cos(360*(\xx+\nn)/(\nn+\mm))},{sin(360*(\xx+\nn)/(\nn+\mm))}) circle (0.025);

            \filldraw[thick,->-=0,fill opacity=0.15] ({\rr*cos(360*(\xx+\nn)/(\nn+\mm))},{\rr*sin(360*(\xx+\nn)/(\nn+\mm))}) circle ({\rr-1});

            \fill[blue] ({\rr*cos(360*(\xx+\nn)/(\nn+\mm))},{\rr*sin(360*(\xx+\nn)/(\nn+\mm))}) circle (0.025);
        }
        \filldraw (0,0) circle (0.025);
    \end{tikzpicture}
    \caption{Reduced lattice of the disk with six marked points on the boundary and three legs carrying matter. The boundary vertices are shown in white, and the out-of-plane legs that carry matter are shown in blue. The matter legs are attached to a bulk vertex and plaquette (shown in gray), which we call a ``lollipop''. This lollipop is connected to the central vertex of the reduced lattice through another bulk leg. }
    \label{fig:reduced}
\end{figure}

For many purposes, it is convenient to reduce the amount of redundancy in the state as much as possible. Because the Hilbert space has support on the edges of $\Lambda$, this corresponds to picking a lattice $\Lambda$ with the minimal possible number of legs.  For the disk with $n$ marked boundary points, the best we can do is $n$ boundary legs and no bulk legs. We can also include legs in the minimal lattice to carry  matter.  These legs must attach to a bulk vertex and plaquette to support an action of the electric and magnetic operators.  Fig.~\ref{fig:reduced} shows a minimal lattice for a disk with six marked points on the boundary and three ``lollipop factors'' in the terminology of \cite{Akers:2024wab,Balasubramanian:2025rcr}  supporting matter legs.   We will call this  minimal structure the {\it reduced lattice}.

Without matter, the physical Hilbert space is easy to read off from the structure of the reduced lattice. There are no plaquettes, so the magnetic constraint is trivially satisfied. There is a single bulk vertex, so there is a single electric constraint to impose. If we orient all the boundary legs inwards, as in Fig.~\ref{fig:reduced}, then we can  simplify the structure of the physical Hilbert space even further. Orientation reversal sends $g \to g^{-1}$ and so the Hilbert space is unaffected by this choice of orientation. 
Because of our convention that the electric operator $A_v(g)$ acts by left multiplication on outflowing legs, and as right multiplication on inflowing legs, we can associate the $V_\pi$ representations of \eqref{eq:peterweylnoncompact} with the outflowing vertex of an edge, and $V_\pi^*$ with the inflowing vertex. Thus, the electric constraint acts only on the $V_\pi^*$ subspace of each boundary leg, and leaves the data in the $V_\pi$ subspaces invariant. Using the compact notation $\vec{\pi} = \pi_1 \otimes \cdots \otimes \pi_n$, we can schematically think of the bulk Hilbert space as
\begin{align}
    \Ha_{\mathrm{phys}}\left(\Sigma_{0,1}^{(n)}\right) = \int_{\widehat{G}}^{\oplus} d\mu(\vec{\pi}) V_{\vec{\pi}} \otimes \Pi_A [ V_{\vec{\pi}}^*] \,. \label{eq:Hphys}
\end{align}
As explained in \cite{Balasubramanian:2025rcr,Akers:2024wab}, the subspace $\Pi_A [ V_{\vec{\pi}}^*]$ is essentially the space of intertwiners from the trivial representation to the $ V_{\vec{\pi}}$ representation. We say ``essentially'' because the trivial representation is non-normalizable with respect to the kinematic Hilbert space inner product, so that strictly speaking there is no fusion channel to the trivial representation. Said differently, the trivial representation is not in the support of the Plancherel measure. But these states are normalizable with respect to the physical inner product, and they are invariant under gauge transformations at the bulk vertex, so we can abuse notation and refer to the states in the $\Pi_A[V_{\vec{\pi}}^*]$ factor in this way without ambiguity. This subspace is completely determined from the group theory data. Therefore, all of the independent information about the states in the physical Hilbert space is located at the boundary vertices, in the form of the $V_\pi$ subspaces. In this sense, the topological tensor networks that we are developing are holographic on the disk. 

We can also gain some insight by considering an alternative tessellation of the disk besides the reduced lattice that we described above. Let $R$ denote a subset of boundary vertices, and $\overline{R}$ denote the complementary subset. Then with respect to this choice, we can define the ``bowtie lattice'' depicted in Fig.~\ref{fig:bowtie}. This is the same as the reduced lattice, but with a single additional bulk leg inserted to separate the $R$ vertices from the $\overline{R}$ vertices. We can think of this as a choice of channel for the intertwiners $\Pi_A [ V_{\vec{\pi}}^*] $ of the reduced lattice to fuse from the $R$ vertices to the $\overline{R}$ vertices. It will be useful to define the auxiliary Hilbert space\footnote{Below, we will refer to $V^*_\omega$ as a Hilbert space of ``gauge fixing edge modes''.}
\begin{align}
    \Ha_R(\omega) = \int_{\widehat{G}}^{\oplus} d\mu(\vec{\pi}) V_{\vec{\pi}} \otimes \Pi_A[V_{\vec{\pi}}^* \otimes V_\omega^*] \,,
\end{align}
and similarly for $\Ha_{\overline{R}}(\omega)$. We will explain the interpretation of $\Ha_R(\omega)$ shortly.
Then one can show that
\begin{align}
    \Ha_{\mathrm{phys}}\left(\Sigma_{0,1}^{(n)}\right) = \int_{\widehat{G}}^{\oplus} d\mu(\omega) \Pi_A[\Ha_R(\omega) \otimes \Ha_{\overline{R}}(\overline{\omega})] \,. \label{eq:HphysAlgebras}
\end{align}
One can think of this decomposition as follows. Gauge invariance implies that physical states transform in the trivial representation of the gauge group. This means that states in the complete Hilbert space $\Ha_{\mathrm{phys}}\left(\Sigma_{0,1}^{(n)}\right)$ can be thought of as labeling fusion channels from the representations at the boundary points to the trivial representation. One way to index these fusion channels is to first catalog the possible intermediate fusion channels from the $R$ subregion to an arbitrary representation $\overline{\omega}$. 
Equivalently, this catalogs the intermediate fusion channels between the representations at the $R$ boundary legs and an additional representation $\omega$. This is what the Hilbert space $\Ha_R(\omega)$ captures. We can think of the $\omega$ representation as living on the endpoints of the central edge of Fig.~\ref{fig:bowtie} after cutting this link open; see Fig.~\ref{fig:factorizedgraph}.
Then, one catalogs the possible intermediate fusion channels from the representations in the $\overline{R}$ subregion to $\omega$. This Hilbert space has a similar geometric interpretation. Finally,  note that Schur's lemma implies that if we pair $\Ha_R(\omega)$ and $\Ha_{\overline{R}}(\overline{\omega})$, then there is a unique fusion channel from each $(\omega, \overline{\omega})$ pair to the trivial representation. Thus, \eqref{eq:HphysAlgebras} holds.

\section{Factorizing the Hilbert space} \label{sec:factorization}

In this section and the remainder of this paper, we will consider states without matter legs and defer discussion of the role of matter legs to \cite{Balasubramanian:2026xyz}. Our goal in what follows will ultimately be to compute entanglement entropy between boundary legs of $\Ha_{\mathrm{phys}}(\Sigma)$.  For concreteness, we will study the case where $\Sigma$ is the disk---other topologies work similarly.

The  naive approach immediately runs into a problem. To explain this problem, let $\Ha = \Ha_R \otimes \Ha_{\overline{R}}$ be some factorization of a Hilbert space $\Ha$ into two subsystems $R,\overline{R}$. Given this factorization, we can start with a globally pure state $\ket{\Psi} \in \Ha$ and trace out the subsystem $\Ha_{\overline{R}}$ to obtain a reduced density matrix $\rho_R$ for the subsystem $R$. This reduced state $\rho_R$ is then used to compute quantum information measures,  such as entanglement entropy, R\'enyi entropies, and so on that quantify correlations between $R$ and $\overline{R}$.

This definition of $\rho_R$ explicitly requires a splitting of $\Ha$ into a tensor product $\Ha_R \otimes \Ha_{\overline{R}}$. However, the integral over representations in  Eq.~\eqref{eq:Hphys} implies that the physical Hilbert space $\Ha_{\mathrm{phys}}(\Sigma)$ does not factorize across the boundary vertices. To circumvent this issue, we will define a map
\begin{align}
    V: \Ha_{\mathrm{phys}}(\Sigma) \to \Hext{R} \otimes \Hext{\overline{R}}
\end{align}
which embeds $ \Ha_{\mathrm{phys}}(\Sigma)$ as a subspace of a factorizing Hilbert space $\Hext{R} \otimes \Hext{\overline{R}}$. We call $\Hext{R}$ the extended Hilbert space of $R$, and similarly for $\overline{R}$. This factorization map will be based on the bowtie lattice $\Lambda_b$, shown in Fig.~\ref{fig:bowtie}.

\subsection{The group basis }

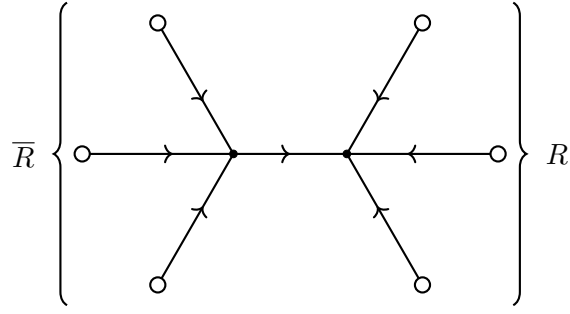
\begin{figure}
    \centering
    \begin{tikzpicture}[scale=2]
        \def\sep{3}
        \def\nn{6}
        \def\hh{0.75}

        \filldraw (0,0) circle (0.025);
        \filldraw (\hh,0) circle (0.025);
        \foreach \xx in {1,...,3}{
            \draw[thick,->-=0.6]  ({\hh+cos(360*(\xx-2)/\nn)},{sin(360*(\xx-2)/\nn)}) -- (\hh,0);
            \filldraw[thick,fill=white] ({\hh+cos(360*(\xx-2)/\nn)},{sin(360*(\xx-2)/\nn)}) circle (0.05);
        }
        \foreach \xx in {4,...,\nn}{
            \draw[thick,->-=0.6]  ({cos(360*(\xx-2)/\nn)},{sin(360*(\xx-2)/\nn)}) -- (0,0);
            \filldraw[thick,fill=white] ({cos(360*(\xx-2)/\nn)},{sin(360*(\xx-2)/\nn)}) circle (0.05);
        }
        \draw[thick,->-=0.5] (0,0) -- (\hh,0);

        \draw [thick,decorate,decoration={brace,amplitude=5pt}]
        (-1-0.1,-1) -- (-1-0.1,1) node[midway,xshift=-1.5em]{$\overline{R}$};

        \draw [thick,decorate,decoration={brace,amplitude=5pt,mirror}]
        (1+\hh+0.1,-1) -- (1+\hh+0.1,1) node[midway,xshift=1.5em]{$R$};

    \end{tikzpicture}
    \caption{The bowtie lattice $\Lambda_b$ between the boundary legs $R$ and $\overline{R}$.}
    \label{fig:bowtie}
\end{figure}

Consider a gauge invariant state $\dket{\psi}$ in the physical Hilbert space in the bowtie lattice presentation $\Lambda_b$ shown in Fig.~\ref{fig:bowtie}. Such a state has an expansion
\begin{align}
    \dket{\psi} = \int d[\vec{g}_R,\vec{g}_{\overline{R}},h] \,\psi(\vec{g}_R, \vec{g}_{\overline{R}}, h) \dket{\vec{g}_{\overline{R}},h,\vec{g}_R} \,,
\end{align}
where $\dket{\vec{g}_{\overline{R}},h,\vec{g}_R}$ is the co-invariant state descending from a representative $\vert\vec{g}_{\overline{R}},h,\vec{g}_R\rangle$ in the pre-Hilbert space $\Ha(\Lambda_b)$. Here, $\vec{g}_R, \vec{g}_{\overline{R}}$ refer to the collection of group elements on each set of  boundary legs, and $h$ is the group element on the additional bulk leg which separates $R$ from $\overline{R}$. This expansion of the state follows from the linearity of the constraints $\Pi_A \Pi_B$ and a resolution of the identity on $\ket{\psi}$. 

The factorization map is defined by its action on $\dket{\psi}$ as 
\begin{align}
    V\dket{\psi} = \int d[\vec{g}_R,\vec{g}_{\overline{R}}, h, k ]\,\psi(\vec{g}_R, \vec{g}_{\overline{R}}, kh) \dket{\vec{g}_{\overline{R}},k}_{\overline{R}} \dket{h,\vec{g}_R}_R\,. \label{eq:Vgroupbasisfactorized}
\end{align}
Because we defined the state $\dket{\psi}$ via the bowtie lattice presentation (which we can always do for the disk topology), this definition of $V$ is unambiguous and well-defined on all of $\Ha_{\mathrm{phys}}(\Sigma)$. If the state is not originally presented in the bowtie lattice, then we can always use the moves defined in Sec.~\ref{sec:themodel} to put the state into this form. 

\begin{figure}
    \centering
    \begin{tikzpicture}[scale=2]
        \def\sep{0.5}
        \def\nn{6}
        \def\hh{2}

        \filldraw (0,0) circle (0.025);
        \filldraw (\hh,0) circle (0.025);
        \foreach \xx in {1,...,3}{
            \draw[thick,->-=0.6]  ({\hh+cos(360*(\xx-2)/\nn)},{sin(360*(\xx-2)/\nn)}) -- (\hh,0);
            \filldraw[thick,fill=white] ({\hh+cos(360*(\xx-2)/\nn)},{sin(360*(\xx-2)/\nn)}) circle (0.05);
        }
        \foreach \xx in {4,...,\nn}{
            \draw[thick,->-=0.6]  ({cos(360*(\xx-2)/\nn)},{sin(360*(\xx-2)/\nn)}) -- (0,0);
            \filldraw[thick,fill=white] ({cos(360*(\xx-2)/\nn)},{sin(360*(\xx-2)/\nn)}) circle (0.05);
        }
        \draw[thick,->-=0.5] (0,0) -- (\hh/2 - \sep/2,0) node[midway,anchor=north] {$k$};
        \draw[thick,->-=0.5] (\hh/2 + \sep/2,0) -- (\hh,0) node[midway,anchor=north] {$h$};
        \filldraw[thick,fill=black] (\hh/2 - \sep/2,0) circle (0.05);
        \filldraw[thick,fill=black] (\hh/2 + \sep/2,0) circle (0.05);

        \node at (0,-1.1) {$\dket{\vec{g}_{\overline{R}},k}_{\overline{R}}$};
        \node at (\hh,-1.1) {$\dket{h,\vec{g}_{R}}_{R}$};

        \draw [thick,decorate,decoration={brace,amplitude=5pt}]
        (-1-0.1,-1) -- (-1-0.1,1) node[midway,xshift=-1.5em]{$\overline{R}$};

        \draw [thick,decorate,decoration={brace,amplitude=5pt,mirror}]
        (1+\hh+0.1,-1) -- (1+\hh+0.1,1) node[midway,xshift=1.5em]{$R$};

    \end{tikzpicture}

    \caption{The state after the Gauss constraint between $R$ and $\overline{R}$ is lifted. The newly exposed legs, with black dots, are the edge modes. Each subgraph represents the entanglement wedges for $\overline{R},R$, respectively. Left $G$ multiplication on the edge modes are a gauge/global symmetry for the $\overline{R}$, $R$, edge modes, respectively. Right $G$ multiplication on the edge modes are a global/gauge symmetry for the $\overline{R}$, $R$ edge modes, respectively. The edge mode legs are therefore crucial for each subregion being gauge-invariantly defined, while simultaneously allowing the Hilbert space between the subregions to factorize because the other multiplication is lifted to a global symmetry.}
    \label{fig:factorizedgraph}
\end{figure}
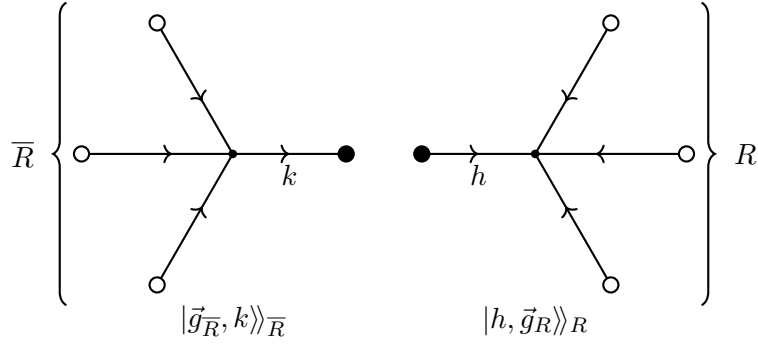

The states $\dket{h,\vec{g}_R}_R$ are a basis of the ``extended Hilbert space'' $\Hext{R}$, which is spanned by the $R$ boundary legs and an additional bulk leg $h$, and similarly for $\dket{\vec{g}_{\overline{R}},k}_{\overline{R}}$.\footnote{See \cite{Mertens:2025ydx} for related work about edge modes in Chern--Simons theories.} The extended Hilbert space also decomposes as in \eqref{eq:Hphys} if we interpret the additional bulk leg as a boundary leg. This justifies the decomposition of the extended Hilbert space given in \eqref{eq:Hextdecomp}.
Denote the partial lattice supporting the extended Hilbert space $\Hext{R}$ by $\Lambda_R$, and the additional bulk leg in $\Lambda_R$ as the corner leg (see Fig.~\ref{fig:factorizedgraph}). 
The states of  $\Hext{R}$ are  invariant under gauge transformations of the bulk vertex of $\Lambda_R$, where the Gauss law remains imposed. Gauge transformations on $\Lambda_R$ act on the group elements $\vec{g}_R$ and $h$ by a right multiplication. The presence of the additional bulk leg which supports $h$ is thus crucial for the gauge invariant definition of the subregion $\Lambda_R$ in the bulk. 

\begin{figure}
    \centering
    \newcommand{\legdisk}[4]{%
        \pgfmathsetmacro{\gap}{asin(#3/#2)}
        \foreach \i in {0,...,\numexpr#1-1\relax}{%
            \pgfmathsetmacro{\ang}{#4 + \i*360/#1}%
            \pgfmathsetmacro{\angnext}{#4 + (\i+1)*360/#1}%
            \draw[thick]
            (\ang+\gap:#2) arc[start angle=\ang+\gap, end angle=\angnext-\gap, radius=#2];

            \draw[color=black!50, thick,->-=0.5] (\ang:#2) -- (0,0);
            \filldraw[black!50,thick,fill=white] (\ang:#2) circle (0.1);

            \fill (\angnext-\gap:#2) circle (0.05);
            \draw[thick,dashed]
            ([shift={(\ang:#2)}]\ang+90:#3)
            arc[start angle=\ang+90, end angle=\ang+270, radius=#3];
            \fill (\ang+\gap:#2) circle (0.05);
        }%
        \fill[color=black!50] (0,0) circle (0.05);%
    }

    \begin{tikzpicture}
        \legdisk{4}{1.5}{0.25}{45}
        \node at (1.5,-1.5) {$\Sigma$};
    \end{tikzpicture}
    \begin{tikzpicture}
        \def\R{1.5}      
        \def\r{0.25}     
        \def\cx{2.1}     
        \def\na{60}      
        \pgfmathsetmacro{\gap}{asin(\r/\R)}
        \coordinate (L)  at (-\cx,0);
        \coordinate (Rc) at (\cx,0);
        \coordinate (M)  at (0,0);
        \coordinate (NTL) at ([shift={(L)}]\na:\R);
        \coordinate (NTR) at ([shift={(Rc)}]180-\na:\R);
        \coordinate (NBL) at ([shift={(L)}]-\na:\R);
        \coordinate (NBR) at ([shift={(Rc)}]\na-180:\R);

        \draw[ultra thick,dashed,blue] (0,0.7) -- (0,-0.7);
        \node[anchor=north,blue] at (0,-0.7) {$\gamma$};

        \draw[thick] ([shift={(L)}]135+\gap:\R) arc[start angle=135+\gap, end angle=225-\gap, radius=\R];
        \draw[thick] ([shift={(L)}]135-\gap:\R) arc[start angle=135-\gap, end angle=\na, radius=\R];
        \draw[thick] ([shift={(L)}]225+\gap:\R) arc[start angle=225+\gap, end angle=360-\na, radius=\R];
        \draw[thick] ([shift={(Rc)}]45-\gap:\R) arc[start angle=45-\gap, end angle=-45+\gap, radius=\R];
        \draw[thick] ([shift={(Rc)}]180-\na:\R) arc[start angle=180-\na, end angle=45+\gap, radius=\R];
        \draw[thick] ([shift={(Rc)}]\na-180:\R) arc[start angle=\na-180, end angle=-45-\gap, radius=\R];

        \draw[thick] (NTL) to[out=-90+\na, in=-90-\na, looseness=1.5] (NTR);
        \draw[thick] (NBL) to[out=90-\na, in=90+\na, looseness=1.5] (NBR);

        \foreach \c/\a in {L/135, L/225, Rc/45, Rc/-45}{
            \fill ([shift={(\c)}]\a+\gap:\R) circle (0.05);
            \fill ([shift={(\c)}]\a-\gap:\R) circle (0.05);
        }

        \draw[color=black!50, thick, ->-=0.5] ([shift={(L)}]135:\R) -- (L);
        \draw[color=black!50, thick, ->-=0.5] ([shift={(L)}]225:\R) -- (L);
        \draw[color=black!50, thick, ->-=0.5] ([shift={(Rc)}]45:\R) -- (Rc);
        \draw[color=black!50, thick, ->-=0.5] ([shift={(Rc)}]-45:\R) -- (Rc);
        \draw[color=black!50, thick, ->-=0.5] (L) -- (M);
        \draw[color=black!50, thick, ->-=0.5] (M) -- (Rc);
        \foreach \c/\a in {L/135, L/225, Rc/45, Rc/-45}{
            \filldraw[thick, fill=white] ([shift={(\c)}]\a:\R) circle (0.1);
        }
        \foreach \p in {L, Rc, M}{ \fill[color=black!50] (\p) circle (0.05); }

        \foreach \c/\a in {L/135, L/225, Rc/45, Rc/-45}{
            \draw[thick, dashed]
            ([shift={(\c)}]\a:\R) ++(\a+90:\r)
            arc[start angle=\a+90, end angle=\a+270, radius=\r];
        }

        \node at (\cx+1.5,-1.5) {$\Sigma$};

    \end{tikzpicture}

    \begin{tikzpicture}
        \begin{scope}[shift={(-1.9,0)}]
            \legdisk{3}{1.5}{0.25}{0}
            \fill[fill=white] (0.75,0) circle (0.2);
            \draw[thick,color=black!50,->-=0.5] (0.5,0) -- (1,0);
            \filldraw[black!50,fill=blue,thick] (1.5,0) circle (0.1);
            \node at (1.5,-1.5) {$\Sigma_{\overline{R}}$};
        \end{scope}
        \begin{scope}[shift={(1.9,0)}]
            \legdisk{3}{1.5}{0.2}{180}
            \filldraw[black!50,fill=blue,thick] (-1.5,0) circle (0.1);
            \node at (1.5,-1.5) {$\Sigma_{R}$};
        \end{scope}
    \end{tikzpicture}

    \caption{Top: Two equivalent presentations of the disk Hilbert space with four marked points, in the reduced lattice gauge and the bowtie lattice gauge, respectively. The physical Hilbert space $\Ha_{\mathrm{phys}}(\Sigma)$ is identical between them. As in Fig.~\ref{fig:factorizedgraph}, the boundary points for $\overline{R}$ are on the left, and the boundary points for $R$ are on the right.  The boundary anchored curve $\gamma$ which separates $\overline{R}$ from $R$ is shown in blue on the right.
    Bottom: The two disks of the extended Hilbert spaces $\Hext{\overline{R}}$ and $\Hext{R}$, respectively, after we cut the disk open along $\gamma$. The degrees of freedom that live on the blue dot descend from the curve $\gamma$, and are the observer edge modes. The gauge fixing edge modes feed into the bulk vertex of each partial disk.
    }
    \label{fig:gluing_disks}
\end{figure}
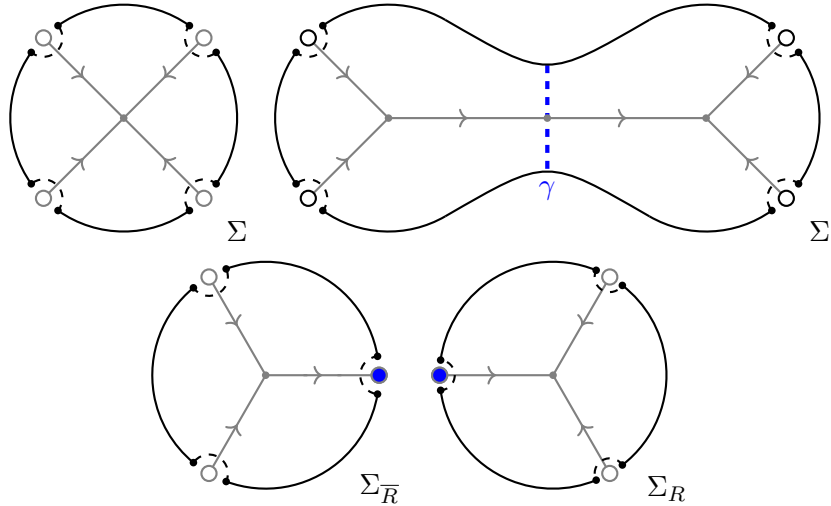

Interestingly, the additional bulk leg of $\Lambda_R$ carries the same Hilbert space as the boundary legs, so we can think of this splitting as arising from artificially placing a finite boundary in the bulk, splitting the two subregions: see Fig.~\ref{fig:gluing_disks}. We can make this more precise as follows. The operators with support on the corner leg can be interpreted as bulk operators with support on a cycle $\gamma$ for which $\Sigma \setminus \gamma = \Sigma_R \sqcup \Sigma_{\overline{R}}$. In other words, $\gamma$ is a boundary anchored curve that separates the $R$ boundary legs from the $\overline{R}$ boundary legs. When we use the lattice deformation moves from Sec.~\ref{sec:themodel} to go to the bowtie lattice, an operator with support on $\gamma$ will be unitarily equivalent to an operator with support on only the bulk leg of the bowtie lattice. When we cut open the disk, the edge modes can be thought of as residual degrees of freedom on $\gamma$ which we must retain in order to glue the two partial disks back together, and retain the full information about the global state. 

Thinking of the cut as an artificial boundary in this way, we must give it a boundary condition in order to define the extended Hilbert space of each partial disk. In principle, we could specify a different boundary condition on either side of the cut $\gamma$, but because we want the cut to be invisible when we glue the state back together, we need to impose the same boundary condition on both sides of the cut.
Additionally, since we want the factorization to be independent of the tessellation $\Lambda$, this boundary condition must be insensitive to deformations of the cut, i.e., it must be a \emph{topological} boundary condition \cite{Kitaev:2011dxc,Kong:2014xyz,Kapustin:2010hk,Beigi:2010htr}. As we review in Appendix~\ref{apx:topvscon}, topological boundary conditions are classified by the set of bulk line operators which are allowed to terminate on them, and this set is determined by the data defining the bulk theory \cite{Kong:2014xyz,KONG2021115607}. The classification is known for finite groups $G$; for the continuous and possibly non-compact groups of this paper, we take the natural generalization of the finite group boundary conditions at face value, as discussed in Appendix~\ref{app:topbcs}.

Cutting open the corner leg and inserting the vertex of Fig.~\ref{fig:factorizedgraph} produces one new boundary vertex on each side of the cut at which the electric constraint is not imposed. This is exactly the condition we imposed at the marked points of the physical boundary in Sec.~\ref{sec:bcs}, so the cut is treated as a pair of new marked points, one for $\Sigma_R$ and one for $\Sigma_{\overline{R}}$. The line operators which can end on it are the charges, labeled by $\pi \in \widehat{G}$. In the language of Appendix~\ref{app:topbcs}, this is the rough boundary condition. Because the rough boundary condition is a topological boundary condition, it is by-definition independent of the details of the lattice near the cut, since any two such lattices are related by the moves of Sec.~\ref{sec:LatticeIndep} away from the cut. Therefore, the factorization map $V$ is independent of our particular presentation of it through the bowtie lattice. 

In Sec.~\ref{sec:entropy_nomatter}, we will discuss the implications of rough boundary conditions at the cut on our final entropy formula. We choose a rough boundary condition on the cut in agreement with the previous literature, which we verify when $G$ is a compact group in Sec.~\ref{sec:entropy_compact}. But in principle, other topological boundary conditions for the cut seem allowed: for instance, the smooth boundary condition, for which fluxes condense rather than charges, would lead to a different factorization map with a different set of edge modes. This alternative factorization map will share the same qualitative features as the rough-boundary factorization map $V$, but with the representations $\pi$ replaced with the corresponding labels of the anyons which can end on the cut for the alternative topological boundary condition. 
We will highlight where our results do and do not depend on the choice of topological boundary condition defining the factorization map as we go, and will return to this choice of boundary conditions in \cite{Balasubramanian:2026xyz}.

Let us return to the physical features of the extended Hilbert space.
As explained above, a right multiplication on the corner leg is a gauge symmetry of the extended Hilbert space for $R$.\footnote{Because the relative orientations of the corner leg are reversed for the extended Hilbert spaces of $R$ and $\overline{R}$, the role of left and right multiplications is similarly reversed between $R$ and $\overline{R}$. This is just a convention.} However, a {\it left} multiplication of $h$ by a group element (leaving $\vec{g}_R$ fixed) is a {\it global} symmetry of $\Hext{R}$. To see this, let us Fourier transform the corner leg $L^2(G)$ into its representation basis via \eqref{eq:peterweylnoncompact}. In this decomposition, we can think of $V_\pi \otimes V_\pi^*$ as the space of matrix elements of the $\pi$ representation, so right multiplication acts on the $V_\pi^*$ index. Because right multiplication is gauged in the bulk, we refer to the degrees of freedom in $V_\pi^*$ (for each $\pi$) as the gauge fixing edge modes. In contrast, left multiplication acts on the $V_\pi$ index, these degrees of freedom do not transform under gauge transformations in the extended Hilbert space. We call these degrees of freedom the {\it observer edge modes}.
Although this splitting of edge modes into matrix elements may seem peculiar to this model, it also occurs in theories of gravity because the charges carried by gravitational solitons are always in the adjoint representation\footnote{Or, in the case of generalized symmetries, the adjoint subcategory.} of the symmetry group of the rest of the theory \cite{McNamara:2021cuo,Harlow:2018tng}, so we expect this kind of splitting to be universal in any theory of gravity.

The physical difference between left and right multiplications is because the Gauss constraint only acts on the degrees of freedom which transform under right multiplications. This exactly matches the role of corner symmetry in  continuum gauge theories and in gravity \cite{Donnelly_2016}, which is how we fixed the convention for the orientation of the corner legs on the extended Hilbert space. The same analysis holds for $\overline{R}$ as well, keeping in mind that bulk leg supporting the edge modes of $\overline{R}$ has the opposite orientation than the bulk leg of $R$ (as is apparent in Fig.~\ref{fig:factorizedgraph}). So for $\overline{R}$, left multiplications are gauged, right multiplications are global symmetries.

Now, consider the simultaneous global symmetry transformation of $V\dket{\psi}$ by a group element $g$, i.e., a left multiplication of $h \to gh$ and a right multiplication $k \to kg^{-1}$. Call this transformation $A_{\eth}(g)$. We can see that
\begin{align}
    A_{\eth}(g)V\dket{\psi} &= \int d[\vec{g}_R \vec{g}_{\overline{R}} h k] \psi(\vec{g}_R, \vec{g}_{\overline{R}}, kh) \dket{ \vec{g}_{\overline{R}},kg^{-1}}_{\overline{R}} \dket{gh,\vec{g}_R}_R 
    \\&= \int d[\vec{g}_R \vec{g}_{\overline{R}} h k] \psi(\vec{g}_R, \vec{g}_{\overline{R}}, kgg^{-1}h) \dket{ \vec{g}_{\overline{R}},k}_{\overline{R}} \dket{h,\vec{g}_R}_R
    \\&= V\dket{\psi}\,.
\end{align}
Thus, even though individual group multiplications on either corner leg can alter states in the image of $V$, the diagonal left multiplication on both sets of edge modes leaves the state invariant.
Geometrically, this corresponds to simultaneously moving the boundary anchored curve $\gamma$ supporting the edge modes within the spacetime, in a correlated manner between $\Sigma_R$ and $\Sigma_{\overline{R}}$, such that the glued surface $\Sigma$ remains invariant. 
This invariance  depends crucially on how the states $\dket{ \vec{g}_{\overline{R}},k}_{\overline{R}} $ and $\dket{h,\vec{g}_R}_R$ are entangled with each other, an entanglement pattern which we will demonstrate below is unique within each superselection sector $\pi$. The entanglement is mediated by the edge modes, so we  see that although the edge modes are ``unphysical'' in the sense that they are introduced solely to factorize the Hilbert space, their structure is uniquely determined by the demand that physical states remain gauge invariant in the factorized Hilbert space. Furthermore, they transform non-trivially under (individual) left multiplication, so they are physical in that sense.

We can also check the action of $V$ on the basis element $\dket{\vec{g}_{\overline{R}}, g, \vec{g}_R} $ of the physical Hilbert space:
\begin{align}
    V\dket{\vec{g}_{\overline{R}}, g, \vec{g}_R} 
    &= \int d[h, k] \delta(g^{-1}kh) \dket{\vec{g}_{\overline{R}}, k}\dket{h, \vec{g}_R} 
    \\&= \int dh \,\dket{\vec{g}_{\overline{R}},h^{-1}} \dket{hg, \vec{g}_R} \label{eq:Vgroupbasis}
    \\&= \left(\int dh A_{\eth}(h) \right) \dket{\vec{g}_{\overline{R}},e} \dket{g,\vec{g}_R} \,.
\end{align}
Here, $e$ is the identity element of the group. The $h$ integral can be identified with $[\Pi_A]_\eth$, the Gauss law projector on the observer edge modes for $R$ and $\overline{R}$, thought of as identifying the bulk vertices introduced by the factorization map, shown in black in Fig.~\ref{fig:factorizedgraph}, and gluing them together. We will call this vertex the corner vertex. The resulting graph after this gluing is just Fig.~\ref{fig:bowtie} with an extra bulk vertex, which we could then apply our graphical moves to return to Fig.~\ref{fig:bowtie} itself. 

To gain more intuition, imagine dropping the Gauss law at the remaining bulk vertices of the factorized bowtie lattice as well, so $\dket{\vec{g}_{\overline{R}},e} \dket{g,\vec{g}_R} \to \ket{\vec{g}_{\overline{R}},e,g,\vec{g}_R}$. This is a state in the pre-Hilbert space  $\Ha(\Lambda_b \sqcup \ell)$ of the bowtie lattice $\Lambda_b$, along with an auxiliary leg $\ell$ supporting the identity element $\ket{e}$. Then, after rearranging the tensor factors of $\Ha(\Lambda_b \sqcup \ell)$, we could think of the factorization map as taking the form
\begin{align}
    V\ket{\psi}= [\Pi_A]_\eth \ket{e} \ket{\psi} \,, \label{eq:factorizationfake}
\end{align}
where $[\Pi_A]_\eth$ is understood to act on $\ket{e}$ and the corner leg of $\Lambda_b$.
The state $\ket{e}$ is the wave function for the observer edge modes of both $R$ and $\overline{R}$, which are introduced to factorize the Hilbert space. This particular choice of edge mode wave function is what ensures that the edge mode group elements agree when gluing $\Lambda_R$ and $\Lambda_{\overline{R}}$ to form the bowtie lattice (plus the additional vertex on the corner leg from the gluing). If we chose a different wave function for the edge modes, say $\ket{g}$, this would introduce a relative ``kink'' at the corner vertex when gluing the states back together, i.e., we would have $\ket{k,h} \to \ket{gh^{-1},h}$ instead of $\ket{k,h} \to \ket{h^{-1},h}$ as we explained above. This relative kink can be interpreted as an electric flux of the $G$ gauge field through the corner vertex.

To see this, consider Maxwell theory on the lattice. Suppose that each subregion has a potential $A_0|_R = 0$ and $A_0|_{\overline{R}} = -c$, so that the electric field normal to the boundary $\partial R$ is $E_\perp = \frac{c}{a}$, where $a$ is the lattice spacing between $R$ and $\overline{R}$. Note that on each subregion, $A_0$ is pure gauge, but the relative difference $E_\perp$ has physical meaning. The wave function $\ket{e}$ is the unique choice which ensures that the edge mode degrees of freedom drop out of the physical Hilbert space after $\Hext{R}$ and $\Hext{\overline{R}}$ are glued back to form the physical Hilbert space. A different wave function for the observer edge modes could then be interpreted as a presence of a physical charge with support on the corner vertex $\eth$.  Furthermore, the map $[\Pi_A]_\eth$ which entangles this wave function with $\ket{\psi}$ is uniquely determined by the requirement that diagonal global transformations on the edge modes act as gauge symmetries on the image of $V$. 

However, this definition of factorization map $V$ is only \emph{projectively} an isometry between $\Ha_{\mathrm{phys}}(\Sigma)$ and $\Hext{R} \otimes \Hext{\overline{R}}$. To see this, consider the matrix element (suppressing the $\vec{g}_R,\vec{g}_{\overline{R}}$ dependence for simplicity)
\begin{align}
    \bra{g} V^\dagger V \ket{h} &= \int d[mn] \braket{m}{n}\braket{mg}{nh} = \int dm \braket{mg}{mh} = \left(\int_G dm \right) \braket{g}{h} \,.\label{eq:groupbasisVoldiverge}
\end{align}
Up to the state independent constant $\int_G dm = \mathrm{Vol}(G)$,  $V^\dagger V$ thus acts as an isometry on the group basis: by linearity, it is a projective isometry on all of $\Ha_{\mathrm{phys}}(\Sigma)$. When $G$ is compact, this definition is sufficient, as we can pick a normalization for the Haar measure of $G$ such that $\mathrm{Vol}(G) = 1$. We will return to the non-compact case after we discuss the action of $V$ in the representation basis.

\subsection{The representation basis}

In the previous subsection, we computed the action of $V$ on a group basis element. However, we can define the action of $V$ on any basis we wish: it will be enlightening to compute its action on the representation basis for the edge modes. For clarity, we will suppress the dependence on the boundary legs, and restore it at the end.

Recall that the representation basis of $L^2(G)$ is defined by its overlap with the group basis as
\begin{align}
    \ket{\pi,ij} = \int dg\braket{g}{\pi,ij}\ket{g} = \int dg \,\pi(g)_{ij} \ket{g} \,,
\end{align}
where $\pi(g)_{ij}$ is the matrix element of the unitary, irreducible $G$ representation $\pi \in \widehat{G}$. For example, if $G=\SU(2)$, then $\pi(g)_{ij}$ is the $ij$-th matrix element of the Wigner D matrix for spin $\pi$. If $G=\R$, then $\pi(g)_{ij}$ is just the plane wave $e^{ikx}$ for a fixed momentum $k \sim \pi$ (the $i,j$ indices are not present because this representation is one dimensional). The set of all $\ket{\pi,ij}$ for a fixed $\pi$ spans the subspace $V_\pi \otimes V_\pi^*$ of the Plancherel decomposition \eqref{eq:peterweylnoncompact}. There is a resolution of the identity
\begin{align}
    \Id_{L^2(G)} = \int_{\widehat{G}} d\mu(\pi) \sum_{i,j} \ketbra{\pi,ij} \,,
\end{align}
where $d\mu(\pi)$ is the Plancherel measure for $G$. Repeatedly using this resolution of the identity on \eqref{eq:Vgroupbasis}  (and leaving the $\vec{g}_R,\vec{g}_{\overline{R}}$ dependence implicit for the moment), we can show that $V\dket{\pi,ab}$ is
\begin{align}
    V\dket{\pi,ab} &= \int d[gh] \int d\mu(\eta,ij;\omega,mn) \eta(h)^*_{ij} \pi(hg)_{ab}\omega(g)^*_{mn} \dket{\eta,ij}\dket{\omega,mn}
    \\&=\int d[gh] \int d\mu(\eta,ij;\omega,mn) \eta(h)^*_{ij} \left(\sum_c \pi(h)_{ac}\pi(g)_{cb} \right) \omega(g)^*_{mn} \dket{\eta,ij}\dket{\omega,mn}
    \\&=\int d\mu(\eta,ij;\omega,mn) \left(\sum_c \braket{\eta,ij}{\pi,ac}\braket{\omega,mn}{\pi,cb}\right)\dket{\eta,ij}\dket{\omega,mn}
    \\&=\sum_c \dket{\pi,ac}\dket{\pi,cb} \,. \label{eq:etamatrixmultstate}
\end{align}
If we restored the dependence on the external legs, we would find that
\begin{align}
    V \dket{\vec{g}_{\overline{R}} ; \pi,ab ; \vec{g}_R} = \sum_c \dket{\vec{g}_{\overline{R}} ; \pi,ac}_{\overline{R}}  \dket{\pi,cb; \vec{g}_R}_R \,. \label{eq:Vcoinvariantrepbasis}
\end{align}
At first, one might worry that this expression is not well-defined, for if the index $c$ could change under a gauge transformation of either the $R$ or $\overline{R}$ vertices, this sum would be gauge-dependent. However, we showed in \cite{Balasubramanian:2025rcr} that after the quotient by null states, the equivalence class $ \dket{\pi,cb; \vec{g}_R}_R$ always has a fixed representation $\pi$ and index $c$. One way to see this is that the $b$ index is contracted into the $R$ vertex, while the $c$ index (associated with the observer edge modes) remains free. The same conclusion holds for $\dket{\vec{g}_{\overline{R}} ; \pi,ac}_{\overline{R}}$ because of our convention for the orientation of the edge modes.

There is an equivalent, more intuitive way to think about this factorization map, which is albeit less mathematically precise. Recall that $\ket{\pi,ab} $ is a basis for $L^2(G)$. To gain some intuition, we can roughly view this state as having a continuous index $\pi$, and two discrete indices $a,b$ for $V_\pi, V_\pi^*$, respectively. This follows from the factorization structure of \eqref{eq:peterweylnoncompact} and the overlap equation \eqref{eq:repbasisoverlap}.
Below, we would like to take the partial trace over the $b$ index, while leaving the $\pi,a$ indices untouched. We can do so as follows.
Split the fixed matrix element state as $\ket{\pi,ab} = \ket{\pi,a}\ket{b_\pi}$.
These partial states are normalized such that
\begin{align}
    \braket{\pi,a}{\omega,c} &= \delta(\pi,\omega) \delta_{ac} \,, \label{eq:fakesplitdelta}\\
    \braket{b_\pi}{d_\omega} &= \delta_{\pi\omega} \delta_{bd} \label{eq:fakesplitkronecker}\,.
\end{align}
Here, $\delta(\pi,\omega)$ means the delta function is normalized with respect to the Plancherel measure, while $\delta_{\pi,\omega}$ is the Kronecker delta function (which equals $1$ if $\pi = \omega$, and vanishes otherwise).
This normalization condition ensures that the $\ket{\pi,ab}$ and $\ket{\pi,a} \ket{b_\pi}$ bases of $V_\pi \otimes V_\pi^*$ have the same norm.  Note that we could have attached the continuous index $\pi$ to the $V^*_\pi$ basis instead: the invariant statement is that $\ket{\pi,ab}$ is a basis for $L^2(G)$, regardless of how we split the state into its tensor factors. But the choice of splitting made above will be convenient for us.

Thus, when projecting to the co-invariant Hilbert space, we can factorize
\begin{align}
    \dket{\pi,ab ; \vec{g}_R}_R = \ket{\pi,a} \dket{b_\pi ; \vec{g}_R}_R \,.
\end{align}
Because of our choice of orientation of the corner legs in $\Lambda_{\overline{R}}$, we 
use the opposite convention on the corner leg of $\overline{R}$, and we 
can see that
\begin{align}
    \dket{\vec{g}_{\overline{R}} ; \pi,ab}_{\overline{R}} = \dket{\vec{g}_{\overline{R}} ; a_\pi}_{\overline{R}} \ket{\pi,b}\,.
\end{align}
Our reason for this opposite convention is so that the state $\dket{\vec{g}_{\overline{R}} ; a_\pi }_{\overline{R}} $ is unit normalized in the gauge fixing edge modes sector, and so the extended Hilbert spaces for $R$ and $\overline{R}$ have the same normalization conventions. In this notation, the factorization map takes the form
\begin{align}
    V \dket{\vec{g}_{\overline{R}} ; \pi,ab ; \vec{g}_R} 
    &= \dket{\vec{g}_{\overline{R}} ; a_\pi}_{\overline{R}} \left(\sum_c \ket{\pi,c}\ket{\pi,c}\right)  \dket{b_\pi; \vec{g}_R}_R \,. \label{eq:middleparen}
\end{align}
The vector inside the middle parenthesis is proportional to the character function $\chi_\pi$ of the representation $\pi$:\footnote{This equality follows from expanding $\chi_\pi(g) = \sum_i \braket{g}{\pi,ii}$, and using a resolution of the identity on $g$.}
\begin{align}
    \ket{\chi_\pi} = \sum_c \ket{\pi,c} \ket{c_\pi} = \int dg \,\chi_\pi(g) \ket{g} \,, \label{eq:characterfndef}
\end{align}
where $\chi_\pi(g) = \tr[\pi(g)]$.\footnote{More precisely, $\chi_\pi(g)$ is the $L^1$ function such that for any compactly supported function $f(g)$, $\tr \left[\int dg f(g) \pi(g) \right] = \int dg f(g) \chi_\pi(g)$. The identification between $\chi_\pi(g)$ and $\tr[\pi(g)]$ is subtle when $G$ is non-compact, because of convergence issues that mean the integral and the trace do not always commute. Nevertheless, such an $L^1$ function always exists \cite{HarishChandra1952,HarishChandra1976,HarishChandra1954complex}.
}
Thus, we will write $\ket{\chi_\pi}'$ for the vector in the middle parenthesis of \eqref{eq:middleparen}, and the prime indicates the ``double delta'' normalization of $\ket{\chi_\pi}'$ compared to the character function $\ket{\chi_\pi}$.
With this definition, the factorization map can be written as
\begin{align}
    V \dket{\vec{g}_{\overline{R}} ; \pi,ab ; \vec{g}_R} &= \dket{\vec{g}_{\overline{R}} ; a_\pi }_{\overline{R}} \ket{\chi_\pi}' \dket{b_\pi ; \vec{g}_R}_R \,. \label{eq:Vrepresentationbasis}
\end{align}
\eqref{eq:Vrepresentationbasis} should be viewed as a heuristic way to understand the factorization map defined by \eqref{eq:Vcoinvariantrepbasis}, as \eqref{eq:Vcoinvariantrepbasis} does not depend on the choice of splitting of $\ket{\pi,ab}$ into its tensor factors. But we find that \eqref{eq:Vrepresentationbasis} gives an intuitive explanation of what the factorization map does to the state, even for non-compact $G$: it introduces a maximally entangled state (as made manifest by \eqref{eq:characterfndef}) into each sector labeled by $\pi$.

Although it may appear that \eqref{eq:Vrepresentationbasis} implies that $V$ maps $\dket{\vec{g}_{\overline{R}} ; \pi,ab ; \vec{g}_R}$ into a Hilbert space with three tensor factors,  this is an artifact of the basis element $\dket{\vec{g}_{\overline{R}} ; \pi,ab ; \vec{g}_R} $ that $V$ acts on in \eqref{eq:Vrepresentationbasis}. The ``state'' $\ket{\chi_\pi}'$ has support on both $\Hext{R}$ and $\Hext{\overline{R}}$, and determines the entanglement of physical states between these two tensor factors within each sector. We can think of $\ket{\chi_\pi}'$ as being the entangled state of the observer edge modes of $R$ and $\overline{R}$. Because the character functions are unique, the entanglement structure within each sector $\pi$ is unique. 

\begin{figure}
    \centering
    \begin{tikzpicture}[thick,scale=1.5]
        \draw (1.4, 0) rectangle (3.0, 0.7);
        \node at (2.2, 0.35) {$\dket{\psi}$};

        \draw[blue]
        (1.9, 0.7) -- (1.9, 1.3) -- (0.3, 1.3) -- (0.3, 3.9) ;

        \draw[red]
        (0.9, 3.9) -- (0.9, 2.1) -- (1.4, 2.1);
        \draw[blue] (1.4, 2.1) -- (1.9, 2.1) -- (1.9, 3.9);
        \filldraw (1.4, 2.1) circle (0.05);
        \node at (1.4, 1.75) {$\ket{\chi_\pi}'$};

        \draw[red] (2.5, 3.9) -- (2.5, 0.7);

        \draw [thick,decorate,decoration={brace,amplitude=5pt}]
        (1.8, 4) -- (2.6, 4)   node[midway,xshift=0.9em,yshift=1.5em]{$V_\pi \otimes \Ha_{R}(\pi)$};
        \draw [thick,decorate,decoration={brace,amplitude=5pt}]
        (0.2, 4) -- (1, 4)   node[midway,xshift=-0.9em,yshift=1.5em]{$\Ha_{\overline{R}}(\overline{\pi}) \otimes V_\pi^*$};
    \end{tikzpicture}
    \caption{The quantum circuit diagram of our factorization map. In this figure, integrate over each representation of the red and blue lines with respect to the Plancherel measure. These integrals prevent the Hilbert space from factorizing between the red and blue legs.}
    \label{fig:Vcirc}
\end{figure}
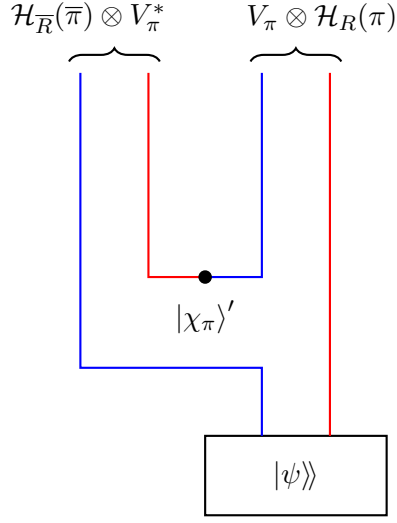

The factorization map $V$ can be given a quantum circuit diagram (Fig.~\ref{fig:Vcirc}). The legs of this figure are associated with a fixed superselection sector $\pi$, and must be integrated with respect to the Plancherel measure. Because of this integral, a red and blue leg must be paired together to form a factorizing Hilbert space. The red line attached to $\dket{\psi}$ is associated with the $R$ boundary legs and the inflowing half of the bulk leg which attaches to this vertex, associated with the gauge fixing edge modes (i.e., the $V_\pi^*$ degrees of freedom of this leg, not the $V_\pi$ degrees of freedom). The blue leg of the character function has support on the observer edge modes of $\Hext{R}$. So the right-most red and blue legs combine to form $\Hext{R}$ after integrating over the representations. The blue line attached to $\dket{\psi}$ is associated with the $\overline{R}$ boundary legs and the outflowing end of the corner leg which attaches to this vertex (i.e., the $V_\pi$ degrees of freedom of this leg, not the $V_\pi^*$ degrees of freedom). These are the gauge fixing edge modes associated with $\overline{R}$. The remaining red leg, associated with the unnormalized character function $\ket{\chi_\pi}'$, corresponds to the observer edge modes of $\overline{R}$.  Together, the left-most red and blue legs form $\Hext{\overline{R}}$. From this perspective, the divergence of $V$ is because of the divergence of the loop in the quantum circuit diagram of $\dbra{\psi}V^\dagger V \dket{\psi}$. 

Said differently, the fact that $V$ is only a projective isometry is related to the normalization of the vector $\ket{\chi_\pi}'$. Indeed, each state $\dket{b_\pi; \vec{g}_R}_R$ is normalizable, so the volume divergence of $V^\dagger V$ must come from the overlap $\braket{\chi_\pi}'$. When $G$ is a compact group, $\ket{\chi_\pi}'$ is related to $\ket{\chi_\pi}$ by a finite normalization constant. When $G$ is non-compact, the proportionality constant is formally divergent.
To understand this divergence in more detail, we can take a partial trace of $\ket{\chi_\pi}'$ over one of its two tensor factors to see that
\begin{align}
    \tr_{V^*_\pi}[\ket{\chi_\pi}'
    \bra{\chi_\omega}'] &= \sum_{a,b} \braket{\pi,a}{\omega,b}\ketbra{\pi,a}{\omega,b} \,,
    \\&= \delta(\pi,\omega)  \sum_{a}\ket{\pi,a}\bra{\pi,a}\,,
    \\&\equiv \delta(\pi,\omega) \Pi_\pi \,. \label{eq:partialtracecharacters}
\end{align}
Recall that $\delta(\pi,\omega)$ is the Plancherel-normalized delta function. $\Pi_\pi $ is an operator on the observer edge mode Hilbert space, and is defined by its algebra
\begin{align}
    \Pi_\pi \Pi_\omega = \delta(\pi,\omega) \Pi_\pi \,. \label{eq:Piproj}
\end{align}
This algebra of projectors $\Pi_\pi$ will serve as our starting point for computing the entropy of the reduced state $\rho_R$ of a boundary subregion. 

Taking trace of both sides of \eqref{eq:partialtracecharacters}, we can see that
\begin{align}
    \braket{\chi_\pi}{\chi_\omega}' = \delta(\pi,\omega) \tr(\Pi_\pi) \,.\label{eq:character_delta_overlap}
\end{align}
The delta function $\delta(\pi,\omega)$ simply removes one of the integrals over representations when taking the partial trace of $V\dket{\psi} \dbra{\psi} V^\dagger$, as we explain in Sec.~\ref{sec:reducedstate}. Thus, the volume divergence of $V$ is equivalent to the fact that $\Pi_\pi$ does not have finite trace with respect to the standard trace on the observer edge modes Hilbert space. Our solution to this, which we explain in Sec.~\ref{sec:algebras}, will be to define a renormalized trace for the operator algebra to which the reduced state belongs.

\subsection{Uniqueness of the factorization map}\label{sec:Arcenter}

We showed in \eqref{eq:Vrepresentationbasis} that within a sector labeled by the representations $\pi$ of the corner leg, the factorization map $V$ uniquely ensures that $V \dket{\vec{g}_{\overline{R}} ; \pi,ab ; \vec{g}_R}$ remains gauge invariant in the factorized Hilbert space. This follows because the character functions are unique. In the group basis, we saw that the factorization map introduces an additional group element $g$ at the bulk splitting surface between $R$ and $\overline{R}$, which can be thought of as the position of a particle on the group manifold of $G$. In continuum Chern--Simons theory, this factorization map was recently shown to be unique in \cite{Mertens:2025ydx}: we have reproduced this result in our model as well.

However, there is an ambiguity in the definition of the factorization map. To see this ambiguity, consider the operator on the physical Hilbert space $\Ha_{\mathrm{phys}}(\Sigma)$ defined by its action
\begin{align}
    \hat{C} \dket{\vec{g}_{\overline{R}} ; \pi,ab ; \vec{g}_R} = C(\pi) \dket{\vec{g}_{\overline{R}} ; \pi,ab ; \vec{g}_R} \,, \label{eq:Crepbasis}
\end{align}
where $C(\pi)$ is the value of a bounded function  $C: \widehat{G} \to \C$ (i.e., $C \in L^\infty(\widehat{G})$).
By linearity, this extends to a definition on all of $\Ha_{\mathrm{phys}}(\Sigma)$. $\hat{C}$ is a physical operator because it maps null states to null states \cite{Balasubramanian:2025rcr}, as one can check explicitly. Because it is a physical operator, it has a representation on the physical Hilbert space regardless of our choice of presentation $\Lambda$. The bowtie lattice we described in Sec.~\ref{sec:themodel} is simply one convenient choice to describe this class of operators.

How does $\hat{C}$ act in the group basis?
We can think of $C(\pi)$ as the Fourier transform of a function
\begin{align}
    C(g) = \int_{\widehat{G}} d\mu(\pi) C(\pi) \chi_\pi(g) \,,
\end{align}
which is constant on conjugacy classes of $G$:
\begin{align}
    C(g) = C(h^{-1} g h) \text{ for all $h \in G$} \,. \label{eq:Fclassfn}
\end{align}
Such functions are called class functions. An example of a class function is the delta function $\delta(g)$, with a Fourier decomposition
\begin{align}
    \delta(g) = \int d\mu(\pi) \chi_\pi(g)\,,
\end{align}
where $\chi_\pi$ is a character in representation $\pi$.
This is the generalization of the familiar statement that the Fourier transform of the delta function is $\hat{\delta}(\pi) = 1$. In the group basis, one can show by Fourier transforming \eqref{eq:Crepbasis} directly that
\begin{align}
    \hat{C} \dket{\vec{g}_{\overline{R}} ; h ; \vec{g}_R}
    = \int_G dk\, C(k) \dket{\vec{g}_{\overline{R}} ; kh ; \vec{g}_R}
    = \int_G dk\, C(k) \dket{\vec{g}_{\overline{R}} ; hk ; \vec{g}_R} \,. \label{eq:Cgroupbasis}
\end{align}
We used the fact that $C$ is a class function to show that these two expressions are equal. 

The fact that $\hat{C}$ acts only on the corner leg separating the two boundary subregions suggests that $\hat{C}$ has support ``between'' the two boundary subregions. We can make this precise as follows.
Consider the algebra $\mathcal{A}_R$ of all operators that can act non-trivially on the boundary vertices associated with the subregion $R$, but trivially on the complementary boundary subregion $\overline{R}$. As stated, $\mathcal{A}_R$ may not be a von Neumann algebra, as it might not be closed under weak limits, i.e., a sequence of matrix elements may not converge to the matrix elements of an operator in the algebra. To address this, we simply complete the algebra by taking the double commutant $\mathcal{A}_R''$.\footnote{The double commutant of an algebra $\mathcal{A}_R$ is constructed by first determining the commutant $\mathcal{A}_R'$, which are the algebra of operators which commute with $\mathcal{A}_R$. The double commutant is then the commutant of the commutant. Intuitively, this completes the algebra because if $\mathcal{O} \not\in \mathcal{A}_R$ but its matrix elements agree with a limit of matrix elements of operators $\mathcal{O}_n \in \mathcal{A}_R$, then if $\mathcal{O}' \in \mathcal{A}_R'$, then $\bra{\psi}[\mathcal{O}',\mathcal{O}]\ket{\psi} = \lim_{n \to \infty} \bra{\psi}[\mathcal{O}',\mathcal{O}_n]\ket{\psi} = 0$.} For notational simplicity, we will let $\mathcal{A}_R = \mathcal{A}_R''$ denote the completed von Neumann algebra. 

Now consider any operator $\mathcal{O} \in \mathcal{A}_R$. It is easy to see that $\mathcal{O}$ and $\hat{C}$ commute, because $\mathcal{O}$ only affects the $\vec{g}_R$ degrees of freedom in \eqref{eq:Crepbasis}. Thus, $\hat{C}$ is an element of the center $\mathcal{Z}(\mathcal{A}_R)$ of the algebra $\mathcal{A}_R$.
In fact, it is not hard to see that these are the \emph{only} central operators of $\mathcal{A}_R$ when $\Sigma$ is a disk. This follows from the Hilbert space decomposition \eqref{eq:HphysAlgebras} of $\Ha_{\mathrm{phys}}(\Sigma_{0,1})$. Operators in $\mathcal{A}_R$ act on each $\Ha_R(\pi)$ factor, and for each factor, the only operator which commutes with all of $\mathcal{A}_R$ is the identity $\Id_{\Ha_R(\pi)}$. Then, the abelian algebra $\mathcal{Z}(\mathcal{A}_R)$ is exactly the algebraic union of the identity of each factor, normalized appropriately to account for the direct integral over all such factors.\footnote{For surfaces of different topology, different boundary conditions on $\Sigma$, or different topological boundary conditions on the cut, the center will be parameterized by other group theoretic data. As an example, see \cite{Balasubramanian:2026xyz} for the edge modes shared between two boundary subregions of the cylinder, each given open boundary conditions.} 

Given any central operator $\hat{C}$ (central with respect to $\mathcal{A}_R$), we can define an alternative factorization map by $V_C = V \hat{C}$, which would factorize the Hilbert space as
\begin{align}
    V_C \dket{\vec{g}_{\overline{R}} ; \pi,ab ; \vec{g}_R} &=  \dket{\vec{g}_{\overline{R}} ; a_\pi}_{\overline{R}}\, C(\pi)\ket{\chi_\pi}'  \dket{b_\pi; \vec{g}_R}_R \,. \label{eq:VCdef}
\end{align}

We saw above that it was crucial that the factorization map behaves as \eqref{eq:VCdef} within each superselection sector in order to retain gauge invariance. Physically, this is because the introduction of the line operator $\hat{C}$ which lives between the factorized regions does not spoil that factorization, and such operators are always central in $\mathcal{A}_R$.
So the family of all possible factorization maps is labeled uniquely by a choice of central operator $\hat{C}$ which is introduced into the state before factorizing the Hilbert space. 

However, the factorization map $V_{\Id} \equiv V$ defined in Sec.~\ref{sec:factorization} is the unique choice of factorization map which does not introduce the defect $\hat{C}$ into the state before we factorize it. In other words, if we reglued the factorized state back into a state in the physical Hilbert space $\Ha_{\mathrm{phys}}(\Sigma)$, then unless we pick the factorization map $V$, the defect $\hat{C}$ will be introduced into the state.  This is the sense in which $V$ is preferred. This will be important when we compute the entropy of the reduced state $\rho_R$, which we define and calculate below. If we use a generalized factorization map $V_C$ rather than the defect-free one $V$, then the degrees of freedom of the operator $\hat{C}$ will contribute to the entropy, which would not be a faithful representation of the entropy shared between the subregions $R$ and $\overline{R}$.

\subsection{The reduced state \texorpdfstring{$\rho_R$}{rhoR}} \label{sec:reducedstate}

We now calculate the reduced state on the $R$ subregion. Explicitly, we will compute\footnote{As we explain in Sec.~\ref{sec:algebras}, the trace on $\Hext{\overline{R}}$ is unique, so there is no ambiguity in this definition.}
\begin{align}
    \rho_R = \Tr_{\overline{R}}[V \dket{\psi}\dbra{\psi} V^\dagger] \,,
\end{align}
where $\Tr_{\overline{R}}$ denotes the trace over the extended Hilbert space of $\overline{R}$. 
Thus, we must trace out a complete basis for $\Hext{\overline{R}}$. This includes both $\dket{\vec{g}_{\overline{R}};a_\pi}_{\overline{R}}$ and half of the vector $\ket{\chi_\pi}'$. We saw in \eqref{eq:partialtracecharacters} that the partial trace of $\ket{\chi_\pi}'$ gives
\begin{align}
    \tr_{V_\pi^*}[\ket{\chi_\pi}' \bra{\chi_\omega}'] = \delta(\pi,\omega) \Pi_\pi \,, \label{eq:characterpartialtrace2}
\end{align}
where $\Pi_\pi$ is the Plancherel normalized projector on the $R$ subregion's observer edge modes defined in \eqref{eq:Piproj}.
It is useful to define the unnormalized coefficients
\begin{align}
    \rho^{\mathrm{un}}_\pi(\vec{g}_R,\vec{h}_R;ab) = \sum_{c,d}\int d[\vec{g}_{\overline{R}},\vec{h}_{\overline{R}}] \psi(\vec{g}_{\overline{R}};\pi,ca;\vec{g}_R) \psi(\vec{h}_{\overline{R}};\pi,db;\vec{h}_R)^*
    \dbraket{\vec{h}_{\overline{R}};d_\pi}{\vec{g}_{\overline{R}};c_\pi}_{\overline{R}} \label{eq:rhoundef}\,,
\end{align}
as well as the normalized density matrix coefficients
\begin{align}
    q_\pi &= \sum_{a,b}\int d[\vec{g}_R,\vec{h}_R]  \rho^{\mathrm{un}}_\pi(\vec{g}_R,\vec{h}_R;ab)\,\dbraket{b_\pi;\vec{h}_{R}}{a_\pi;\vec{g}_{R}}_R \,, \label{eq:qpidef}\\
    \rho_\pi(\vec{g}_R,\vec{h}_R;ab) &= \rho^{\mathrm{un}}_\pi(\vec{g}_R,\vec{h}_R;ab)/ q_\pi \,.
\end{align}
By explicitly plugging in \eqref{eq:qpidef} into \eqref{eq:rhoundef}, it is easy to see that 
\begin{align}
    \int d\mu(\pi) \, q_\pi = \dnorm{\psi}\,, \label{eq:qpi_normalized}
\end{align}
so $q_\pi$ integrates to one (with respect to the Plancherel measure) if and only if the global state $\dket{\psi}$ is normalized in the co-invariant inner product.

Furthermore, we define the partial density matrix
\begin{align}
    \rho_\pi = \sum_{a,b}\int d[\vec{g}_R,\vec{h}_R]  \rho_\pi(\vec{g}_R,\vec{h}_R;ab) \dket{a_\pi;\vec{g}_{R}}\dbra{b_\pi;\vec{h}_{R}}_R \,. \label{eq:rhopidef}
\end{align}
By construction, the reduced density matrix is normalized for every $\pi$. In other words, $\tr_{\Ha_R(\pi)}[\rho_\pi]=1$ for every sector.\footnote{Here, $\tr_{\Ha_R(\pi)}[\cdots]$ is the unique trace for the algebra $\mathcal{B}(\Ha_R(\pi))$ of bounded operators on $\Ha_R(\pi)$. As we will review below, this trace is unique precisely because $\mathcal{B}(\Ha_R(\pi))$ has a trivial center: all central operators in $\mathcal{A}_R$ come from relative rescalings within each sector labeled by $\pi$.}
Then by explicitly expanding $\dket{\psi}$ in the $\dket{\vec{g}_{\overline{R}} ; h ; \vec{g}_R} $ basis, using the factorization map action \eqref{eq:Vgroupbasisfactorized}, and plugging in the expressions \eqref{eq:characterpartialtrace2} and \eqref{eq:rhopidef} to the resulting trace formula, it is straightforward to show that
\begin{align}
    \rho_R = \int d\mu(\pi) q_\pi \rho_\pi \otimes \Pi_\pi \,. \label{eq:rhoRfirst}
\end{align}
This reduced state $\rho_R$ is an operator on the extended Hilbert space $\Hext{R}$, and its form can be understood schematically as follows. The partial trace over the $\overline{R}$ extended Hilbert space must be performed sector by sector, with the sectors labeled by irreps $\pi$, as explained in \eqref{eq:Hextdecomp}. The partial trace over the observer edge modes produces a crucial factor $\delta(\pi,\omega)$ which reduces the double integral over representations (one each for the bra and the ket) to a single integral, as well as producing the projector $\Pi_\pi$ acting on the observer edge modes of each sector. Within a fixed sector, $\rho_\pi$ is the normalized reduced density matrix, and $q_\pi$ is the probability density (normalized with respect to the Plancherel measure) for the reduced state $\rho_R$ to be found in the $\pi$ representation. Integrating over all the sectors with respect to the Plancherel measure, we obtain our result. In Sec.~\ref{sec:algebras}, we will discuss a more algebraic approach to obtaining the same result more directly, without the intermediate step of tracing out the $\overline{R}$ extended Hilbert space.

We can compare this form of the density matrix to the literature when $G$ is compact. In that case, it is known \cite{Donnelly:2011hn} that
\begin{align}
    \rho_R = \sum_\pi p_\pi \left[\rho_\pi \otimes \frac{\Id_{V_\pi}}{d_\pi} 
    \right] \,,
\end{align}
where $\sum_\pi p_\pi = 1$, and $\Id_{V_\pi}$ is the identity matrix on the $\pi$ observer edge modes.
Noting that
\begin{align}
    \frac{\Id_{V_\pi}}{d_\pi} \frac{\Id_{V_\omega}}{d_\omega} = \frac{\delta_{\pi\omega}}{d_\pi} \frac{\Id_{V_\pi}}{d_\pi} \,,
\end{align}
and comparing to the algebraic relation \eqref{eq:Piproj}, we can see that $\Pi_\pi = \Id_{V_\pi} / d_\pi$ when $G$ is compact. This is a consequence of the observer edge modes being maximally entangled within each sector. Thus, the family of projectors $\Pi_\pi$ satisfying the algebra \eqref{eq:Piproj} is the natural generalization of the maximally mixed state when $G$ is non-compact. 

To ensure that $\rho_R$ has unit trace, we must be careful because the naive trace of $\Pi_\pi$ is divergent, as we noted near \eqref{eq:character_delta_overlap}. To deal with this, in Sec.~\ref{sec:algebras} we will explain how to define a renormalized trace and its algebraic origin.

\section{The renormalized trace and operator algebras} \label{sec:algebras}

In this section, we will discuss how to renormalize the trace on the factorized Hilbert space $\Hext{R}$ so that the factorization map $V$ is an isometry, even when $G$ is non-compact. 

In the theory of von Neumann algebras, a state $\phi$ is defined as a linear map from operators $\mathcal{O}$ in the algebra to the complex numbers, such that the value assigned to a positive semi-definite operator is positive. Additionally, a state is normalized such that the identity map evaluates to 1. If $\phi(\Id)=\infty$, but $\phi$ still satisfies all the other properties listed above, then we instead call $\phi$ a weight.

We will first consider the case where $\mathcal{O}$ is an element of an algebra with  a trivial center, i.e., it is a {\it factor}. There are three qualitative families of factors, type I, type II, and type III, which differ in their finiteness properties \cite{Sorce:2023fdx}. We will be most interested in type I factors, because the algebra of operators acting on  topological tensor networks can be decomposed into a direct integral of type I factors.
In simple systems such as the harmonic oscillator, we can think of the algebraic state $\phi_\rho$ as taking the expectation value of $\mathcal{O}$ with respect to a density matrix $\rho$:
\begin{align}
    \phi_\rho (\mathcal{O}) = \tr(\rho \mathcal{O}) \,.
\end{align}
Such states are called normal. In more complicated systems, such as a topological tensor network, the algebraic approach is useful because we can bypass the separate definition of the density matrix $\rho$ and the trace $\tr$, as expectation values are simply defined via the map $\phi_\rho$. This freedom allows us to rigorously define a renormalized trace for many algebras by first defining a well-defined state $\phi_\rho$ and reverse engineering the definition of trace and density matrix, a procedure that works so long as the algebra is not a type III factor. Given a definition of trace on the algebra, states $\phi_\rho$ can be associated with density matrices $\rho$. If the algebra is type II, then this is the best we can do: there are no pure states on a type II algebra. If the algebra is type I, then we can find special density matrices which are rank 1, i.e., pure states. 

\begin{table}
    \centering
    \begin{tabular}{|c|c|}
        \hline
        Name & Object    \\
        \hline
        \hline
        $\mathcal{B}_R$ & Bounded operators on the extended Hilbert space $\Hext{R}$. \\
        & Line operators that can end on the $R$ vertices or the cut.  \\
        & Unique trace defined in \eqref{eq:BRtrace}. 
        \\\hline
        $\mathcal{A}_R$ & Bounded operators on the $R$ boundary vertices. \\
        & Line operators that can only end on the $R$ vertices. \\
        & Trace determined by defect-free factorization map in \eqref{eq:physicaltrace}. \\\hline
        $\mathcal{Z}(\mathcal{A}_R)$ & Central operators of $\mathcal{A}_R$. \\
        & Operators within $\mathcal{A}_R$ which only act on the edge modes. \\
        & Trace determined by microcanonical flatness of the edge modes in \eqref{eq:centertrace}. \\\hline
        $\mathcal{E}$ & Operator valued weight $\mathcal{E}: \mathcal{B}_R \to \mathcal{A}_R$. \\
        & Embeds $\mathcal{A}_R \subset \mathcal{B}_R$ and renormalizes $\Tr_{\mathcal{A}_R}$. \\
        & Used to prove the uniqueness of our renormalization procedure for the trace. \\\hline
        $\mathcal{F}$ & Operator valued weight $\mathcal{F}:  \mathcal{A}_R \to \mathcal{Z}(\mathcal{A}_R)$. \\
        & Embeds $\mathcal{Z}(\mathcal{A}_R) \subset \mathcal{A}_R$ and renormalizes $\Tr_{\mathcal{Z}(\mathcal{A}_R)}$, \\
        & Used to determine the edge contribution to the entropy of $\rho_R$. \\\hline
    \end{tabular}
    \caption{Table summarizing the mathematical objects appearing in this section, as well as their purpose and their relationships to each other.}
    \label{tab:algebras}
\end{table}

If a von Neumann algebra has a non-trivial center, we can decompose it into many subalgebras which do have trivial centers, and each subalgebra will be type I, II, or III. In simple settings, these factors will be labeled by a discrete set of data, so the total von Neumann algebra is a direct sum of  factors. In more complicated settings, the labels of these factors are continuous, and the algebra is instead a direct integral over the factors. A central operator $\hat{C}$ acts proportionally to the identity within each factor, and the entire action of $\hat{C}$ is determined by the sector-dependent proportionality constant.

The definition of the trace within a type I or II factor\footnote{A type III factor has no notion of trace.} is unique, possibly up to a choice of normalization if the algebra acts on an infinite dimensional Hilbert space. However, if a von Neumann algebra has a non-trivial center (so that it can be decomposed into a non-trivial direct sum/integral of multiple factors), there is a priori nothing which fixes the relative normalization of this factor-wise unique trace. 
Thus, the trace of an operator is not uniquely defined if it is part of an operator algebra with a non-trivial center \cite{Sorce:2023fdx} without fixing this ambiguity first.

In Table~\ref{tab:algebras}, we provide a short summary of the mathematical concepts we develop in this section. We will attempt to make the remaining sections fairly self-contained, so a reader who is  not interested in the details may wish to skip to Sec.~\ref{sec:entropy_nomatter} and refer to this table when needed.

\subsection{An example: \texorpdfstring{$G = \R$}{G=R}} \label{sec:Rexample}

Before we derive our formula for the renormalized trace for general gauge groups $G$, we will first explain how to proceed in an example, by defining the reduced density matrix when $G=\R$. We use this example to explain the  properties that will generalize to arbitrary non-compact transformable groups.
When $G=\R$, $\widehat{G} = \R$ as well, and the representations $\pi$ are just momenta $k$. Each representation Hilbert space $V_k$ is one dimensional because $\R$ is abelian, so $\ket{k,ab} \equiv \ket{k}$. Furthermore, each $V_k$ is spanned by the plane waves $ \ket{k} = \int dx\, e^{ikx} \ket{x}$. 

For the moment we will suppress the dependence of $\dket{\psi}$ on the boundary legs, and focus on the behavior of the edge modes. In this case, the factorization map is
\begin{align}
    V\ket{\psi} = \int dk \,\psi(k) \ket{k}_{\overline{R}}\ket{k}_R \,.
\end{align}
The reduced state is therefore
\begin{align}
    \rho_R = \int dk |\psi(k)|^2 \ketbra{k} \,.
\end{align}
Let $\mathcal{A}_R$ be the von Neumann algebra of all operators on $L^2(\R)$ which are diagonal in the momentum basis,\footnote{In particular, this is the algebra $L^\infty(\R)$ of bounded functions of the momentum operator $\hat{k}$. To complete $L^\infty(\R)$ and recover the operators missing from this subalgebra, one must add the position operator $\hat{x}$ and complete the resulting algebra.} of which the reduced state $\rho_R$ is a special case. Associated with $\mathcal{A}_R$ is a choice of trace, which is not necessarily the same as the trace defined for a generic operator acting on $L^2(\R)$. Indeed, if we defined $\Tr$ to be the usual trace of bounded operators on $L^2(\R)$, then we would find that 
\begin{align}
    \Tr(\rho_R) = \int dk\, |\psi(k)|^2 \braket{k} = \mathrm{Vol}(\R) \int dk\, |\psi(k)|^2 \,. \label{eq:tracediv_reals}
\end{align}
As expected, this has a volume divergence from the momentum-space delta function. On the other hand, if we {\it defined} the trace on $\mathcal{A}_R$ to be
\begin{align}
    \Tr_{\mathcal{A}_R}(\rho_R) = \int dk \,|\psi(k)|^2 \,, \label{eq:Rrenormalizedtrace}
\end{align}
then $\rho_R$ is properly normalized if $\ket{\psi}$ was. Furthermore, this definition of the trace is linear in $\rho$ and satisfies $\Tr_{\mathcal{A}_R}(\rho \sigma) = \Tr_{\mathcal{A}_R}(\sigma \rho)$ for any positive operators in $\mathcal{A}_R$.  So this is a legitimate definition of a trace on $\mathcal{A}_R$ \cite{Sorce:2023fdx}. As a heuristic way of understanding the trace in \eqref{eq:Rrenormalizedtrace} we can say that we have defined 
\begin{align}
    \Tr_{\mathcal{A}_R}(\ketbra{k}) = 1 \,.
\end{align}

There is of course an ambiguity in this definition of trace. If $C(k)$ is any positive function of $k$, then 
\begin{align}
    \Tr^C_{\mathcal{A}_R}(\rho_R) = \int dk \, C(k) \, |\psi(k)|^2
\end{align}
also satisfies all the requirements. This is the central ambiguity of the trace we mentioned in the introduction to this section. Fixing  $C(k)$, and therefore the trace on $\mathcal{A}_R$, requires additional physical input beyond the algebra itself.

For completeness, we will now restore the dependence on the boundary legs and show that a similar definition of the trace continues to normalize the state for the full tensor network. By equation \eqref{eq:Vgroupbasisfactorized}, in the group basis, the factorized state $V\dket{\psi}$ takes the form
\begin{align}
    V \dket{\psi} = \int d[\vec{x}_R ,\vec{y}_{\overline{R}},a,b] \psi(\vec{x},\vec{y}, a+b) \dket{\vec{y}, b}_{\overline{R}} \dket{a,\vec{x}}_R \,.
\end{align}
In the representation basis for the edge modes, the state instead takes the form
\begin{align}
    V \dket{\psi} = \int d[\vec{x}_R ,\vec{y}_{\overline{R}},k] \psi(\vec{x},\vec{y}, k) \dket{\vec{y}, k}_{\overline{R}} \dket{k,\vec{x}}_R \,,
\end{align}
where, e.g., $\dket{k,\vec{x}}_R$ is the image of $\ket{k}\ket{\vec{x}}$ under the Gauss law projector on the $R$ bulk vertex, and $\psi(\vec{x},\vec{y}, k)$ is the Fourier transform of $\psi(\vec{x},\vec{y}, a)$ in the last variable. Indeed, this matches \eqref{eq:Vcoinvariantrepbasis} because $V_k$ is one dimensional, so there are no indices to sum over.

Now we perform the partial trace over the $\overline{R}$ degrees of freedom, leaving the reduced state
\begin{align}
    \rho_R = \int d[\vec{x}_R, \vec{y}_R, \vec{m}_{\overline{R}},\vec{n}_{\overline{R}}, k,q] \left( \psi(\vec{x},\vec{m}, k)\psi^*(\vec{y},\vec{n}, q)\right)  \dket{k,\vec{x}} \dbra{q,\vec{y}} \dbraket{\vec{n},q}{\vec{m}, k} \,.
\end{align}
Using the definition of the co-invariant inner product, one can show that 
\begin{align}
    \dbraket{\vec{n},q}{\vec{m}, k} = \delta(k-q)\int dx \, e^{ikx} \langle{\vec{n}}|{\vec{m} + x\cdot \vec{1}}\rangle \,. \label{eq:Rexamplebraket}
\end{align}
Therefore, simplifying notation by defining
\begin{align}
    \rho(\vec{x},\vec{y} ; k) = \int d[\vec{m},b]\psi(\vec{x},\vec{m}, k)\psi^*(\vec{y},\vec{m} + b\cdot \vec{1}, k)e^{ikb}\,,
\end{align}
the reduced state is 
\begin{align}
    \rho_R &= \int d[\vec{x},\vec{y},k,q] \rho(\vec{x},\vec{y} ; k) \delta(k-q) \dket{k,\vec{x}} \dbra{\vec{y}, q} \,, \\
    &= \int d[\vec{x},\vec{y},k] \rho(\vec{x},\vec{y} ; k) \dket{k,\vec{x}} \dbra{\vec{y}, k} \,.
\end{align}
However, this operator has a divergent trace because the momenta $k$ are forced to be the same: this leads to a $\delta(k=0)$ volume divergence from \eqref{eq:Rexamplebraket} when we evaluate the trace of $\rho_R$. Thus, we renormalize by defining 
\begin{align}
    \Tr_{\mathcal{A}_R}(\rho_R) = \int d[\vec{x},\vec{y},k] \rho(\vec{x},\vec{y} ; k) \int da \, e^{ika} \langle{\vec{y}}|{\vec{x} + a\cdot \vec{1}}\rangle \,.
\end{align}
With this definition, one can verify $\rho_R$ has trace 1 if $\dket{\psi}$ was normalized in the co-invariant inner product as a special case of \eqref{eq:qpi_normalized}. Indeed, the extra terms in the trace are just \eqref{eq:Rexamplebraket} without the divergent $\delta(k=0)$, so this definition is the same as the ``naive'' definition of the trace but without the volume divergence.

\subsection{The algebra \texorpdfstring{$\mathcal{B}_R$}{BR} and its trace}

Recall that the extended Hilbert space $\Hext{R}$ can be decomposed as
\begin{align}
    \Hext{R} &= \int d\mu(\pi)\, \Ha_R(\pi) \otimes V_\pi \,,\\
    \Ha_R(\pi) &= \int_{\widehat{G}}^{\oplus} d\mu(\vec{\omega}) V_{\vec{\omega}} \otimes \Pi_A[V_{\vec{\omega}}^* \otimes V_\pi^*] \,.
\end{align}
Therefore, a resolution of the identity on $ \Hext{R} $ takes the form
\begin{align}
    \Id_{\mathrm{ext}} = \int d\mu(\pi) \sum_{\alpha, i} \ketbra{\pi; \alpha, i}\,,
\end{align}
where $\alpha$ is an index for $\Ha_R(\pi)$, $i$ is an index for $V_\pi$, and $\braket{\pi; \alpha, i}{\omega;\beta,j} = \delta(\pi,\omega) \delta_{\alpha \beta} \delta_{i j}$.
Using two resolutions of the identity, the most general operator on the extended Hilbert space takes the form
\begin{align}
    \mathcal{O} & = \int d\mu(\pi,\omega) \sum_{\alpha,\beta,i,j} \mathcal{O}(\pi,\omega)_{\beta, j}^{\alpha, i} \ketbra{\pi ; \alpha, i}{\omega ; \beta, j}  \,,
\end{align}
where $\mathcal{O}(\pi,\omega)_{\beta, j}^{\alpha, i}$ are a set of coefficients.

We refer to the set of all bounded operators $\mathcal{O}$ on the extended Hilbert space as $\mathcal{B}_R$. This is the analog of ``all bounded operators on $L^2(\R)$'' from Sec.~\ref{sec:Rexample}.
The algebra of bounded operators on a Hilbert space is always a type I factor, and so the trace on $\mathcal{B}_R$ is uniquely defined (up to an overall normalization constant) by the sum over the diagonal matrix elements of $\mathcal{O}$. Explicitly,
\begin{align}
    \Tr_{\mathcal{B}_R}[\mathcal{O}] = \int d\mu(\pi) \sum_{\alpha,i} \mathcal{O}(\pi,\pi)_{\alpha, i}^{\alpha, i}\,. \label{eq:BRtrace}
\end{align}
If we take the expression for $\rho_R$ from \eqref{eq:rhoRfirst}, plug it into this definition of the trace, and recall that the sum over the $\alpha$ index produces a factor of $\tr_{\Ha_R(\pi)}[\rho_\pi] = 1$, we find that 
\begin{align}
    \Tr_{\mathcal{B}_R}[\rho_R] = \int d\mu(\pi) q_\pi \tr_{V_\pi}[\Pi_\pi] =\int d\mu(\pi) q_\pi \cdot \dim(V_\pi) \delta(\pi,\pi) = \infty \,. \label{eq:tracevoldiv}
\end{align}
For compact gauge groups $G$, $\dim(V_\pi) \delta(\pi,\pi) = \mathrm{Vol}(G)$, so this divergence is only an issue for non-compact groups, as explained in the example $G=\R$ near \eqref{eq:tracediv_reals}. This is precisely the volume divergence \eqref{eq:groupbasisVoldiverge} which made the factorization map $V$ only a projective isometry.

\paragraph{Relation to the boundary Hilbert space.} To better understand the extended Hilbert space, it is convenient to use the Gauss law at the bulk vertex of $\Lambda_R$ to gauge fix the group element on the corner leg to the identity. Recall that a gauge transformation at that vertex acts as $h \to h \alpha^{-1}$, $\vec{g}_R \to  \vec{g}_R \cdot \alpha^{-1}$, so a gauge invariant wave function satisfies $\psi(h,\vec{g}_R) = \psi(e, \vec{g}_R \cdot h^{-1})$ and is determined by its restriction to $h = e$. The extended Hilbert space is therefore equivalent to the \emph{unconstrained} Hilbert space of the boundary legs of $R$:
\begin{align}
    \Hext{R} \cong \bigotimes_{\ell \in R} L^2(G) \,.
    \label{eq:gaugefixed}
\end{align}
Gauge fixing the corner leg trades the constraint for the corner degrees of freedom, which is the lattice statement of the familiar fact that the extended Hilbert space of a subregion is the one with the Gauss law relaxed at the entangling surface \cite{Donnelly:2011hn,Donnelly_2016}. In the present model this relaxation is nothing new: the vertex created at the cut is one more vertex at which the electric constraint is not imposed, on exactly the same footing as the $n$ marked points of Sec.~\ref{sec:bcs}, so the corner is treated as an additional marked point of the physical boundary. It is also clear from this presentation that $\mathcal{B}_R$ is isomorphic to the complete set of bounded operators on the $R$ boundary vertices.

\subsection{The algebra \texorpdfstring{$\mathcal{A}_R$}{AR} and its trace} \label{sec:A_Ralg}

As explained in Sec.~\ref{sec:Arcenter}, we refer to the algebra of operators in the physical Hilbert space that act only on the $R$ boundary vertices as $\mathcal{A}_R$. This algebra has a faithful representation on the extended Hilbert space as the commutant of the operators that act only on the edge modes. Intuitively, the physical operators on the full disk are always superpositions and products of boundary anchored line operators. The operators with support  only on the $R$ vertices, when the complementary region $\overline{R}$ has been traced out, will then never intersect the corner leg.

As represented on the extended Hilbert space, the most general operator in $\mathcal{A}_R$ is therefore
\begin{align}
    \mathcal{O}_{\mathcal{A}_R} = \int d\mu(\pi) \, \mathcal{O}(\pi) \otimes \Pi_\pi\,,
\end{align}
where $\mathcal{O}(\pi)$ is an operator from $\Ha_R(\pi)$ to itself. In other words, it is diagonal in the $\pi$ label, and central on the observer edge modes in each sector. 
This is the analog of the operators ``diagonal in the $k$ basis'' from Sec.~\ref{sec:Rexample}.
Thus, the reduced state $\rho_R$ is always an element of the $\mathcal{A}_R$ subalgebra of the full algebra $\mathcal{B}_R$ of all operators on the extended Hilbert space. This means that if we can define a renormalized trace for just this subalgebra, then it is this trace that will enforce that $\rho_R$ is properly normalized.

A trace on $\mathcal{A}_R$ is given by
\begin{align}
    \Tr_{\mathcal{A}_R}[\mathcal{O}] = \int d\mu(\pi) \tr_{\Ha_R(\pi)}[\mathcal{O}(\pi)] \,, \label{eq:physicaltrace}
\end{align}
where $d\mu(\pi)$ is the Plancherel measure. 
This is a trace because it is a linear map from $\mathcal{A}_R$ to complex numbers that satisfies the necessary algebraic properties, such as $\Tr_{\mathcal{A}_R}[\mathcal{O}_1 \mathcal{O}_2] = \Tr_{\mathcal{A}_R}[\mathcal{O}_2 \mathcal{O}_1]$ (cyclicity of the trace) and $\Tr_{\mathcal{A}_R}[\mathcal{O}_1^\dagger \mathcal{O}_1] \geq 0$ (positive operators have positive trace) for any operators $\mathcal{O}_1, \mathcal{O}_2 \in \mathcal{A}_R$.

However, as explained above, $\mathcal{A}_R$ has a large center, generated by the operators $\hat{C}$ of \eqref{eq:Crepbasis}. Thus, a choice of trace for $\mathcal{A}_R$ is equivalent to a choice of central operator $\hat{C}$ to insert into the example trace given above:
\begin{align}
    \Tr_{\mathcal{A}_R}^{C} [\mathcal{O}] = \Tr_{\mathcal{A}_R}[\hat{C} \mathcal{O}] = \int d\mu(\pi) \, C(\pi) \tr_{\Ha_R(\pi)}[\mathcal{O}(\pi)] \,. \label{eq:CasMeasure}
\end{align}
For this to be a legitimate trace, we need $C(\pi) \in \R^+$ for all representations $\pi$.
In other words, for any choice of measure over the representations, we can define a trace of $\mathcal{A}_R$. So the question ``which trace?'' is equivalent to the question ``which measure on $\widehat{G}$?'', and the answer cannot come from the algebra alone: it requires  additional physical input.

The obvious source of such input (restricting the
unique trace on the ambient algebra $\mathcal{B}_R$) is not available to us, because of the volume divergence \eqref{eq:tracevoldiv}. The principle we will use instead follows from an observation that turns the two ambiguities we have encountered (the choice of trace on $\mathcal{A}_R$, and the choice of factorization map $V_C = V \hat{C}$ from Sec.~\ref{sec:Arcenter}) into a single one. Recall from \eqref{eq:character_delta_overlap} that the failure of $V$ to be an isometry is controlled entirely by the norm of the observer edge mode state $\ket{\chi_\pi}' $, and that $V_C$ multiplies this state by $C(\pi)$. Then if we define $\rho_R^C$ to be the reduced state descending from the factorization map $V_C$, we can use the same chain of arguments as in Sec.~\ref{sec:reducedstate} to show that
\begin{align}
    \rho_R^C &= \Tr_{\overline{R}}[V_C \dket{\psi}\dbra{\psi}V_C^\dagger]\,,
    \\&=\int d\mu(\pi) \, q_\pi \, \rho_\pi \otimes
    |C(\pi)|^{2}\,\cdot\,\Pi_\pi \,.
    \label{eq:rhoRC}
\end{align}
Comparing with \eqref{eq:CasMeasure}, and using \eqref{eq:qpi_normalized}, we can see that
\begin{align}
    \def\xx{5pt}
    \Tr^{|C|^{-2}}_{\mathcal{A}_R}\left[\rho_R^C\right] = \int d\mu(\pi) \,
    \underbrace{|C(\pi)|^{-2}}_{\text{trace}} \hspace{\xx}\cdot\hspace{-\xx}
    \underbrace{|C(\pi)|^{2}}_{\text{factorization map}} \hspace{-\xx}\cdot\hspace{\xx} q_\pi = \int d\mu(\pi) \, q_\pi = 1 \,.
    \label{eq:rhoRCnorm}
\end{align}
More explicitly,
\begin{align}
    V_C \text{ is an isometry} \qquad \Longrightarrow \qquad \Tr_{\mathcal{A}_R} = \Tr^{|C|^{-2}}_{\mathcal{A}_R} \,. \label{eq:VCisometry}
\end{align}
For each factorization map, there is exactly one trace which makes it an isometry.\footnote{The converse is only partially true: a choice of trace only fixes the factorization map up to a unitary $\hat{U} \in \mathcal{Z}(\mathcal{A}_R)$, because $\hat{U}$ will act as a phase in each sector. So, strictly speaking, the defect-free factorization map is a stronger assumption than is needed to fix the trace of $\mathcal{A}_R$.} We may therefore trade the problem of choosing a trace for the problem of choosing a factorization map, which is a question we can
answer on physical grounds.

In fact, we already answered this question in Sec.~\ref{sec:Arcenter}: the factorization maps $V_C$ with $\hat{C} \neq \Id$ insert a defect at the corner before cutting the state open. Demanding that no such defect is introduced then requires using the factorization map $V$ itself. Through \eqref{eq:VCisometry}, this means that the physical choice for the trace $\Tr_{\mathcal{A}_R}$ is given by \eqref{eq:physicaltrace}, which integrates the representations using the Plancherel measure $d\mu(\pi)$. This fixes the ambiguity in the trace, and we will use this trace for $\mathcal{A}_R$ in the remainder of this paper. More carefully, this only fixes the trace up to a final, overall scale factor, corresponding to the choice of scale for the trace of $\mathcal{B}_R$. Rescaling $\Tr_{\mathcal{A}_R} \to \kappa^{-1}\Tr_{\mathcal{A}_R}$, which can be interpreted as a rescaling of the Haar measure for $G$ by $dg \to \kappa \, dg$, this remaining factor can be fixed by e.g. demanding the maximal compact subgroup $K \subset G$ has unit volume. As we will see below, this final ambiguity only affects the entropy of a state $\rho_R$ up to the state independent constant $\ln(\kappa)$, which we will discuss in more detail in Sec.~\ref{sec:entropy_nomatter}.

\subsubsection*{The relation between $\mathcal{B}_R$, $\mathcal{A}_R$, and the edge modes}

It is worth asking what operators $\mathcal{A}_R$ is missing relative to $\mathcal{B}_R$, since this is what distinguishes an algebra with a large center, and hence an ambiguous trace, from a factor with a unique one. 

To answer this, it is convenient to use the presentation \eqref{eq:gaugefixed} of $\Hext{R}$. In this presentation, the corner symmetry $A_{\eth}(g)$ of Sec.~\ref{sec:factorization} acts as the diagonal translation $U(g): \psi(\vec{g}_R) \mapsto \psi(\vec{g}_R \cdot g)$. We call the algebra of all such translations $\mathcal{A}_{\mathrm{edge}}$. Two identifications follow immediately. First,
\begin{align}
    \mathcal{A}_R = \mathcal{A}_{\mathrm{edge}}' \,, \label{eq:ARcommutant}
\end{align}
that is, $\mathcal{A}_R$ is exactly the algebra of operators invariant under the corner symmetry. Second, taking commutants once more,
\begin{align}
    \mathcal{A}_R' = \mathcal{A}_{\mathrm{edge}}'' \cong \int_{\widehat{G}}^{\oplus} d\mu(\pi) \,
    \mathcal{B}(V_\pi) \,. \label{eq:edgealgebra}
\end{align}
Here, $\mathcal{B}(V_\pi)$ is the algebra of all bounded operators acting sector by sector on the observer edge modes; this particular algebra is also called the group von Neumann algebra of $G$ \cite{Dixmier1977,takesaki2003theory,Haar}. Its center, and therefore the center of $\mathcal{A}_R$, is $L^\infty(\widehat{G})$, i.e., the algebra of bounded functions over the superselection labels under pointwise multiplication, which is exactly the family of central operators $\hat{C}$ of \eqref{eq:Crepbasis} that made the trace ambiguous. 

Now, $\mathcal{A}_R$ and $\mathcal{A}_R'$ together generate only the operators that are block diagonal in $\pi$, i.e., 
\begin{align}
    \mathcal{A}_R \vee \mathcal{A}_R' = \int_{\widehat{G}}^\oplus d\mu(\pi) \mathcal{B}(\Ha_R(\pi) \otimes V_\pi) \,,  
\end{align}
where $\mathcal{A}_R \vee \mathcal{A}_R'$ is the algebraic union of $\mathcal{A}_R $ and $\mathcal{A}_R'$. What is missing from $\mathcal{A}_R \vee \mathcal{A}_R'$ is precisely the operators which move between superselection sectors. In the gauge fixed presentation \eqref{eq:gaugefixed} these are supplied by the multiplication operators
\begin{align}
    (M_f \psi)(\vec{g}_R) = f(g_1) \, \psi(\vec{g}_R) \,, \qquad f \in L^\infty(G)
\end{align}
acting on any one boundary leg $g_1$. Undoing the gauge fixing identifies $M_f$ as an open line operator running from that boundary vertex to the corner, with $f$ the bounded function smearing the holonomy it measures. So the operators missing from $\mathcal{A}_R$ are the line operators which end on the edge modes, and they carry an $L^\infty(G)$ worth of data. 

The line operators which stretch from the corner to a boundary vertex do not commute with the boundary anchored line operators, so when we combine $\mathcal{A}_R$ and $L^\infty(G)$ to form $\mathcal{B}_R$, we must account for the non-commutation between them. Let $U_1(g)$ denote the left group action on the boundary leg that $L^\infty(G)$ acts on: this can be thought of as a line operator which is anchored on either side of the boundary leg. Then the operators in $\mathcal{A}_R$ and $L^\infty(G)$ satisfy the commutation relation
\begin{align}
    U_1(g) M_f U_1(g)^{-1} = M_{\alpha_g(f)} \,, \qquad  \alpha_g(f)(k) = f(g^{-1}k) \,.
\end{align}
Every other operator within $\mathcal{A}_R$ commutes with $L^\infty(G)$.

With this commutation relation understood, we conclude that $\mathcal{B}_R = \mathcal{A}_R \vee L^\infty(G)$. The reason this abstract construction is clarifying is that it says that the edge modes introduced by the factorization map $V$ are the minimal set that does the job: exactly one copy of $L^\infty(G)$ at the corner is required to kill the center of $\mathcal{A}_R$ and factorize the Hilbert space. This is the precise sense in which $V$ adds no more structure than factorization demands. Additionally, the idea that additional degrees of freedom are required to define the algebra of a subregion in quantum gravity has appeared in many recent works \cite{Chandrasekaran:2022cip,Kudler-Flam:2023qfl,Chandrasekaran:2022eqq,Witten:2023qsv,Witten:2023xze,Chen:2024rpx,Witten:2021unn,Jensen2023,AliAhmad:2023etg,Faulkner:2024gst,DeVuyst:2024khu,Fewster:2024pur,Abdalla:2025gzn,Harlow:2025pvj,Soni:2023fke,Akers:2024wab,Klinger:2023tgi,Klinger:2023auu,Krishnan:2023fnt,Gomez:2022eui,Ciambelli:2021nmv,Balasubramanian:2023dpj,Klinger:2026tws}. In these works, the extra degrees of freedom are often called ``observers'' or ``edge modes'' (or, sometimes, ``quantum reference frames''), which is our reason for the terminology ``observer edge modes''.

\subsection{The center \texorpdfstring{$\mathcal{Z}(\mathcal{A}_R)$}{Z(A\_R)} and its trace}
\label{sec:centertrace}

There is one more algebra for which  we need to define the trace  before computing the entropy of $\rho_R$: the center $\mathcal{Z}(\mathcal{A}_R)$. In the previous subsection, we fixed the trace on $\mathcal{A}_R$ in \eqref{eq:physicaltrace} by demanding that the associated factorization map introduce no defect at the corner.
Unfortunately, however, the resulting trace $\Tr_{\mathcal{A}_R}$ diverges for the operators in the center $\mathcal{Z}(\mathcal{A}_R)$,\footnote{This is because elements in the center $\mathcal{Z}(\mathcal{A}_R)$ must satisfy $\mathcal{O}(\pi) = \Id_{\Ha_R(\pi)}$, whose trace diverges in an infinite dimensional Hilbert space.} so to define a trace on the center, we must renormalize the trace one more time.

Recall from Sec.~\ref{sec:Arcenter} that the center $\mathcal{Z}(\mathcal{A}_R)$ is generated by the operators $\hat{C}$ of \eqref{eq:Crepbasis}, which act as multiplication by a bounded function $C: \widehat{G} \to \C$ on each superselection sector. As an algebra, then, $\mathcal{Z}(\mathcal{A}_R) \cong L^\infty(\widehat{G})$ under pointwise multiplication. 
Physically, $\mathcal{Z}(\mathcal{A}_R)$ consists of the gauge invariant boundary anchored line operators that sit between $R$ and $\overline{R}$. In the bowtie lattice presentation of the physical Hilbert space, such operators can always be deformed to only act on the corner leg, so it will commute with any operator in $\mathcal{A}_R$. Sliding this line operator past the $R$ vertex of the bowtie lattice, the line operator can equivalently be represented as acting on only the $R$ boundary legs (or, by deforming the line operator in the opposite direction, the $\overline{R}$ boundary legs). Thus, $\mathcal{Z}(\mathcal{A}_R) \subset \mathcal{A}_R$, even though we can equivalently think about the line operators generating $\mathcal{Z}(\mathcal{A}_R)$ as acting on the corner.

Because the algebra $\mathcal{Z}(\mathcal{A}_R)$ is abelian, a trace on it is simple to construct: cyclicity is immediate, and the only remaining requirement is that positive operators have positive trace. Thus every trace on $\mathcal{Z}(\mathcal{A}_R)$ takes the form
\begin{align}
    \Tr^\nu_{\mathcal{Z}(\mathcal{A}_R)}[\hat{C}] = \int_{\widehat{G}} d\nu(\pi) \, C(\pi) \label{eq:centertrace}
\end{align}
for some positive measure $\nu$ on $\widehat{G}$, and conversely every such measure defines a trace. 

In Sec.~\ref{sec:A_Ralg}, we used the physical principle that the factorization map should not introduce any defects into the global state to determine that the analogous trace defining-measure for $\mathcal{A}_R$ should be taken to be the Plancherel measure. But $\mathcal{Z}(\mathcal{A}_R)$ is \emph{precisely} the algebra of those defects, so the same argument does not work here. Instead, we need another physical principle to determine the preferred choice of trace on $\mathcal{Z}(\mathcal{A}_R)$.

The physical input we will use relies on a structural difference between $\mathcal{A}_R$ and its center. Recall from \eqref{eq:Cgroupbasis} that in the group basis, $\hat{C}$ acts on the corner leg by convolution with the class function $C(g)$, and recall from \eqref{eq:Fclassfn} that a class function satisfies $C(g) = C(h^{-1} g h)$ for every $h \in G$. Thus the center resolves the group elements at the corner only up to conjugacy in $G$: an operator in $\mathcal{Z}(\mathcal{A}_R)$ cannot tell two group elements in the same conjugacy class apart. The algebra $\mathcal{A}_R$ has no such restriction, because on each boundary leg of $R$ we are free to act with an arbitrary group multiplication. So $\mathcal{A}_R$ is an algebra of functions on $G$, while $\mathcal{Z}(\mathcal{A}_R)$ is an algebra of functions on the space of conjugacy classes of $G$.

The Plancherel measure is characterized as the unique measure on the unitary dual $\widehat{G}$ which is conjugate to the Haar measure on $G$: it is the unique measure making the Fourier transform \eqref{eq:peterweylnoncompact} between the group basis and the representation basis of $L^2(G)$ unitary. It is therefore the measure appropriate to an algebra that resolves all of $G$, which is why it appeared for $\mathcal{A}_R$. By the same logic, the measure appropriate to $\mathcal{Z}(\mathcal{A}_R)$ should be the one dual to the flat measure on the space of conjugacy classes, since that is the space the center actually acts on. This condition fixes the preferred  measure for the trace of $\mathcal{Z}(\mathcal{A}_R)$. We will call the resulting measure over the representations the ``microcanonical measure'' $d\pi$, for reasons we explain below.

It is worth seeing why the two measures differ at all, which is easy to see in the case of compact $G$. Recall that the character functions define vectors $\ket{\chi_\pi} = \int dg \, \chi_\pi(g) \ket{g}$ in $L^2(G)$, and that these vectors span the class functions. If $G$ is compact then $\widehat{G}$ is discrete, and the character functions are  orthonormal,
\begin{align}
    \braket{\chi_\pi}{\chi_\omega} = \int_G dg \, \chi_\pi(g)^* \chi_\omega(g) = \delta_{\pi \omega} \,,
\end{align}
in the normalization convention $\mathrm{Vol}(G) = 1$. Notice that the right hand side is a Kronecker delta, not the Plancherel-normalized delta function $\delta(\pi,\omega) = \delta_{\pi\omega}/\mu(\pi)$ that appears in \eqref{eq:repbasisoverlap}. In other words, with respect to the pairing that the center sees, the characters are already an orthonormal basis, so for class functions the dual measure on $\widehat{G}$ is simply the counting measure. The two normalizations differ by exactly $\mu(\pi) = d_\pi$, and this factor is the entire discrepancy between the Plancherel measure and the microcanonical one for compact groups.

For non-compact $G$, the characters $\chi_\pi(g)$ are distributions rather than $L^2$ functions, so this pairing requires more care, but the prescription it suggests survives. We present the details explicitly in Appendix~\ref{app:microcanonical}, but briefly, the divergence of $\Tr_{\mathcal{A}_R}[\hat{C}]$ can be tracked down to the infinite volume of some conjugacy classes in non-compact groups which must be renormalized separately for each conjugacy class. 

With probability one, a conjugacy class of a non-compact group has a representative which lies in a unique ``Cartan subgroup'' of $G$.\footnote{This is the definition of a ``regular element''. Regular elements are always dense in $G$. Non-regular elements will not play a role in our analysis.} Compact groups have a unique Cartan subgroup (up to conjugation in $G$) called the maximal torus of $G$: for $\SU(2)$, the maximal torus is given by rotations around the $\hat{z}$ axis (any other choice of axis is related to $\hat{z}$ by conjugation in $\SU(2)$) used to define the magnetic spin quantum number, and therefore the irreducible representations of $\SU(2)$. For non-compact groups, $G$ generally has many inequivalent Cartan subgroups, and the conjugacy classes of $G$ stratify into which Cartan subgroup they are conjugate to.\footnote{There is also a quotient by the Weyl group; see Appendix~\ref{app:microcanonical} for more details.} The uniform measure over conjugacy classes, then, is given by the disjoint union of the Haar measure of each Cartan subgroup of $G$, component by component. This component-by-component renormalization is an example of ``renormalized group averaging'', an idea recently developed in \cite{Alonso-Monsalve:2025lvt}. The microcanonical measure $d\pi$, then, is the Fourier dual to this component-wise Haar measure over Cartan subgroups.

The relationship between the Plancherel measure $d\mu(\pi)$ and the microcanonical measure $d\pi$ is well known in the math literature \cite{HarishChandra1952,HarishChandra1976,HarishChandra1954complex}. For example, when $\pi$ labels a representation in the principal series, it is given by
\begin{align}
    \frac{d\mu(\pi)}{d\pi} = |c(\pi)|^{-2} \equiv \mu(\pi) \,, \label{eq:cfunction}
\end{align}
where $c(\pi)$ is Harish-Chandra's $c$-function \cite{Helgason2000GGA}, essentially defined through \eqref{eq:cfunction} by its role in the relative weighting between the Plancherel measure and the microcanonical measure, though it also has an independent definition in terms of products over the roots of the Lie algebra of $G$. $\mu(\pi)$, then, serves as a density of states relative to the microcanonical measure $d\pi$, in the sense that $d\mu(\pi) = \mu(\pi) d\pi$ in the same way that $dN = \rho(E) dE$. For representations not in the principal series, $\frac{d\mu(\pi)}{d\pi}$ is simply defined to be the relative weighting given to each representation by the two measures. 

Thus, the preferred choice of trace for $\mathcal{Z}(\mathcal{A}_R)$ is given by 
\begin{align}
    \Tr_{\mathcal{Z}(\mathcal{A}_R)}[\hat{C}] = \int_{\widehat{G}} d\pi \, C(\pi) \,.
\end{align}
This trace over $\mathcal{Z}(\mathcal{A}_R)$, and especially the difference between the Plancherel and microcanonical measures, will play an important role in our interpretation of the various pieces of the entropy of $\rho_R$. 

\subsection{Operator valued weights: how to embed \texorpdfstring{$\mathcal{Z}(\mathcal{A}_{R}) \subset \mathcal{A}_R \subset \mathcal{B}_R$}{Z(A\_R) in A\_R in B\_R} consistently}
\label{sec:ovw}

At this point we have defined three different operator algebras and defined their traces. The algebras are nested,\footnote{Recall that $\mathcal{Z}(\mathcal{A}_R)$ can either be thought of as the algebra of gauge invariant, boundary anchored line operators that split $R$, $\overline{R}$ and act only on the edge modes of the extended Hilbert space. Because of the topological invariance of the model, there is an equivalent presentation of $\mathcal{Z}(\mathcal{A}_R)$ as the center of $\mathcal{A}_R$, which explains why it is a subalgebra despite acting solely on the edge modes.
}
\begin{align}
    \mathcal{Z}(\mathcal{A}_R) \subset \mathcal{A}_R \subset \mathcal{B}_R \,,
    \label{eq:nesting}
\end{align}
but the traces are not, because each was fixed by a distinct physical principle. As we saw above, the reduced state $\rho_R$ is an element of $\mathcal{A}_R \subset \mathcal{B}_R$, but $\Tr_{\mathcal{B}_R}[\rho_R] = \infty$ by \eqref{eq:tracevoldiv}. Similarly, the central operators $\hat{C}$ are elements of $\mathcal{A}_R$, but $\Tr_{\mathcal{A}_R}[\hat{C}] = \infty$, because a central operator acts as the identity on each $\Ha_R(\pi)$. So passing from one algebra to the next in \eqref{eq:nesting} is not simply a matter of forgetting some operators: it also requires a renormalization of the trace, and we had to use different physical principles for both of these renormalizations.

However, because the algebras are nested, we would like to be able to say that the entropy of a state on $\mathcal{B}_R$ contains a contribution from the exchange of operators in $\mathcal{A}_R$, and similarly between the algebra $\mathcal{A}_R$ and its center. Such a statement is only meaningful if we can define a map between the algebras which not only embeds one algebra within the other, but also translates between the traces on each algebra (i.e., the map should implement the renormalization of the trace). Our goal in this subsection is to construct these maps (called operator valued weights) and to show that they are unique. In the next section, by comparing the entropy of $\mathcal{A}_R$ and $\mathcal{Z}(\mathcal{A}_R)$ using this map, we will be able to unambiguously determine the way that the area operator emerges in our model.

It is worth first understanding why the standard tool for this job is unavailable. For concreteness, we will momentarily focus on the embedding of $\mathcal{A}_R$ within $\mathcal{B}_R$, but similar considerations will apply to the embedding of $\mathcal{Z}(\mathcal{A}_R)$ within $\mathcal{A}_R$.
The standard object relating an algebra to a subalgebra is a \emph{conditional expectation}: a positive map $E: \mathcal{B}_R \to \mathcal{A}_R$ which restricts to the identity on $\mathcal{A}_R$, is bimodular over it (see \eqref{eq:bimodular}), and preserves the unit, $E(\Id_{\mathcal{B}_R}) = \Id_{\mathcal{A}_R}$ (this is called boundedness).\footnote{See \cite{Gesteau:2023hbq} for more details about the relationship between conditional expectations and the renormalization group.} Suppose such a map were compatible with the traces, which means that $\Tr_{\mathcal{B}_R} = \Tr_{\mathcal{A}_R} \circ\, E$. Because $E$ acts as the identity on $\mathcal{A}_R$, this would imply $\Tr_{\mathcal{B}_R}\vert_{\mathcal{A}_R} = \Tr_{\mathcal{A}_R}$. But we have just seen that $\Tr_{\mathcal{B}_R}$ is identically infinite on the positive elements of $\mathcal{A}_R$, so no such map can exist for us. Crucially, the obstruction is precisely the volume divergence \eqref{eq:tracevoldiv} that forced us to renormalize the trace in the first place.

The resolution is to keep every property of a conditional expectation except boundedness. Concretely, we look for a map $\mathcal{E}$ which is positive and $\mathcal{A}_R$-bimodular,
\begin{align}
    \mathcal{E}(\mathcal{O}_1 \, \hat{X} \,\mathcal{O}_2) = \mathcal{O}_1 \, \mathcal{E}(\hat{X}) \, \mathcal{O}_2 \,, \qquad \mathcal{O}_1,\mathcal{O}_2 \in \mathcal{A}_R \,, \quad \hat{X} \in \mathcal{B}_R\,,
    \label{eq:bimodular}
\end{align}
but which is permitted to send a bounded operator $\hat{X}$ to an unbounded operator $\mathcal{E}(\hat{X})$.\footnote{More precisely, $\mathcal{E}$ maps the positive cone of $\mathcal{B}_R$ into the \emph{extended} positive cone of $\mathcal{A}_R$, which is the enlargement of the positive cone that accommodates unbounded elements. This is the technical device that makes the definition sensible; see \cite{Haagerup:1979ovwI,takesaki2003theory} for details.} Such a map is called an \emph{operator valued weight} \cite{Haagerup:1979ovwI,Haagerup:1979ovwII,takesaki2003theory}. 
We will additionally demand that $\mathcal{E}$ be {\it normal}, {\it semifinite}, and {\it faithful} \cite{Haagerup:1979ovwI,Haagerup:1979ovwII}. These are the same three conditions we impose on a trace, and this is not a coincidence: a canonical example of an operator valued weight for finite matrix algebras is the partial trace of a tensor factor, and imposing these three conditions will ensure that $\mathcal{E}$ has similarly well-behaved properties in this more general setting.

\emph{Normality} is a continuity requirement. If a positive operator is assembled as an increasing limit of simpler pieces, then normality says that $\mathcal{E}$ of the limit is the limit of the pieces. In other words, $\mathcal{E}$ commutes with monotone limits. This is what allows us to evaluate $\mathcal{E}$ one superselection sector at a time and then reassemble the answer with the direct integral over $\widehat{G}$, and it is also what guarantees that $\mathcal{E}$ takes normal states of $\mathcal{B}_R$ to normal states of $\mathcal{A}_R$, i.e., density matrices to density matrices.

\emph{Semifiniteness} says that the divergence we are permitting is not everywhere. A map which returned $\infty$ on every nonzero positive operator would satisfy all of the algebraic conditions above and carry no information at all, so we ask instead that the operators on which $\mathcal{E}$ is finite be weakly dense in $\mathcal{B}_R$.\footnote{Weakly dense means dense in the weak operator topology on $\Hext{R}$, i.e., the topology in which a sequence of operators converges when all of its matrix elements do.} We will see that $\mathcal{E}(\Id_{\mathcal{B}_R})$ does diverge, and semifiniteness is the statement that every operator in $\mathcal{B}_R$ can still be reached as a limit of operators that $\mathcal{E}$ handles perfectly well.

\emph{Faithfulness} says that no positive operator is invisible to $\mathcal{E}$: if $\hat{X} \geq 0$ and $\mathcal{E}(\hat{X}) = 0$, then $\hat{X} = 0$. This is what makes $\mathcal{E}$ a translation between the two algebras rather than a lossy projection onto one of them.

\paragraph{The map $\mathcal{E}: \mathcal{B}_R \to \mathcal{A}_R$.} Recall from Sec.~\ref{sec:algebras} that a general element of $\mathcal{B}_R$ is
\begin{align}
    \hat{X} = \int d\mu(\pi,\omega) \sum_{\alpha,\beta,i,j} X(\pi,\omega)_{\beta, j}^{\alpha, i} \ketbra{\pi ; \alpha, i}{\omega ; \beta, j}\,,
\end{align}
where $\alpha$ indexes $\Ha_R(\pi)$ and $i$ indexes the observer edge modes $V_\pi$, while a general element of $\mathcal{A}_R$ is $\mathcal{O} = \int d\mu(\pi) \, \mathcal{O}(\pi) \otimes \Pi_\pi$. Comparing the two, an element of $\mathcal{B}_R$ carries three pieces of data that an element of $\mathcal{A}_R$ does not: the matrix elements off-diagonal in $\pi$, and the dependence on the two edge mode indices $i$ and $j$. The map which discards exactly this data and nothing else is
\begin{align}
    \mathcal{E}(\hat{X}) = \int d\mu(\pi) \left(\sum_{\alpha,\beta,i} X(\pi,\pi)^{\alpha, i}_{\beta, i} \ketbra{\alpha_\pi}{\beta_\pi}\right) \otimes \Pi_\pi \,,
    \label{eq:ovwdef}
\end{align}
where $\ket{\alpha_\pi}$ is an orthonormal basis for $\Ha_R(\pi)$, and the object in parentheses is read as an operator on $\Ha_R(\pi)$. In other words, $\mathcal{E}$ decoheres $\hat{X}$ in the superselection label $\pi$, traces over the observer edge modes within each sector, and reinstates the factor $\Pi_\pi$, which is the operator that $\Tr_{\mathcal{A}_R}$ assigns unit trace. Bimodularity \eqref{eq:bimodular} is immediate from this description. The reason is that an operator in $\mathcal{A}_R$ is block diagonal in $\pi$ and acts proportionally to the identity on the $i$ index, so it passes through both the decoherence and the partial trace untouched.

Taking the $\mathcal{A}_R$ trace \eqref{eq:physicaltrace} of \eqref{eq:ovwdef} and comparing with the $\mathcal{B}_R$ trace \eqref{eq:BRtrace},
\begin{align}
    \Tr_{\mathcal{A}_R}\left[\mathcal{E}(\hat{X})\right]
    = \int d\mu(\pi) \sum_{\alpha, i} X(\pi,\pi)^{\alpha, i}_{\alpha, i}
    = \Tr_{\mathcal{B}_R}[\hat{X}] \,,
    \label{eq:tracecomposition}
\end{align}
so we can see that
\begin{align}
    \Tr_{\mathcal{B}_R} = \Tr_{\mathcal{A}_R} \circ\,\, \mathcal{E} \,.
    \label{eq:BtoA}
\end{align}
The two traces were defined by different prescriptions, and differ by an infinite renormalization, and yet they are related by a map which simply forgets the degrees of freedom that distinguish the two algebras. 

Note that $\mathcal{E}$ really is an operator valued weight and not a conditional expectation, as promised. To see this, we apply \eqref{eq:ovwdef} to the identity element of $\mathcal{B}_R$, which gives
\begin{align}
    \mathcal{E}(\Id_{\mathcal{B}_R}) = \int d\mu(\pi) \, \dim(V_\pi) \,\delta(\pi,\pi) \, \Id_{\Ha_R(\pi)} \otimes \Pi_\pi \,,
\end{align}
whose coefficient in each sector is the divergent factor of \eqref{eq:tracevoldiv}. The unboundedness of $\mathcal{E}$ is therefore not a technical aside; it \emph{is} the group volume divergence, relocated from the trace into the map itself. 

Notice that \eqref{eq:BtoA} implies that if $\hat{X}$ is trace class in $\mathcal{B}_R$, then $\mathcal{E}(\hat{X})$ is trace class in $\mathcal{A}_R$, and the traces of $\hat{X}$ and $\mathcal{E}(\hat{X})$ agree across their respective algebras. Therefore, normalized states of $\mathcal{B}_R$ are sent to normalized states of $\mathcal{A}_R$. More concretely, consider a normalized state $\hat{\rho}_R \in \mathcal{B}_R$ with matrix elements $\rho(\pi,\omega)^{\alpha, i}_{\beta, j}$, and define
\begin{align}
    q_\pi \equiv \sum_{\alpha, i} \rho(\pi,\pi)^{\alpha, i}_{\alpha, i} \,, \qquad
    \rho_\pi \equiv q_\pi^{-1} \sum_{\alpha,\beta,i} \rho(\pi,\pi)^{\alpha, i}_{\beta, i} \ketbra{\alpha_\pi }{\beta_\pi} \,.
    \label{eq:qpifromB}
\end{align}
Here, $\ket{\alpha_\pi}$ is an orthonormal basis for $\Ha_R(\pi)$. Then \eqref{eq:ovwdef} reads
\begin{align}
    \mathcal{E}(\hat{\rho}_R) = \int d\mu(\pi) \, q_\pi \, \rho_\pi \otimes \Pi_\pi \,,
    \qquad \int d\mu(\pi) \, q_\pi = 1\,,
\end{align}
which is exactly the form \eqref{eq:rhoRfirst} of the reduced state we obtained in Sec.~\ref{sec:factorization} by tracing out the extended Hilbert space of $\overline{R}$, with $q_\pi$ and $\rho_\pi$ playing the same roles. In other words, the structure we found there was not an artifact of the factorization map. It is what the unique trace-compatible restriction from $\mathcal{B}_R$ to $\mathcal{A}_R$ does to any state.

Finally, we check that $\mathcal{E}$ satisfies the three conditions we asked for (normality, semifiniteness, faithfulness). Normality is immediate from \eqref{eq:ovwdef}, because the sum over the edge mode index $i$ is the supremum of its finite partial sums, and a monotone limit in the integrand of a direct integral is a monotone limit of the result. Semifiniteness follows by exhibiting the operators on which $\mathcal{E}$ is finite: if $\hat{X}$ is supported on a set of representations of finite Plancherel measure and is trace class on the observer edge modes in each sector, then $\mathcal{E}(\hat{X})$ is a bounded element of $\mathcal{A}_R$, and such operators are weakly dense in $\mathcal{B}_R$.

Faithfulness is the only one of the three that requires an argument, because one might reasonably worry that some positive operator is discarded entirely when we decohere $\hat{X}$ over the representations $\pi$. To see that $\mathcal{E}$ is faithful, suppose $\hat{X} \geq 0$ and $\mathcal{E}(\hat{X}) = 0$. Each diagonal block $X(\pi,\pi)$ is then a positive operator whose partial trace over $V_\pi$ vanishes, so the block itself vanishes for almost every $\pi$. The Cauchy--Schwarz inequality applied to the quadratic form $\bra{\psi} \hat{X} \ket{\sigma}$ then forces the blocks off-diagonal in $\pi$ to vanish as well, and so $\hat{X} = 0$. Crucially, this argument uses the positivity of $\hat{X}$: $\mathcal{E}$ certainly annihilates purely off-diagonal self-adjoint operators, but no such operator is positive, and faithfulness is only ever a condition on the positive cone.

\paragraph{Uniqueness.} So far we have exhibited  a map $\mathcal{E}$ with the properties we wanted. But we have not yet shown that it is the only such map, and it is the uniqueness of $\mathcal{E}$ that makes our renormalization procedure canonical rather than merely consistent.
To demonstrate that $\mathcal{E}$ is unique, we use a theorem due to Haagerup \cite{Haagerup:1979ovwI,Haagerup:1979ovwII}.
Haagerup's theorem shows that given a von Neumann subalgebra $\mathcal{A} \subset \mathcal{B}$ and a choice of (normal, semifinite, faithful) trace on each algebra, there is at most one normal semifinite faithful operator valued weight $\mathcal{E} : \mathcal{B} \to \mathcal{A}$ satisfying $\Tr_{\mathcal{B}} = \Tr_{\mathcal{A}} \circ \,\,\mathcal{E}$. Equivalently, once one of the two traces and $\mathcal{E}$ are fixed, the other trace is determined, because any two operator valued weights for the same inclusion differ by a positive element affiliated to the center of $\mathcal{A}$, which is the same ambiguity as a change of trace.

This is the statement we were after. In Sec.~\ref{sec:A_Ralg} we found a family of candidate traces $\Tr^C_{\mathcal{A}_R}$ labeled by a positive function $C(\pi)$ on $\widehat{G}$, and we fixed $C = 1$ by demanding that the associated factorization map introduce no defect at the corner. Haagerup's theorem says that this single choice also uniquely fixes the operator valued weight $\mathcal{E}$, and therefore fixes what it means to restrict a state of $\mathcal{B}_R$ to $\mathcal{A}_R$. There is no further freedom or ambiguity hiding in the restriction procedure. Had we instead chosen the trace $\Tr^C_{\mathcal{A}_R}$, the unique compatible weight would have been $\mathcal{E}_C(\hat{X}) = \hat{C}^{-1} \mathcal{E}(\hat{X})$, and the reduced state would have been \eqref{eq:rhoRC}, consistently with \eqref{eq:VCisometry}. But again, demanding that the associated factorization map introduces no defect at the corner fixes $C=1$. So the uniqueness of $\Tr_{\mathcal{B}_R}$, the defect-free condition, and the uniqueness of the operator valued weight $\mathcal{E}$ combine to prove that our definition of $\Tr_{\mathcal{A}_R}$ is also unique.

As an aside, one can use the same language to bypass the edge modes entirely, by defining an operator valued weight $\mathcal{E}_\Sigma: \mathcal{B}(\Ha_{\mathrm{phys}}(\Sigma)) \to \mathcal{A}_R$ directly from the algebra of the whole surface to the algebra of the subregion, without ever passing through the factorization map, the extended Hilbert space or the edge modes directly. This is essentially what our factorization map together with the renormalized trace accomplished explicitly, and it is closer to the more algebraic perspective of \cite{Dong:2018seb,Casini:2013rba,Klinger:2023auu,Klinger:2023tgi,Riello:2021lfl,Lin:2018bud,Soni:2015yga}. However, the two approaches are equivalent, and the same ambiguities and renormalization procedures must be adopted even in the more global approach. From our perspective, the extended Hilbert space approach is simply a tool that makes the physical meaning of these steps more transparent.

\paragraph{The map $\mathcal{F}: \mathcal{A}_R \to \mathcal{Z}(\mathcal{A}_R)$.} The same construction applies one level down in \eqref{eq:nesting}. Because $\mathcal{Z}(\mathcal{A}_R) \cong L^\infty(\widehat{G})$ is abelian, the bimodularity condition \eqref{eq:bimodular} carries no information, for a central operator is a number in each sector and numbers commute past everything. An operator valued weight from $\mathcal{A}_R$ to its center is therefore determined by a single positive density $\lambda(\pi)$,
\begin{align}
    \mathcal{F}(\mathcal{O}) = \int d\mu(\pi) \, \lambda(\pi) \, \tr_{\Ha_R(\pi)}[\mathcal{O}(\pi)] \, \Id_{\Ha_R(\pi)} \otimes \Pi_\pi \,,
    \label{eq:TZgeneral}
\end{align}
which traces out everything except the superselection label and weighs the result by $\lambda(\pi)$ sector by sector. To determine $\lambda(\pi)$, we impose compatibility with the two traces we have already fixed, $\Tr_{\mathcal{A}_R} = \Tr_{\mathcal{Z}(\mathcal{A}_R)} \circ\, \mathcal{F}$. Using $\Tr_{\mathcal{A}_R}[\mathcal{O}] = \int d\mu(\pi) \tr_{\Ha_R(\pi)}[\mathcal{O}(\pi)]$ from \eqref{eq:physicaltrace} and $\Tr_{\mathcal{Z}(\mathcal{A}_R)}[\hat{C}] = \int d\pi \, C(\pi)$ from \eqref{eq:centertrace}, this reads
\begin{align}
    \int d\mu(\pi) \, \tr_{\Ha_R(\pi)}[\mathcal{O}(\pi)] = \int d\pi \, \lambda(\pi) \, \tr_{\Ha_R(\pi)}[\mathcal{O}(\pi)] \,,
\end{align}
for every $\mathcal{O} \in \mathcal{A}_R$, which fixes
\begin{align}
    \lambda(\pi) = \frac{d\mu(\pi)}{d\pi} = \mu(\pi) \,.
    \label{eq:lambdaisRN}
\end{align}
The weight relating $\mathcal{A}_R$ to its center is therefore \emph{not} simply the naive partial trace. It carries a factor of the Radon--Nikodym derivative $\mu(\pi)$ between the Plancherel and microcanonical measures. Therefore, setting $\lambda(\pi) = \mu(\pi)$, we see that $\mathcal{F}$ satisfies the desired trace-preservation property
\begin{align}
    \Tr_{\mathcal{A}_R} = \Tr_{\mathcal{Z}(\mathcal{A}_R)} \circ \,\, \mathcal{F} \,. \label{eq:AtoZ}
\end{align}
Applied to the reduced state $\rho_R$ of \eqref{eq:rhoRfirst}, for which $\tr_{\Ha_R(\pi)}[\rho_\pi] = 1$, we find
\begin{align}
    \mathcal{F}(\rho_R) = \int d\mu(\pi) \, \big(\mu(\pi) \, q_\pi\big) \, \Id_{\Ha_R(\pi)} \otimes \Pi_\pi \,,
    &&
    \Tr_{\mathcal{Z}(\mathcal{A}_R)}[\mathcal{F}(\rho_R)] = \int d\pi \, \mu(\pi) \, q_\pi = 1 \,,
    \label{eq:TZonrho}
\end{align}
where the last equality is just $d\mu(\pi) = \mu(\pi) d\pi$ together with the normalization \eqref{eq:qpi_normalized} of $q_\pi$. So the restriction of $\rho_R$ to the center is the probability distribution over superselection labels, as we would expect, and it is normalized with respect to the microcanonical measure that Sec.~\ref{sec:centertrace} argued the center should carry.

\section{Entropy} \label{sec:entropy_nomatter}

In this section, we will compute the entropy of the state $\rho_R$ for the algebra $\mathcal{A}_R$. We will find that
\begin{align}
    S(\rho_R) &= H[\,p_\pi] +\langle \hat{A} \rangle_\rho + S_{\mathrm{bulk}}(\rho_R) \,.
    \label{eq:resultsummary}
\end{align}

The first term in \eqref{eq:resultsummary} is 
\begin{align}
    H[\,p_\pi] \equiv -\int d\pi \, p_\pi \ln(p_\pi) \,, \qquad p_\pi \equiv \mu(\pi) \, q_\pi \,, \qquad \int d\pi \, p_\pi = 1 \,. \label{eq:Hdef}
\end{align}
This is the differential entropy of the probability density $p_\pi$ \cite{Cover:2005lom}.\footnote{The differential entropy can be interpreted as (minus) the relative entropy between the classical probability density $p_\pi$ and the (unnormalizable) measure $d\pi$ (the microcanonical measure over representations), with an infinite constant, namely the entropy of $d\pi$, $S(d\pi)$, renormalized away. We can write this as $H[\,p_\pi] = -S_{\mathrm{rel}}[\,p_\pi d\pi \,||\, d\pi]
    \sim - \int d\pi \, p_\pi \ln(p_\pi d\pi/d\pi)$.}
When the representations $\pi$ have a continuous spectrum (which they always will when $G$ is non-compact), this term is not sign definite. We will comment more on this point below.

The expression $\langle \hat{A} \rangle_\rho$ in the second term is the expectation value of a state independent, central operator that we will define below in \eqref{eq:areaoperator}.   
The spectrum of this operator is completely fixed by the gauge group $G$ and the topological boundary condition we imposed at the cut $\gamma$; it does not depend on the tessellation, or on the number of marked points $n$. In contrast, the contribution $S_{\mathrm{bulk}}(\rho_R)$ (defined in \eqref{eq:Sbulkdef}) depends in a detailed way on the choice of boundary conditions that we impose on $\Sigma$.

Let us briefly compare this entropy formula to the generalized entropy in gravity, and defer a more detailed comparison to \cite{Balasubramanian:2026xyz} when we include the matter legs in the tensor network.
Consider a two dimensional holographic CFT, and let $\ket{\psi}$ be a global state of this CFT with a semiclassical bulk dual that is well-described in the low energy limit of three-dimensional gravity. In AdS/CFT, the generalized entropy \cite{Ryu_2006,Hubeny:2007xt,Faulkner:2013ana,Engelhardt:2014gca} of a reduced state $\sigma_R \sim \tr_{\overline{R}}[\ketbra{\psi}]$ is defined by
\begin{align}
    S(\sigma_R) \;=\; \underset{\gamma \,\sim\, R}{\mathrm{min\;ext}} \left[ \frac{A_\gamma}{4G_N} + S_{\mathrm{bulk}}\bigl(\rho_{\mathrm{int}(R \cup\gamma)}\bigr) \right] \,. \label{eq:QESformula}
\end{align}
Here, $A_\gamma$ is the area of a boundary anchored curve $\gamma$ which is homologous to $R$, $S_{\mathrm{bulk}}\bigl(\rho_{\mathrm{int}(\gamma \cup R)}\bigr) $ is the von Neumann entropy of all matter fields in the bulk subregion $\mathrm{int}(\gamma \cup R)$. The $\mathrm{min\;ext}$ then indicates that we are instructed to first extremize the sum of these two terms over all possible curves $\gamma$, and then minimize over all possible extremal curves. 

The generalized entropy formula \eqref{eq:QESformula} is a good approximation to the entanglement entropy shared between a boundary subregion $R$ and its complement when the gravitational path integral which computes $\Tr(\sigma^n_R)$ is dominated by a single, replica symmetric saddle point for $n \gtrsim 1$ \cite{Lewkowycz_2013,Dong:2016hjy}.

Comparing \eqref{eq:resultsummary} with \eqref{eq:QESformula}, the two entropy formulas seem to have two qualitative differences. First, \eqref{eq:resultsummary} carries the differential entropy contribution $H[\,p_\pi]$. This difference is not important, because a semi-classical state will have $H[\,p_\pi] \sim \mathcal{O}(\ln(G_N))$ \cite{Almheiri:2016blp,Harlow:2016vwg,Akers:2018fow,Dong:2023xxe}, and therefore will be subleading in \eqref{eq:QESformula}. The second, more substantial difference between the two entropy formulas is that \eqref{eq:resultsummary} does not contain a minimization over possible cuts $\gamma$ through the tensor network. Indeed, because of the topological symmetry of the model, the boundary anchored cut $\gamma$ which separates $R$ and $\overline{R}$ is only \emph{defined} up to homotopy within $\Sigma$, and for the disk is therefore unique. 

We will  argue in \cite{Balasubramanian:2026xyz} that when the measure of the gravitational path integral has been taken into account, the eigenvalues of $\hat{A} $ are $\ell/4 G_N$, where $\ell$ is the minimal geodesic length of a boundary curve which separates $R$ from $\overline{R}$.
Thus, despite first appearances, \eqref{eq:resultsummary} seems to implicitly have the extremization structure characteristic of holographic entropy formulas, at least when the tensor network does not have any matter legs.

\subsection{Deriving the entropy formula}

Using the expression for the state $\rho_R$ derived in \eqref{eq:rhoRfirst}, which we reproduce here for convenience,
\begin{align}
    \rho_R &= \int d\mu(\pi) \, q_\pi \rho_\pi \otimes \Pi_\pi\,, \\
    \Tr_{\mathcal{A}_R}[\rho_R] &= \int d\mu(\pi) q_\pi = 1 \,,
\end{align}
we can compute the entropy $S(\rho_R)$ using the formula
\begin{align}
    S(\rho_R) = - \Tr_{\mathcal{A}_R}[\rho_R \ln \rho_R] = - \partial_n\Tr_{\mathcal{A}_R}[\rho_R^n]_{n=1} \,. \label{eq:entropy_partialn}
\end{align}
Explicitly,
\begin{align}
    \Tr_{\mathcal{A}_R}[\rho_R^n] 
    &=  \Tr_{\mathcal{A}_R}\left[\int \prod_{i=1}^n d\mu(\pi_i) q_{\pi_i} \rho_{\pi_i} \otimes \Pi_{\pi_i} \right]\,, \\
    &= \Tr_{\mathcal{A}_R}\left[\int d\mu(\pi) q_\pi^n \rho_\pi^n \otimes \Pi_\pi \right]\,, \\
    &= \int d\mu(\pi) q_\pi^n \tr_{\Ha_R(\pi)}[\rho_\pi^n ] \,.
\end{align}
We used the projection property \eqref{eq:Piproj} to reduce the number of integrals when moving from the first to the second line, and the definition of the trace \eqref{eq:physicaltrace} in moving from the second to the third. Inserting this expression into \eqref{eq:entropy_partialn},
\begin{align}
    S(\rho_R) &= - \int d\mu(\pi) q_\pi \ln(q_\pi) + \int d\mu(\pi) q_\pi S(\rho_\pi) \,, \label{eq:SrhoR_first}\\
    S(\rho_\pi) &= - \tr_{\Ha_R(\pi)}[\rho_\pi \ln(\rho_\pi)] \,.
\end{align}

Thus, the entropy $S(\rho_R)$ has two contributions: an average entropy within each superselection sector $\pi$, and a classical Shannon entropy measuring the uncertainty about which sector we are in. We must be careful in interpreting  the Shannon piece: because $\pi$ is a continuous label in general, we should actually interpret this piece of the entropy as measuring the relative entropy between the probability measure $p_\pi d\pi = q_\pi d\mu(\pi)$ and the (unnormalizable) microcanonical measure $d\pi$, in the sense of \eqref{eq:Hdef}. To understand this term in more detail, it is helpful to invoke the operator valued weight $\mathcal{F}$ from the last section. Using the trace-preserving property \eqref{eq:AtoZ}, as well as the fact $\ln(\mathcal{F}(\rho_R)) \in \mathcal{Z}(\mathcal{A}_R)$ and the bimodularity of $\mathcal{F}$, we can see that
\begin{align}
    S(\rho_R) - S(\mathcal{F}(\rho_R)) &= -\Tr_{\mathcal{A}_R}[\rho_R \ln(\rho_R)] + \Tr_{\mathcal{Z}(\mathcal{A}_R)}[\mathcal{F}(\rho_R) \ln(\mathcal{F}(\rho_R))] \,,\label{eq:rhoFrho}\\
    &= -\Tr_{\mathcal{A}_R}[\rho_R \ln(\rho_R)] + \Tr_{\mathcal{A}_R}[\rho_R \ln(\mathcal{F}(\rho_R))]\,,
    \\&\equiv - S_{\mathrm{rel}}(\rho_R || \mathcal{F}(\rho_R))\,.\label{eq:SrelrhoFrho}
\end{align}
Here, $S_{\mathrm{rel}}(\rho_R || \mathcal{F}(\rho_R))$ is the relative entropy between the state $\rho_R$ and the weight $\mathcal{F}(\rho_R)$ for $\mathcal{A}_R$. We are careful to not call $\mathcal{F}(\rho_R)$ a state for $\mathcal{A}_R$: while it is indeed a valid state for the center $\mathcal{Z}(\mathcal{A}_R)$, it is not normalizable as a state for $\mathcal{A}_R$. This is the reason we had to renormalize the trace. 
Nevertheless, the relative entropy $S_{\mathrm{rel}}(\rho_R || \mathcal{F}(\rho_R))$ is defined as above and is finite. 

Defining $p_\pi = \mu(\pi) q_\pi$, which is the probability density for the state $\rho_R$ to be in the representation $\pi$ with respect to the microcanonical measure $d\pi$ over the representations, we can explicitly compute that
\begin{align}
    S(\mathcal{F}(\rho_R)) &= -\int d\pi \, p_\pi \ln(p_\pi) \equiv  H[\,p_\pi]\,,\\
    -S_{\mathrm{rel}}(\rho_R || \mathcal{F}(\rho_R)) &= \int d\mu(\pi) q_\pi \ln(\mu(\pi)) + \int d\mu(\pi) q_\pi S(\rho_\pi) \,.
\end{align}
Because of the additional contribution in the relative entropy, it is convenient to define the central, state independent operator
\begin{align}
    \hat{A} = \int d\mu(\pi) \, \ln(\mu(\pi)) \, \Id_{\Ha_R(\pi)} \otimes \Pi_\pi \,. \label{eq:areaoperator}
\end{align}
With this definition, the additional contribution to the relative entropy can be written as
\begin{align}
    \int d\mu(\pi) q_\pi \ln(\mu(\pi)) = \Tr_{\mathcal{A}_R}[\rho_R \hat{A} ] \equiv \langle \hat{A} \rangle_\rho \,.
\end{align}
The entropy of $\rho_R$ then decomposes as
\begin{align}
    S(\rho_R) = H[\,p_\pi] + \langle \hat{A}\rangle_\rho + \int d\mu(\pi) q_\pi S(\rho_\pi) \,. \label{eq:SrhoR_gen}
\end{align}
Thus, $S(\rho_R)$ actually splits into \emph{three} natural pieces. Note, however, that the first two terms combine to give $- \int d\mu(\pi) q_\pi \ln(q_\pi)$, so \eqref{eq:SrhoR_first} and \eqref{eq:SrhoR_gen} agree, as expected.

Note that this three-way split is  meaningful because we fixed the traces on $\mathcal{A}_R$ and $\mathcal{Z}(\mathcal{A}_R)$ in Sec.~\ref{sec:A_Ralg} and Sec.~\ref{sec:centertrace}. The individual terms in \eqref{eq:SrhoR_gen} refer to the measures $d\mu(\pi)$ and $d\pi$ which define those two traces, whereas the total \eqref{eq:SrhoR_first} does not. One might worry that the area operator is therefore an artifact of a normalization convention. The reason the area operator is meaningful is that both traces descend from the \emph{same} Haar measure on $G$: the Plancherel measure $d\mu(\pi)$ is dual to $dg$ itself, while the microcanonical measure $d\pi$ is dual to the Haar measures on the Cartan subgroups of $G$, whose normalizations are fixed relative to $dg$ by the Weyl integration formula (see Appendix~\ref{app:microcanonical}). A rescaling $dg \to \kappa \, dg$ therefore rescales the two traces uniformly,
\begin{align}
    d\mu(\pi) \to \kappa^{-1} \, d\mu(\pi) \,, \qquad d\pi \to \kappa^{-1} \, d\pi \,, \label{eq:haarshift}
\end{align}
so that the Radon--Nikodym derivative $\mu(\pi) = d\mu(\pi)/d\pi$ of \eqref{eq:cfunction} is invariant. Indeed, $\mu(\pi)$ has an independent definition in terms of the Lie algebra of $G$ \cite{Helgason2000GGA}, which has no dependence on the normalization of the Haar measure at all. Thus, the spectrum $\ln (\mu(\pi))$ of $\hat{A}$, and with it the area term $\langle \hat{A} \rangle_\rho$ in the entropy, does not depend on the normalization of the Haar measure. With this in mind, let us now discuss each of the three contributions to the entropy of $\rho_R$ in more detail.

\subsection{The differential entropy contribution}

The first term, $H[\,p_\pi]$, can be interpreted as the Shannon entropy that arises from the classical uncertainty over which superselection sector the mixed state $\rho_R$ is in, measured relative to the microcanonical measure $d\pi$. This contribution to $S(\rho_R)$ is not positive definite when $G$ is non-compact, which is easiest to see by comparing with the compact case. When $G$ is compact, so $\widehat{G}$ is discrete, $d\pi$ is the counting measure and $p_\pi$ is an honest probability distribution,\footnote{If $G$ is compact and continuous, then \eqref{eq:HcompactG} still holds, but then there is no maximum entropy state.} so that 
\begin{align}
    H[\,p_\pi]_{\mathrm{compact}\,\, G} = -\sum_\pi p_\pi \ln(p_\pi) \geq 0 \label{eq:HcompactG}
\end{align}
as usual. However, when $\widehat{G}$ is continuous, which is always the case for some representations when $G$ is non-compact, there is no minimum entropy state with respect to which we can define $H=0$. For example, let $p^\epsilon_\pi$ be the probability density described by an indicator function with support of measure $\epsilon$ with respect to $d\pi$ (and height $\epsilon^{-1}$ so that $p^\epsilon_\pi$ is properly normalized) around any representation with a continuous label $\pi_0$. Then it is easy to see that
\begin{align}
    H[\,p^\epsilon_\pi] = -\ln(\epsilon^{-1}) \,.
\end{align}
Because we can take the limit $\epsilon \to 0$ due to the continuity of the representation label $\pi$, there is no minimum distribution over the representations. Relatedly, the differential entropy $H[\,p_\pi]$ \emph{does} depend on the normalization of the Haar measure, and under a $dg \to \kappa\, dg$, it shifts as
\begin{align}
    H[\,p_\pi] \,\to\, H[\,p_\pi] - \ln(\kappa) \,.
\end{align}
This means that the total entropy $S(\rho_R)$ transforms as $S(\rho_R) \to S(\rho_R) - \ln(\kappa) $ as well.
By taking $\kappa$ to be arbitrarily large, we can make $S(\rho_R)$ as negative as we like. We can think about this ambiguity as a scheme dependence which arises from the renormalization procedure we used to define the entropy. Indeed, as noted above, the differential entropy can be interpreted as a renormalized relative entropy, and this ambiguity is a reflection of this. However, because this ambiguous constant is state independent, the entropy \emph{difference} between any two states is independent of this choice of scale $\ln(\kappa)$, so these are the physical quantities of this model. We will discuss this further in Sec.~\ref{sec:bulkvsbdy}. 

An ambiguity in the generalized entropy of this kind is qualitatively similar to what has been found in other investigations in quantum gravity \cite{Witten:2021unn,Chandrasekaran:2022eqq,Chandrasekaran:2022cip,Jensen2023,Penington:2023dql,Kudler-Flam:2023qfl}, where the entropy of a subregion is likewise defined up to a state independent constant $S_0$. That  said, the mechanism is not quite the same. In those works, the subtraction is forced because the algebra of the subregion is a type II factor, on which the trace is defined only up to an overall rescaling, and for which there are no pure states. Here, $\mathcal{A}_R$ is instead a direct integral of type I factors. But after imposing the defect-free condition to determine the preferred trace on $\mathcal{A}_R$, the remaining ambiguity in the trace corresponds to the normalization of the Haar measure, which cannot be fixed canonically for non-compact groups. 

\subsection{The area contribution} \label{sec:areaop_nomatter}

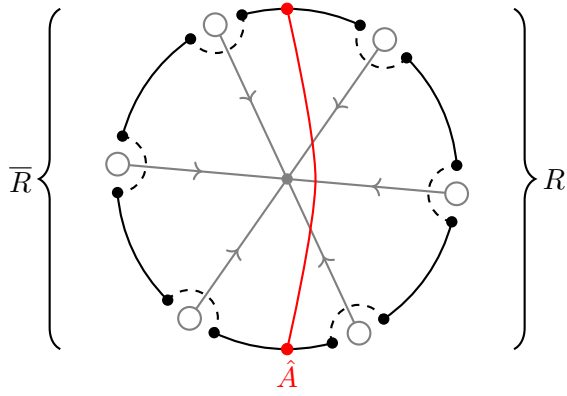
\begin{figure}
    \centering
    \begin{tikzpicture}[scale=1.5]
        \newcommand{\legdisk}[4]{%
            \pgfmathsetmacro{\gap}{asin(#3/#2)}
            \foreach \i in {0,...,\numexpr#1-1\relax}{%
                \pgfmathsetmacro{\ang}{#4 + \i*360/#1}%
                \pgfmathsetmacro{\angnext}{#4 + (\i+1)*360/#1}%
                \draw[thick]
                (\ang+\gap:#2) arc[start angle=\ang+\gap, end angle=\angnext-\gap, radius=#2];

                \draw[color=black!50, thick,->-=0.5] (\ang:#2) -- (0,0);
                \filldraw[black!50,thick,fill=white] (\ang:#2) circle (0.1);

                \fill (\angnext-\gap:#2) circle (0.05);
                \draw[thick,dashed]
                ([shift={(\ang:#2)}]\ang+90:#3)
                arc[start angle=\ang+90, end angle=\ang+270, radius=#3];
                \fill (\ang+\gap:#2) circle (0.05);
            }%
            \fill[color=black!50] (0,0) circle (0.05);%
        }
        \def\xx{0.5}
        \legdisk{6}{1.5}{0.25}{55}
        \draw[red,thick] plot[smooth] coordinates {(0,1.5)  (0.25,0) (0,-1.5)};
        \filldraw[red] (0,1.5) circle (0.05);
        \filldraw[red] (0,-1.5) circle (0.05);
        \node[red,anchor=north] at (0,-1.5) {$\hat{A}$};
        \draw[decorate, thick,decoration={brace, amplitude=8pt, mirror}] (1.5+\xx,-1.5) -- (1.5+\xx,1.5) node[midway,xshift=15] {$R$};
        \draw[decorate, thick,decoration={brace, amplitude=8pt}] (-1.5-\xx,-1.5) -- (-1.5-\xx,1.5)  node[midway,xshift=-15] {$\overline{R}$};

    \end{tikzpicture}
    \caption{The area operator $\hat{A}$ on the disk. Because $\hat{A}$ is a topological operator (i.e., it is a physical operator that commutes with the gauge constraints) we can think of it as acting on either the $R$ or $\overline{R}$ boundary legs by deforming the line operator slightly to the left. Because $\hat{A}$ can be deformed such that it acts either only on the $R$ legs or not at all, it is central with respect to $\mathcal{A}_R$. }
    \label{fig:areaop}
\end{figure}

The second term, $\langle \hat{A} \rangle_\rho$, is the expectation value in the state $\rho_R$ of the central operator $\hat{A}$ defined in \eqref{eq:areaoperator} and shown in Fig.~\ref{fig:areaop}. There are two features which distinguish it from the other two contributions.

The first feature is that the operator $\hat{A}$ is fixed by data from the gauge group alone. It is built entirely from the density of states $\mu(\pi) = d\mu(\pi)/d\pi$ of \eqref{eq:cfunction}, so its spectrum $\ln(\mu(\pi))$ is determined by $G$, and is defined for every $\pi \in \widehat{G}$. In particular, $\hat{A}$ is independent of the graph $\Lambda$ used to construct the physical Hilbert space, and the choice of state $\dket{\psi}$ within it.  What it does depend on, besides $G$, is the boundary condition imposed at the cut. Here, we will now explain in what way the topological boundary condition on the cut affects the area operator.

Recall from Sec.~\ref{sec:factorization} that to define the factorization map, we had to give the cut a boundary condition, and that we chose a topological one so that the details of the lattice near the cut are irrelevant.\footnote{If we instead imposed a non-topological boundary condition such as \cite{Dong2024,Delcamp:2016eya}, then the entropy would depend on the details of the lattice near the cut.} Specifically, we chose the rough boundary condition \cite{Bravyi:1998sy}, on which the anyons that condense are the pure charges, labeled by $\pi \in \widehat{G}$. Suppose we had instead chosen a different topological boundary condition at the cut, for instance the smooth boundary condition, on which the anyons that condense are the pure fluxes, labeled by conjugacy classes $[g] \in [G]$, or one of the boundary conditions which mix charges and fluxes. The derivation of the entropy formula would go through essentially unchanged, because everything it used was fixed by the topological symmetry of the model: the result would still be independent of the microscopic details of the lattice, and it would still split into three terms. What would change is the label set of the superselection sectors, which would consist of the anyons that condense on the cut rather than the irreducible representations $\pi$, so that the integrals in each of the three terms would run over this alternative label set. The spectrum of the area operator would change accordingly, to the ratio of the measures defining the traces on the new algebra $\widetilde{\mathcal{A}}_R$ and its center $\mathcal{Z}(\widetilde{\mathcal{A}}_R)$. We leave the explicit determination of these alternative spectra for future work. In \cite{Balasubramanian:2026xyz} we will also explain the choice of topological boundary condition appropriate to three-dimensional gravity, which makes the spectrum of $\hat{A}$ equal to $\ell/4G_N$, with $\ell$ the length of a boundary anchored bulk geodesic separating $R$ from $\overline{R}$.

The second feature is that the spectrum of $\hat{A}$ does not depend on the number of marked points $n$ defining the open boundary conditions on $\Sigma$. When we tune the boundary theory to criticality and send $n \to \infty$, $\hat{A}$ is therefore the same operator, up to the choice of the two boundary points at which it is anchored. Its expectation value $\langle \hat{A} \rangle_\rho$ of course depends on the state through the distribution $q_\pi$, so a sensible continuum limit requires a sequence of states $\rho_R^{(n)}$ whose distributions $q^{(n)}_\pi$ converge as $n \to \infty$. The integral of the spectrum of $\hat{A}$ against the limiting distribution is then the CFT answer for the contribution of $\hat{A}$ to the entropy.

Finally, we note that $\hat{A}$ is a positive operator if and only if $\mu(\pi) \geq 1$ for all representations $\pi$, up to a set of measure zero. In other words, $\hat{A} \geq 0$ precisely when the density of states relative to the microcanonical measure is everywhere at least one, which by the argument given below \eqref{eq:SrhoR_gen} is a statement about $G$ and not about a choice of normalization for the Haar measure. 

A state independent operator, supported on the surface which separates the two subregions, whose expectation value contributes universally to the entropy that they share, is precisely what we mean by an area operator, and we identify $\hat{A}$ as such. We will make this identification more precise in \cite{Balasubramanian:2026xyz}, where we argue that once the measure of the three-dimensional gravitational path integral is accounted for, the eigenvalues of $\hat{A}$ are $\ell/4G_N$, with $\ell$ the length of the minimal bulk geodesic homologous to $R$.

\subsection{The bulk entropy contribution}

The third term is the average of the entropies $S(\rho_\pi)$ of the states within each superselection sector, weighted by the probability $q_\pi d\mu(\pi)$ of finding $\rho_R$ in that sector. We will denote it as
\begin{align}
    S_{\mathrm{bulk}}(\rho_R) \equiv \int d\mu(\pi) \, q_\pi \, S(\rho_\pi) \,. \label{eq:Sbulkdef}
\end{align}
This term is always non-negative. The reason for this is that each $\rho_\pi$ is a normalized density matrix on $\Ha_R(\pi)$, and $\tr_{\Ha_R(\pi)}$ is the unique trace on the type I factor $\mathcal{B}(\Ha_R(\pi))$, which has a trivial center. There is therefore nothing to renormalize sector by sector, and $S(\rho_\pi) \geq 0$ for every $\pi$ by the usual argument. As with the other two terms, the sectors being averaged over are the ones that the boundary condition at the cut allows.

Unlike $\hat{A}$, however, this term is sensitive to the details of the model. The particular Hilbert space $\Ha_R(\pi)$ depends on how many boundary points $R$ contains, for example. In general, $S_{\mathrm{bulk}}(\rho_R)$ will contribute extensively in the number of boundary points to leading order, and the coefficient of this contribution will not be universal. Independently of this, however, $S_{\mathrm{bulk}}(\rho_R)$ has the same structure, independent of the details of the renormalization group flow to the CFT in the $n\to\infty$ limit: it will consist of the average of the sector-wise entropy over all of the sectors allowed by the boundary conditions imposed at the cut.  

\subsection{The case of compact \texorpdfstring{$G$}{G}} \label{sec:entropy_compact}

It is instructive to specialize \eqref{eq:SrhoR_gen} to a compact gauge group, where each of the three terms can be compared with known results in lattice gauge theory. For concreteness, take $\mathrm{Vol}(G)=1$. Then $\widehat{G}$ is discrete, the microcanonical measure is the counting measure, and the density of states is simply the dimension of the representation,
\begin{align}
    \mu(\pi) = d_\pi \,, \qquad \int d\pi \to \sum_\pi \,, \qquad \int d\mu(\pi) \to \sum_\pi d_\pi \,.
\end{align}
The observer edge modes are finite dimensional, and as we saw below \eqref{eq:rhoRfirst}, the projector $\Pi_\pi = \Id_{V_\pi}/d_\pi$ is the maximally mixed state on $V_\pi$. The reduced state \eqref{eq:rhoRfirst} therefore takes the familiar form
\begin{align}
    \rho_R = \sum_\pi p_\pi \left[\rho_\pi \otimes \frac{\Id_{V_\pi}}{d_\pi}\right] \,, \qquad p_\pi = d_\pi q_\pi \,, \qquad \sum_\pi p_\pi = 1 \,,
\end{align}
while the area operator \eqref{eq:areaoperator} becomes
\begin{align}
    \hat{A} = \sum_\pi \ln(d_\pi) \, \Id_{\Ha_R(\pi)} \otimes \Id_{V_\pi} \,,
\end{align}
whose spectrum is $\ln(d_\pi)$. The three terms of \eqref{eq:SrhoR_gen} then read
\begin{align}
    S(\rho_R) = - \sum_\pi p_\pi \ln(p_\pi) + \sum_\pi p_\pi \ln(d_\pi) + \sum_\pi p_\pi S(\rho_\pi) \,. \label{eq:SrhoR_compact}
\end{align}
This is precisely the standard decomposition of the entanglement entropy of a subregion in lattice gauge theory \cite{Akers:2024wab,Dong2024,Donnelly:2011hn,Casini:2013rba,Donnelly:2014gva,Soni:2015yga}: a classical Shannon term over the superselection sectors, a term counting the entanglement of the edge modes on the entangling surface, and the average entropy within the sectors.

Three features of the general case become transparent in this limit. First, the initial two terms of \eqref{eq:SrhoR_compact} are separately non-negative. The reason is that $p_\pi$ is an honest probability distribution over a discrete set, so $-\sum_\pi p_\pi \ln(p_\pi) \geq 0$, while $d_\pi \geq 1$ for every irreducible representation, so $\hat{A} \geq 0$. In other words, the sign indefiniteness discussed above is entirely a non-compact phenomenon.

Second, no renormalization is required. The trace \eqref{eq:BRtrace} on $\mathcal{B}_R$ is finite whenever $\mathrm{Vol}(G) < \infty$, so the factorization map $V$ is an isometry without modification, and the usual extended Hilbert space treatment applies unchanged. The machinery of Sec.~\ref{sec:algebras} is only needed once $G$ is non-compact.

Finally, the distinction between the Plancherel and the microcanonical measures, which is the  content of the area term, reduces here to the familiar factor of the dimensions of the representations $d_\pi$. The Plancherel measure weights each irreducible representation by $d_\pi$ because it resolves all of the group $G$, while the counting measure appropriate to class functions weights each representation once. The fact that $\hat{A}$ has spectrum $\ln(d_\pi)$ is then the algebraic version of the statement that the edge modes at the entangling surface carry a $d_\pi$ dimensional Hilbert space. For non-compact $G$, $\ln(\mu(\pi))$ is the correct replacement for $\ln(d_\pi)$, and if we interpret $\mu(\pi)$ as a density of states for each of the representations, the compact and non-compact cases have similar interpretations.

\subsection{Bulk versus boundary entropy} \label{sec:bulkvsbdy}

In AdS/CFT, there are two distinct notions of entropy associated with a boundary subregion $R$. The first is the von Neumann entropy $S(\hat{\rho}_R)$ of the reduced boundary state $\hat{\rho}_R$ of the boundary subregion $R$. The second is the bulk generalized entropy $S_{\mathrm{gen}}(\rho_R)$ of the entanglement wedge $E(R)$ which is dual to the same boundary subregion \cite{Ryu_2006,Faulkner:2013ana,Hubeny:2007xt,Engelhardt:2014gca}. Entanglement wedge recovery \cite{Harlow:2016vwg,Jafferis_2016,Dong:2016eik} then asserts that these two notions of entropy coincide numerically, and so $S(\hat{\rho}_R) = S_{\mathrm{gen}}(\rho_R)$. So, should we think of the entropy $S(\rho_R)$ computed above as the analog of the bulk or the boundary entropy?
To answer this, we  first note that $\rho_R$ is a state for the algebra of boundary anchored line operators $\mathcal{A}_R$, rather than the complete set of bounded operators on the extended Hilbert space $\mathcal{B}_R$ (see Sec.~\ref{sec:algebras}).
The two algebras $\mathcal{B}_R$ and $\mathcal{A}_R$ have rather different characters, and the difference is what allows us to interpret the entropy \eqref{eq:SrhoR_gen} holographically.  

In the gauge fixed presentation \eqref{eq:gaugefixed}, the extended Hilbert space $\Hext{R}$ factorizes completely across the boundary vertices of $R$, and $\mathcal{B}_R$ is the algebra of all bounded operators on those vertices. It therefore has an intrinsically boundary definition: nothing in its construction refers to the interior of $\Sigma$. The algebra $\mathcal{A}_R$, by contrast, is generated by gauge invariant, boundary anchored line operators, and it has a large center. The Hilbert space generated by acting with $\mathcal{A}_R$ on a reasonable choice of vacuum state (a cyclic and separating vector \cite{Dixmier1977}) does not factorize across the boundary vertices, precisely because of that center. In this sense, we should think of $\rho_R$ as an intrinsically bulk object, a state for a subalgebra defined by bulk objects, while a state for $\mathcal{B}_R$ is a boundary state. The operator valued weight $\mathcal{E}$ of Sec.~\ref{sec:ovw} is the dictionary between the two: $\mathcal{A}_R$ plays the role of the code algebra \cite{Harlow:2016vwg}, and $\mathcal{E}$ is the restriction of a boundary state to it.

Accordingly, let $\hat{\rho}_R$ denote any normalized state for $\mathcal{B}_R$ whose restriction to the code algebra is the bulk state,
\begin{align}
    \mathcal{E}(\hat{\rho}_R) = \rho_R \,. \label{eq:preimage}
\end{align}
Repeating the argument of \eqref{eq:rhoFrho}--\eqref{eq:SrelrhoFrho} one level up in the nesting \eqref{eq:nesting}, using the bimodularity \eqref{eq:bimodular} of $\mathcal{E}$ and the trace compatibility \eqref{eq:BtoA} in place of \eqref{eq:AtoZ}, we find
\begin{align}
    S(\hat{\rho}_R) = S(\rho_R) - S_{\mathrm{rel}}(\hat{\rho}_R || \mathcal{E}(\hat{\rho}_R)) \,. \label{eq:bulkvsbdy}
\end{align}

Then suppose that we could find a state $\hat{\rho}^*_R$ in the family \eqref{eq:preimage} for which the relative entropy vanishes. If $S_{\mathrm{rel}}(\hat{\rho}_R || \mathcal{E}(\hat{\rho}_R)) \geq 0$, with equality if and only if $\hat{\rho}^*_R = \mathcal{E}(\hat{\rho}^*_R)$, we would then be entitled to call $\hat{\rho}^*_R$ the boundary state dual to $\rho_R$, define a bulk to boundary map by $\mathcal{E}^{-1}(\rho_R) \equiv \hat{\rho}^*_R$, and to read off from \eqref{eq:bulkvsbdy} that the boundary entropy equals the bulk entropy. This is the structure of holographic quantum error correction with a center, in which the area operator sits in the center of the code algebra and the entropy of the boundary state splits into an area term and a bulk term \cite{Harlow:2016vwg}.

\subsubsection{Compact \texorpdfstring{$G$}{G}}

For the moment, suppose $G$ is compact. We define
\begin{align}
    e^{S_0} \equiv \Tr_{\mathcal{B}_R}[\rho_R] \,. \label{eq:S0def}
\end{align}
By \eqref{eq:tracevoldiv}, $e^{S_0} = \mathrm{Vol}(G)$ is independent of the state $\rho_R$ we chose to define it. Equivalently, by \eqref{eq:ovwdef}, $e^{S_0}$ is the factor by which $\mathcal{E}$ fails to act as the identity on the subalgebra $\mathcal{A}_R$,
\begin{align}
    \mathcal{E}(\mathcal{O}) = e^{S_0} \, \mathcal{O} \,, \qquad \mathcal{O} \in \mathcal{A}_R \,. \label{eq:EonA}
\end{align}
These definitions of $S_0$ are equivalent, since $\Tr_{\mathcal{B}_R}[\rho_R] = \Tr_{\mathcal{A}_R}[\mathcal{E}(\rho_R)]$ by \eqref{eq:BtoA}. So $S_0$ measures exactly the failure of $\mathcal{E}$ to be a conditional expectation in the sense of Sec.~\ref{sec:ovw}: the map $\mathcal{E}$ restricts to the identity on $\mathcal{A}_R$, and is bounded, precisely when $S_0 = 0$. The quantity $S_0$ also carries the scheme dependence of Sec.~\ref{sec:entropy_nomatter}: under a rescaling $dg \to \kappa \, dg$ of the Haar measure, $\rho_R$ is rescaled along with $\Tr_{\mathcal{A}_R}$, so that
\begin{align}
    S_0 \to S_0 + \ln(\kappa) \,, \qquad S(\rho_R) \to S(\rho_R) - \ln(\kappa) \,,
\end{align}
while $\Tr_{\mathcal{B}_R}$, being the ordinary trace on the boundary Hilbert space, is not rescaled at all.

When $S_0$ is finite, \eqref{eq:EonA} hands us the boundary state $\hat{\rho}_R^*$ we are looking for. The operator $e^{-S_0}\rho_R$ is a normalized state of $\mathcal{B}_R$, and we may take
\begin{align}
    \hat{\rho}^*_R = e^{-S_0} \rho_R \,,
\end{align}
which satisfies \eqref{eq:preimage} by \eqref{eq:EonA}. Its entropy follows from bimodularity and \eqref{eq:BtoA}, which give $\Tr_{\mathcal{B}_R}[\rho_R \ln(\rho_R)] = \Tr_{\mathcal{A}_R}[\mathcal{E}(\rho_R \ln(\rho_R))] = e^{S_0} \Tr_{\mathcal{A}_R}[\rho_R \ln(\rho_R)]$, so that
\begin{align}
    S(\hat{\rho}^*_R) = S(\rho_R) + S_0 \,, \qquad S_{\mathrm{rel}}(\hat{\rho}^*_R || \mathcal{E}(\hat{\rho}^*_R)) = - S_0 \,. \label{eq:recoveredentropy}
\end{align}
Both sides of the first equation behave as they should. The left hand side is the von Neumann entropy of an honest density matrix on the boundary Hilbert space, so it is non-negative and independent of the normalization of the Haar measure. The right hand side is a sum of two separately scheme dependent quantities whose $\ln(\kappa)$ shifts cancel. Comparing with the discussion below \eqref{eq:haarshift}, we recognize $S_0$ as the state independent constant, up to which the entropy of a subregion in gravity is defined \cite{Witten:2021unn,Chandrasekaran:2022eqq,Chandrasekaran:2022cip,Jensen2023,Penington:2023dql,Kudler-Flam:2023qfl}. Here it acquires a sharp definition through \eqref{eq:S0def}: $e^{S_0}$ is the number of boundary states per bulk state. If we demand that the relative entropy $S_{\mathrm{rel}}(\hat{\rho}^*_R || \mathcal{E}(\hat{\rho}^*_R)) $ is strictly non-negative, then we must set $S_0=0$.

When $S_0 = 0$, so that $\mathcal{E}$ is a genuine conditional expectation, the dual state is $\hat{\rho}^*_R = \rho_R$ itself, read as an element of the larger algebra, and the boundary entropy exactly equals the bulk entropy. In that case, the relative entropy in \eqref{eq:bulkvsbdy} is non-negative and vanishes only if $\hat{\rho}_R^* = \rho_R$, so $\hat{\rho}^*_R$ is unique. It is worth seeing this explicitly. Take $G$ compact and any state of $\mathcal{B}_R$ which satisfies \eqref{eq:preimage} and is of the product form $\hat{\rho}_R = \int d\mu(\pi) \, q_\pi \, \rho_\pi \otimes \tau_\pi$, with $\tr_{V_\pi}[\tau_\pi] = 1$. Then a short computation gives
\begin{align}
    S_{\mathrm{rel}}(\hat{\rho}_R || \mathcal{E}(\hat{\rho}_R)) = \langle \hat{A} \rangle_\rho - \int d\mu(\pi) \, q_\pi \, S(\tau_\pi)  \,.
\end{align}
Because $S(\tau_\pi) \leq \ln(d_\pi)$, with equality only for $\tau_\pi = \Pi_\pi$, we have that $	S_{\mathrm{rel}}(\hat{\rho}_R || \mathcal{E}(\hat{\rho}_R)) \geq 0$. The bulk entropy is therefore the maximum of the boundary entropy over all boundary states which restrict to $\rho_R$, and the maximum is attained precisely at the dual state $\hat{\rho}^*_R = \rho_R$. 

\subsubsection{Non-compact \texorpdfstring{$G$}{G}}

When $G$ is non-compact, $S_0 = \infty$ by \eqref{eq:tracevoldiv}, and the construction above fails. Moreover, it fails for a reason which no choice of $\hat{\rho}_R$ can repair. Every normalized state for $\mathcal{B}_R$ is a density matrix on the boundary Hilbert space, so $S(\hat{\rho}_R) \geq 0$, and \eqref{eq:bulkvsbdy} then bounds
\begin{align}
    S_{\mathrm{rel}}(\hat{\rho}_R || \mathcal{E}(\hat{\rho}_R)) \leq S(\rho_R) \,.
\end{align}
But we saw in Sec.~\ref{sec:entropy_nomatter} that $S(\rho_R)$ is not bounded below when $\widehat{G}$ is continuous: the differential entropy of a distribution concentrated on a set of measure $\epsilon$ is $-\ln(\epsilon^{-1})$, which we can make as negative as we like. For any such state, the equation $S_{\mathrm{rel}} = 0$ has no solution at all within the family \eqref{eq:preimage}, and there is no boundary state dual to $\rho_R$. Equivalently, relative entropy with respect to $\mathcal{E}(\hat{\rho}_R)$ is not positive definite, which is another way of saying that $\mathcal{E}$ is an operator valued weight and not a conditional expectation. By \eqref{eq:recoveredentropy}, the failure of monotonicity is by exactly $S_0$.

\paragraph{What would restore recovery.} The obstruction to defining a holographic bulk-to-boundary map is therefore not that the effective theory of this paper is inconsistent, but that it does not by itself supply a zero of the entropy. Anything which renders $S_0$ finite removes the obstruction. From Sec.~\ref{sec:entropy_compact}, the way to make $S(\rho_R)$ bounded below is to make $\widehat{G}$ discrete, so that $H[\,p_\pi]$ becomes an honest Shannon entropy as in \eqref{eq:HcompactG}. This is the same condition. For example, consider the case $G = \R$ of Sec.~\ref{sec:Rexample}. Discretizing $\widehat{G}$ at an average spacing $2\pi\Delta$ with respect to $dk$ is the same as compactifying $\R \to \mathrm{U}(1)$ to circumference $1/\Delta$, under which
\begin{align}
    e^{S_0} = \mathrm{Vol}(G) = \frac{1}{\Delta} \,, \label{eq:S0spacing}
\end{align}
and the differential entropy is shifted to $H[\,p_\pi] + \ln(\Delta^{-1})$, which is the shift by $S_0$ in \eqref{eq:recoveredentropy}. So a discrete spectrum for the area operator, a finite $S_0$, and a boundary entropy with a zero are three descriptions of the same deformation.

Note that this deformation cannot be implemented by simply replacing a non-compact group $G$ with a different, compact group $G'$. The physics of the model, including the spectrum of the area operator, is fixed by $G$, so we are not free to trade one gauge group for another and expect the same effective description at small Newton's constant $G_N$. What is required is a deformation which simultaneously discretizes $\widehat{G}$ and makes $S_0$ finite, while leaving the effective theory intact. The resulting structure is presumably not another group.

The gap $\Delta$ that this deformation introduces must also be very small. Positivity of the boundary entropy in \eqref{eq:recoveredentropy} requires $S_0 \geq -S(\rho_R)$ for every state, that is, $S_0$ must be at least as large as the most negative value the effective theory assigns to $S(\rho_R)$. Since the scale of the entropy is set by the area term, which we argue in \cite{Balasubramanian:2026xyz} has eigenvalues $\ell/4G_N$ after accounting for the measure of the path integral, this requires
\begin{align}
    S_0 \sim \frac{1}{G_N} \,, \qquad \Delta \sim e^{-1/G_N} \,,
\end{align}
using \eqref{eq:S0spacing}. In other words, in order for the effective theory to remain accurate to all orders in $G_N$ while still admitting a boundary dual, the underlying discreteness must be non-perturbatively small. Discreteness of this size, arising from effects invisible at any order in perturbation theory, is a recurring theme in gravity \cite{Saad:2018bqo,Penington:2019npb,Almheiri:2019psf,Penington:2019kki,Almheiri:2019qdq,Iliesiu:2024cnh,Akers:2025ynh,Balasubramanian:2022lnw,Balasubramanian:2022gmo,Chandra:2022fwi,Balasubramanian:2024lqk,Balasubramanian:2026azk}. 

Finally, it is worth being precise about what these effects are and are not needed for. Nothing in the effective theory itself is lost without them. The map $\mathcal{E}$ is faithful, so no bulk information is discarded in passing to the boundary algebra, and every difference of entropies is finite and independent of the scheme. As we said above, what the effective theory cannot supply on its own is the zero of the entropy, and with it the normalizability of the boundary state dual to $\rho_R$. If some effect makes $S_0$ finite, then $\hat{\rho}^*_R = e^{-S_0} \rho_R$ is a genuine boundary state, the boundary entropy is the bulk entropy plus $S_0$, and holographic recovery of the bulk state is restored. $e^{S_0}$ can then be interpreted as counting the boundary states that the bulk description does not resolve on its own.

\section{String nets and topological entanglement entropy}
\label{sec:topEE}

Above, we derived the entropy formula \eqref{eq:resultsummary} for topological tensor networks with arbitrary transformable gauge groups $G$.
When $G$ is a finite group, however, our model is the same as Kitaev's quantum double model \cite{Kitaev:1997wr}, and the entanglement entropy of its states is a familiar object in condensed matter theory \cite{Kitaev:2005dm,Levin:2006arx,Flammia:2009axf,Hamma:2004vdz}: in particular, the entropy of a quantum state in Kitaev's double model contains a universal term, the topological entanglement entropy, which is similar to the ``area term'' in our entropy formula \eqref{eq:resultsummary}. 

In this section, we will make this connection more precise by showing that the expectation value of the area operator agrees sector by sector with the topological entanglement entropy, up to a state independent constant. For finite $G$, this constant is $-\ln(|G|)$; for continuous and/or non-compact groups, this constant diverges, and is absorbed by the renormalization constant $S_0$ we explained in Sec.~\ref{sec:entropy_nomatter}.
To show this, it will be convenient to restate \eqref{eq:resultsummary} in a form which refers only to the anyon content of the theory, and not to the group used to construct it. This restatement is worth having for its own sake, because it separates what in our formula is a consequence of topological symmetry from what is a consequence of group theory. It is also the form of the result one would need in order to replace $\mathrm{Rep}(G)$ by an input which does not come from a group at all, which is what a comparison with three-dimensional gravity requires, as we discuss in Sec.~\ref{sec:CTV}.

\subsection{The quantum double model and Turaev--Viro theory} \label{sec:doublemodel}

Recall from Secs.~\ref{sec:themodel} and \ref{sec:LatticeIndep} that our topological tensor networks are generalizations of string nets \cite{Levin_2005,kirillov2011stringnet}: the legs of the graph $\Lambda$ are labeled by the objects of a tensor category, its vertices by their intertwiners, and two labeled graphs define the same state whenever they are related by a fixed set of local moves.  The latter is why  the resulting Hilbert space depends only on the surface $\Sigma$ that $\Lambda$ tessellates. In our case, the input category is $\widehat{G} = \mathrm{Rep}(G)$, the intertwiners are the states of $\Pi_A[V_{\vec{\pi}}^*]$ in \eqref{eq:Hphys}, and the local moves are moves 1 and 2 of Sec.~\ref{sec:LatticeIndep}, which is exactly the content of the lattice independence we established there. A string net in the usual sense differs from ours in requiring the input to be a fusion category with finitely many simple objects (representations).

When $G$ is a finite group, this construction is Kitaev's quantum double model \cite{Kitaev:1997wr}, which places a copy of $L^2(G)$ on each edge of a lattice and imposes two constraints: an electric constraint at each vertex, requiring invariance under simultaneous group multiplication on the incident edges, and a magnetic constraint on each plaquette, requiring the holonomy around it to be trivial. These are the same two constraints we imposed in Sec.~\ref{sec:themodel}, and the double model is the special case of our construction where $G$ is finite. Its physical Hilbert space is that of the Dijkgraaf--Witten theory for $G$ \cite{Dijkgraaf:1989pz,Hu:2012wx}, and equivalently that of a different TQFT called Turaev--Viro theory built from $\mathrm{Rep}(G)$ \cite{TuraevViro1992,BarrettWestbury1996,kirillov2011stringnet,Buerschaper_2009}. When $G$ is not a finite group, the TQFT which is equivalent to our model is not yet known.

The equivalence between the quantum double model and Turaev--Viro theory  explains the ``doubling'' of the symmetry that we met in Sec.~\ref{sec:themodel}. Turaev--Viro theory takes as input a fusion category $\mathcal{C}$ (the representations $\pi$ of $G$ and their braiding/fusion rules) and produces a three-dimensional TQFT; the anyons of that TQFT are not the objects of $\mathcal{C}$ itself but the objects of its Drinfeld center $Z(\mathcal{C})$ \cite{kirillov2011stringnet,kirillov2010corners}. For $\mathcal{C} = \mathrm{Rep}(G)$ the Drinfeld center is the same as the representations of the quantum double $D[G]$, which is why the theory carries both electric and magnetic operators. 

The entanglement entropy of the double model is most naturally stated in terms of its excitations, the anyons of $D[G]$, so we describe these next. An anyon is a point defect on a Cauchy slice $\Sigma$: a puncture around which the flat connection can have a nontrivial holonomy, which we call its flux, and which can carry a charge measured by a line operator encircling it. In the double model for finite $G$, a defect of this kind is labeled by a conjugacy class $[g] \subset G$ together with an irreducible representation $\lambda$ of the centralizer $C_G(g)$, the first datum recording the flux and the second the charge; these pairs are exactly the simple objects of the Drinfeld double $D[G]$ described above.

We can now state the connection between our main result  \eqref{eq:resultsummary} and the entanglement entropy of a finite string net. Let $\gamma$ be a curve which separates $\Sigma$ into two pieces, and suppose the state has probability $p_a$ of being in the $a$ sector within each piece, so that $\sum_a p_a = 1$. Then the entanglement entropy across $\gamma$ contains a universal contribution which does not depend on the geometry of $\gamma$, and the entropy takes the form\footnote{Note that the final term $S(\rho_a)$ will generally depend on the boundary conditions imposed at $\partial \Sigma$, such as the number of marked points within each subregion.} \cite{Kitaev:2005dm,Levin:2006arx,Dong:2008ft,Bonderson:2017osr}
\begin{align}
    S(\rho_R) \;=\; \underbrace{- \sum_a p_a \ln (p_a)}_{H[\,p_a]} \;+\; \underbrace{\sum_a p_a \ln \bigl(\mathcal{S}_1^{\,a}\bigr)}_{\langle \hat{A}\rangle + \mathrm{constant}} \;+\; \sum_a p_a \, S(\rho_a) \,. \label{eq:stringnetentropy}
\end{align}
Here, $\mathcal{S}_1^{\,a}$ is the matrix element of the modular $S$ transformation between the anyon $a$ and the vacuum line, and it is equal to the quantum dimension of $a$ divided by the total quantum dimension of the theory,
\begin{align}
    \mathcal{S}_1^{\,a} = \frac{d_a}{\mathcal{D}} \,, \qquad \mathcal{D}^2 = \sum_a d_a^2 \,. \label{eq:S1adef}
\end{align}
When only the trivial anyon $a=1$ contributes, the universal term reduces to $-\ln(\mathcal{D})$. This is the topological entanglement entropy of Kitaev and Preskill and of Levin and Wen \cite{Kitaev:2005dm,Levin:2006arx}.\footnote{This term is sometimes isolated by taking sums and differences of adjacent regions so that the extensive parts of the entropy cancel. However, the topological entanglement entropy still contributes even when the extensive piece $\sum_a p_a \, S(\rho_a)$ is present.} Note that when $G$ is a finite group, $\mathcal{D} = |G|$. When $G$ is not a finite group, $\mathcal{D}$ is infinite: for a general transformable group $G$, $\mathcal{D} = |G| = \delta(e) \mathrm{Vol}(G)$, which diverges unless $G$ is a finite group. 
This is the divergence that renormalized entropy removes, leaving behind the $S_0$ ambiguity \eqref{eq:S0def}.

Equation \eqref{eq:stringnetentropy} is written with respect to the flat sum over anyons, in which each label is counted once. Equation \eqref{eq:SrhoR_gen} was instead written with respect to two different measures on the labels: the microcanonical measure $d\pi$, which counts each representation once, and the Plancherel measure $d\mu(\pi) = \mu(\pi) d\pi$, which weights it by the density of states. Suppose we do the same on the anyon labels, defining
\begin{align}
    q_a = \frac{p_a}{\mathcal{S}_1^{\,a}} \,, \qquad \sum_a p_a = \sum_a \mathcal{S}_1^{\,a} q_a = 1 \,,
\end{align}
so that the flat sum $\sum_a$ plays the role of $\int d\pi$ and the weighted sum $\sum_a \mathcal{S}_1^{\,a}$ plays the role of $\int d\mu(\pi)$. Then the first two terms of \eqref{eq:stringnetentropy} combine, and the entropy becomes
\begin{align}
    S(\rho_R) = \sum_a \mathcal{S}_1^{\,a} \Bigl[ - q_a \ln (q_a) + q_a S(\rho_a) \Bigr] \,, \label{eq:stringnetentropy2}
\end{align}
which is exactly the form of \eqref{eq:SrhoR_first} under the substitution $\int d\mu(\pi) \to \sum_a \mathcal{S}_1^{\,a}$. In other words, $\mathcal{S}_1^{\,a}$ occupies precisely the position that $\mu(\pi)$ occupies in our formulas: it is the density of states of the label $a$, measured relative to the flat measure which counts each label once.\footnote{The factor of $\frac{1}{\mathcal{D}}$ is absorbed into the difference between the finite sum and the continuous integral.}

The identification can be made more precise when $G$ is a finite group. In the quantum double model, the pure charge anyons are those with trivial flux, $a = (e,\pi)$ with $\pi$ an irreducible representation of $G$. Their quantum dimension is $d_\pi$, and the total quantum dimension of $D[G]$ is $\mathcal{D} = |G|$, so that
\begin{align}
    \mathcal{S}_1^{\,(e,\pi)} = \frac{d_\pi}{|G|} = \frac{\mu(\pi)}{|G|} \,. \label{eq:S1purecharge}
\end{align}
The right hand side is the Plancherel measure of $\pi$, in the normalization of the Haar measure in which $G$ has volume $1$ used in Sec.~\ref{sec:entropy_compact}, divided by the volume of the gauge group. So for a finite group, the eigenvalue $\ln(\mu(\pi))$ of our area operator \eqref{eq:areaoperator} agrees with the topological entanglement entropy of the $(e,\pi)$ sector, up to the state independent constant $-\ln(|G|)$ which diverges for general transformable gauge groups. Because this divergence is state independent, it is absorbed into the $S_0$ scheme dependence upon renormalization.

The agreement even extends to which sectors appear, which is a sharper check. Our formula \eqref{eq:SrhoR_gen} runs over $\widehat{G}$ alone: a single representation label, and no flux. Equation \eqref{eq:stringnetentropy} runs over all anyons which can end on the cut $\gamma$. In Sec.~\ref{sec:factorization} we gave the cut the rough boundary condition, on which only the pure charges $(e,\pi)$ condense (see Appendix~\ref{app:topbcs} for this statement in the language of anyon condensation), so a boundary anchored cut on the disk can only see anyons of this form. The restriction to a single representation label in Sec.~\ref{sec:factorization} was therefore a consequence of the rough boundary condition on the cut, as we explained in Sec.~\ref{sec:areaop_nomatter}. A different topological boundary condition on the cut changes only which anyons the cut can see, and not the form of any of the results above (see Sec.~\ref{sec:areaop_nomatter}).\footnote{Furthermore, the spectrum of the area operator is independent of the choice of boundary conditions we impose on the boundary of $\Sigma$.}

This is the reformulation we were after. The invariant content of the area operator is not that its spectrum is $\ln(\mu(\pi))$ for $\pi \in \widehat{G}$. It is that the eigenvalue attached to the superselection sector labeled by an anyon $a$ is
\begin{align}
    \langle \hat{A} \rangle_a \sim \ln \bigl( \mathcal{S}_1^{\,a} \bigr) \,, \label{eq:areainvariant}
\end{align}
with $a$ running over whichever anyons the cut in question is able to support, and $\sim$ means equality up to the state independent constant explained above. 
The details of the gauge group entered our derivation only through two places: it determined the anyon spectrum, and, together with the boundary conditions, it determined which part of that spectrum a boundary anchored cut can see. 
Written as \eqref{eq:areainvariant}, the result no longer refers to the detailed group theory of $G$, and it only depends on the physical properties (such as the anyon content) of the theory. 

One immediate consequence concerns cuts which are not boundary anchored. If the cut $\gamma$ that we use to factorize the Hilbert space is not boundary anchored, we expect all the anyons of the theory to contribute to the spectrum of the area operator. This is the statement that the sectors shared between the two boundary components of a cylinder are labeled by all of the possible anyon labels $([g],\lambda)$ \cite{Zhang:2011jd,Delcamp:2016eya}, with
\begin{align}
    \ln \bigl( \mathcal{S}_1^{\,([g],\lambda)} \bigr) = \ln \left( \frac{|[g]| \, d_\lambda}{|G|} \right) \,. \label{eq:finiteGquantumdim}
\end{align}
Here, $|[g]|$ is the number of elements in the conjugacy class, and $d_\lambda$ is the dimension of the $\lambda$ representation. 

In \cite{Balasubramanian:2026xyz}, we construct the same class of closed ribbon operators for an arbitrary transformable group and show that they carry exactly this pair of labels; the pure charges of the main text sit inside the label space as the subset with trivial flux. We therefore expect that the analogous spectrum \eqref{eq:areainvariant} should hold for a more general transformable group. Settling this would require constructing the factorization map along a closed cut and repeating the analysis of Sec.~\ref{sec:algebras} for it, which we leave for future work.

\subsection{Towards gravity} \label{sec:CTV}

We will now discuss how our results connect to three-dimensional gravity in more detail. To this end, we set $G=\SL(2,\R)$ in this section.
One of the main reasons that three-dimensional gravity and Chern--Simons theory are not the same quantum theory (which we review in more detail in \cite{Balasubramanian:2026xyz}) is that, to all orders in perturbation theory, gravity can be thought of as a subtheory of $\SL(2,\R) \times \SL(2,\R)$ Chern--Simons theory obtained by restricting to the connected component of the Chern--Simons phase space on which the spacetime metric is invertible. The tensor networks of this paper quantize the full Chern--Simons theory, because we allow every unitary irreducible representation of $\SL(2,\R)$ to appear on every leg. It is natural to ask what would change in the entropy formula \eqref{eq:resultsummary} if we imposed invertibility of the metric, and in particular what would happen to the spectrum of the area operator $\hat{A}$.

Imposing invertibility replaces the input data of the model altogether, and none of the intermediate steps in the derivation of the area operator $\hat{A}$ survive such a replacement verbatim, because each of them referred to the specific gauge group $G$. However, the anyon formulation \eqref{eq:areainvariant} of the area spectrum is different: it refers only to the anyon content of the theory and to which anyons the cut can see. Following \cite{Donnelly_2016,Lin:2017uzr,Jafferis:2019wkd,Balasubramanian:2023dpj,Balasubramanian:2025rcr, Hartman:2025cyj, Mertens:2025ydx}, we therefore propose that \eqref{eq:areainvariant} remains true when the input $\mathrm{Rep}(G)$ to our tensor network model is no longer given by the representation theory of a classical group. The question then becomes what to substitute for $\mathrm{Rep}(G)$.

\paragraph{Conformal Turaev--Viro theory.}

The string net construction reviewed in Sec.~\ref{sec:doublemodel} describes a TQFT with finitely many anyons. We saw at the end of Sec.~\ref{sec:LatticeIndep} that our tensor networks do not satisfy this assumption: for any continuous and/or non-compact group $G$, the Hilbert space of the torus is infinite dimensional, so the TQFT associated with our networks, if it exists, has infinitely many anyons. This left open the question of whether such a TQFT exists at all.

Recently, a state sum construction was proposed which breaks the same finiteness assumption in the same way. Conformal Turaev--Viro (CTV) theory \cite{Hartman:2025cyj,Hartman:2025ula} replaces the finite input category of Turaev--Viro theory with the representation category of the modular double of the quantum group $\mathcal{U}_q(\sl(2,\R))$ \cite{Faddeev:1999fe,Ponsot:1999uf,Ponsot:2000mt}, whose representations are labeled by a continuous Liouville momentum $P \in \R_+$ and whose $6j$ symbols are known explicitly \cite{Ponsot:2000mt,Teschner:2003em,Teschner:2012em}. The resulting theory is the doubled version of Virasoro TQFT \cite{Collier_2023,Collier:2024mgv}, in the same way that ordinary Turaev--Viro theory is the doubled version of Chern--Simons theory: its anyons are labeled by pairs of Liouville momenta $(P_+,P_-)$, and its states on a surface are pairs of Virasoro conformal blocks. The existence of CTV theory is therefore evidence that the string net-like constructions of this paper, which have a continuum of anyons for the same reason, define genuine TQFTs, with $\mathrm{Rep}(G)$ playing the role that the modular double plays in CTV theory.

Because the partition functions of Virasoro TQFT reproduce the gravitational path integral on a fixed topology \cite{Collier_2023}, CTV theory quantizes only the subspace of the $\SL(2,\R) \times \SL(2,\R)$ Chern--Simons phase space on which the metric is invertible \cite{Witten:1988hc}, whereas our networks, which allow every representation of $\SL(2,\R)$ on every leg, quantize the full Chern--Simons phase space in the large level limit. At the level of the Hilbert space, the restriction to invertible metrics should be implemented by a single substitution: replace the input category $\mathrm{Rep}(\SL(2,\R))$ on the legs of the tensor network by the representations of the modular double of $\mathcal{U}_q(\sl(2,\R))$, with the deformation parameter $q$ fixed by Newton's constant. Because the area spectrum \eqref{eq:areainvariant} refers only to the anyon content of the theory and not to the group, this substitution should leave the structure of our entropy formula intact and changes only the labels $a$ and the density of states $\mathcal{S}_1^{\,a}$. We will investigate this substitution, and its consequences for the spectrum of the area operator, in \cite{Balasubramanian:2026xyz}.

\section{Discussion} \label{sec:discussion}
	
In \cite{Balasubramanian:2025rcr}, we constructed a class of tensor networks which prepare states of topological field theories with continuous and/or non-compact gauge groups. The Hilbert space of physical states prepared by these tensor networks on a surface $\Sigma$ is defined as equivalence classes under shifts by states that are null under the gauge constraints, and is equipped with a co-invariant inner product. The TQFTs which these tensor networks prepare are not known when $G$ is not a finite group. Nevertheless, we showed that the resulting physical Hilbert space is independent of the lattice used to define the tensor networks in the bulk of $\Sigma$, so the Hilbert spaces are indeed topological invariants of $\Sigma$. Furthermore, recent constructions of TQFTs which relax analogous finiteness assumptions of traditional TQFTs \cite{Collier_2023,Hartman:2025cyj} suggest that a first principles construction of these TQFTs indeed exists. It would be interesting to directly construct these theories. 

In this paper, we computed the entanglement entropy shared between a subregion $R$ of the boundary of $\Sigma$ and its complement $\overline{R}$ in these tensor networks. To do so, we defined a factorization map $V$ which embeds the physical Hilbert space into a product $\Hext{R} \otimes \Hext{\overline{R}}$. The factorization map required the introduction of {\it edge modes} supported on cuts separating the networks into two parts. These edge modes play a similar role to recently proposed observer models in quantum gravity \cite{Abdalla:2025gzn,Harlow:2025pvj}, quantum reference frames \cite{Fewster:2024pur}, and the embedding map of the corner symmetry proposal \cite{Ciambelli:2021vnn,Ciambelli:2021nmv,Freidel:2021cjp,Ciambelli:2022cfr,Ciambelli:2024qgi,Balasubramanian:2023dpj}. It would be interesting to understand these connections in more detail. Here, we simply note that the structure of the edge modes introduced by the factorization map were determined by the boundary conditions we imposed along the cut $\gamma$, which we took to be topological so that the resulting entropy does not depend on the details of the lattice near $\gamma$.

Using this factorization map, we then defined the reduced state $\rho_R$ for the boundary subsystem we wish to compute the entropy of, along with a renormalized trace which ensured that $\rho_R$ is properly normalized.
The resulting entropy formula \eqref{eq:SrhoR_gen} for the reduced state $\rho_R$ split into three pieces: a differential entropy $H[\,p_\pi]$ measuring the classical uncertainty about which superselection sector $\rho_R$ occupies, an average $S_{\mathrm{bulk}}(\rho_R)$ of the entropies within each sector, and the expectation value $\langle \hat{A} \rangle_\rho$ of a state independent operator. In analogy to AdS/CFT, we call $\hat{A}$ the area operator.
When $G$ is a finite group, the eigenvalues of $\hat{A}$ are the sector-wise topological entanglement entropies of Kitaev and Preskill and of Levin and Wen \cite{Kitaev:2005dm,Levin:2006arx}, up to a state independent constant which diverges in the limit $G$ that becomes continuous and/or non-compact. In this sense, the expectation value of the area operator in our model is the natural generalization of the topological entanglement entropy for finite groups, generalized to a theory with infinitely many anyons. 

In a companion paper \cite{Balasubramanian:2026xyz}, we will study the entropy of topological tensor networks with added matter, and define a procedure for averaging over this matter, even though its Hilbert space may be infinite dimensional. We will argue that if we properly account for the invertibility of the metric, the result reproduces the quantum extremal surface formula in gravity for small $G_N$, at least for boundary subregions of time symmetric spacetimes. We will also see that the same approach reproduces the entropy between the two asymptotic boundaries of rotating BTZ black holes (which are not static spacetimes), and provides a generalization of the quantum extremal surface formula that applies to compact bulk regions.

\paragraph{Acknowledgments:} We thank Xi Dong, Don Marolf, Wayne Weng, Chris Akers, Yimu Bao, Jon Sorce, Elba Alonso-Monsalve, Dan Sehayek, Leo Shaposhnik, Alexander Jahn, Wissam Chemissany, Sami Kaya, and Juan Maldacena for helpful discussions.
CC is supported by the National Science Foundation Graduate Research Fellowship under Grant No. DGE-2236662. 
CC was also supported in part by grant NSF PHY-2309135 to the Kavli Institute for Theoretical Physics (KITP). 
VB was supported in part by the DOE through DE-SC0013528 and QuantISED grant DE-SC0020360.  Part of this work was performed at the Aspen Center for Physics, which is supported by National Science Foundation grant PHY-2210452.

\appendix

\section{Boundary conditions} \label{apx:topvscon}

In canonical quantization, the physical Hilbert space $\Ha_{\mathrm{phys}}(\Sigma)$ assigned to a spatial surface $\Sigma$ depends on the boundary conditions imposed at each component of $\partial \Sigma$. In the main text, we impose \emph{open} boundary conditions on the boundary components of $\Sigma$. An open boundary condition is defined by composing \emph{topological} boundary conditions, in which $n$ rough segments alternate with $n$ smooth segments at the lattice scale.\footnote{As we define in Appendix~\ref{app:topbcs}, rough boundaries come from dropping the electric constraints, and smooth boundaries come from dropping the magnetic constraints. The $n$ smooth boundaries ensure that the boundary vertices are not identified with each other.} 
If we alter the boundary conditions on $\Sigma$ by defining a boundary Hamiltonian that is tuned to criticality and take the number of boundary points $n \to \infty$, then the open boundary condition instead flows to a \emph{conformal} boundary condition. This is the boundary condition which leads to a full conformal field theory on the boundary of $\Sigma$.

In this appendix, we will review topological boundary conditions in Appendix~\ref{app:topbcs}, describe open boundary conditions in this language in Appendix~\ref{app:openbcs}, and review conformal boundary conditions in Appendix~\ref{app:confbcs}.

\subsection{Topological boundary conditions} \label{app:topbcs}

A topological boundary condition is one that preserves the complete topological invariance of the bulk: a small deformation of a boundary component carrying a topological boundary condition does not change the underlying state. Such a boundary condition is characterized operationally by which bulk line operators are allowed to terminate on the boundary. Recall that the bulk excitations of the theories in Sec.~\ref{sec:themodel} are anyons, i.e., Wilson and 't Hooft lines with generically nontrivial mutual braiding. A topological boundary condition selects a maximal set of mutually transparent bulk lines (meaning they braid trivially with each other, though perhaps not with the remaining anyons), and allows exactly those lines to end on the boundary without leaving behind a localized excitation. One then says that these anyons \emph{condense} on the boundary \cite{Bravyi:1998sy,Kitaev:2011dxc,Bais:2008ni}.

For concreteness, let us first discuss the case where the bulk gauge group on each leg of the tensor network is $G = \Z_2$. This model is known as the toric code \cite{Kitaev:1997wr}, and it admits exactly two topological boundary conditions \cite{Bravyi:1998sy}. The toric code has four anyons: the trivial anyon $1$, the electric charge $e$, the magnetic flux $m$, and the dyonic charge $\epsilon$. Relaxing the electric constraint at boundary vertices but continuing to impose the magnetic constraint at the plaquettes produces the \emph{rough} boundary, on which bulk Wilson lines may end. One then says that the electric charge $e$ condenses. Relaxing the magnetic constraint at boundary plaquettes but continuing to impose the electric constraint at the boundary vertices produces the \emph{smooth} boundary, on which 't~Hooft lines end: one then says that the flux $m$ condenses. There is no topological boundary condition that allows the dyonic anyon $\epsilon$ to end, because $\epsilon$ has fermionic statistics: it is not transparent with respect to itself.

One can generalize the toric code by replacing the gauge group $\Z_2$ with any other finite group $G$. This is known as Kitaev's quantum double model \cite{Kitaev:1997wr}.\footnote{The TQFT associated to Kitaev's double model (including the toric code) is the $G$ Dijkgraaf--Witten theory \cite{Dijkgraaf:1989pz} with the same gauge group.} The anyons of the double model are labeled by a conjugacy class $[g] \subset G$ (the flux) together with an irreducible representation $\lambda$ of the centralizer $C_G(g)$ (the charge). The rough and smooth boundary conditions of the toric code extend directly: relaxing the electric constraint at the boundary vertices gives the rough boundary, on which the pure charges $(e,\pi)$, $\pi \in \widehat{G}$, condense, and relaxing the magnetic constraint at the boundary plaquettes gives the smooth boundary, on which the pure fluxes $([g],1)$ condense. In each case, the condensing anyons form a maximal set of mutually transparent lines; such a set is called a Lagrangian algebra, and its anyons are sometimes called Lagrangian anyons \cite{Kapustin:2010hk,Kong:2014xyz,KONG2021115607}.\footnote{The name refers to a Lagrangian subalgebra of the boundary category, not to a local action density.}

In the language of $G$ gauge theory (the TQFT description of the double model), these two boundary conditions have familiar names. The rough boundary is the \emph{Dirichlet} boundary condition for the gauge field \cite{Wen:2023otf,Wen:2024udn}: the holonomy of the gauge field along the boundary is fixed to be trivial, so that the gauge symmetry is completely broken there and the group $G$ survives as a \emph{global} symmetry acting on the boundary degrees of freedom. Wilson lines, which measure the charges of this global symmetry, can end on such a boundary. The smooth boundary is the \emph{Neumann} boundary condition for the gauge field \cite{Wen:2023otf,Wen:2024udn}: the gauge field is left free on the boundary, so that $G$ remains gauged there and only gauge invariant combinations of the boundary degrees of freedom survive, while the flux through the boundary is unconstrained and 't~Hooft lines can end on it.  The complete set of topological boundary conditions of Kitaev's double model has been classified \cite{Beigi:2010htr,Kitaev:2011dxc,Kong:2014xyz}: they are labeled by a subgroup $K \subseteq G$ together with a second cohomology class on $K$, with $K = \{e\}$ the rough boundary and $K = G$ the smooth one, and the intermediate entries partially break the gauge symmetry on the boundary.\footnote{More generally, the possible topological boundary conditions for TQFTs with finitely many anyons have been classified \cite{Kong:2014xyz,KONG2021115607}, not just the TQFTs associated with the double model.} We will not need the detailed classification in this paper. The point is that the possible topological boundary conditions for a TQFT with finitely many anyons are rigid, and are classified by the group theoretic data (or, more generally, a modular tensor category) intrinsic to the model.

The rough boundary condition is exactly what we imposed on the cut $\gamma$ in Sec.~\ref{sec:factorization}, applied to a transformable group $G$: the electric constraint is dropped at the new vertex on each side of the cut, and the line operators which can end there are the pure charges, labeled by $\widehat{G}$. The situation is much less understood for the TQFTs of Sec.~\ref{sec:themodel}, whose gauge groups are continuous and possibly non-compact. There, the topological boundary conditions have not been classified, and we will not attempt a complete classification here. Instead, we take the rough and smooth boundary conditions of the quantum double model at face value and impose their natural extension. Of course, we expect that, as in the quantum double model, the extended rough and smooth boundary conditions are two entries in a longer list, and that the missing entries would enlarge the space of allowed boundary conditions on the cut defining the factorization map. 

\subsection{Open boundary conditions} \label{app:openbcs}

\begin{figure}
    \centering
    \begin{tikzpicture}[scale=1.35]
        \def\L{4}
        \foreach \i in {0,...,3}{
            \foreach \j in {0,...,3}{
                \fill[blue!10] (\i,\j) rectangle (\i+1,\j+1);
                \node[blue!60!black,font=\small] at (\i+0.5,\j+0.5) {$B_p$};
            }
        }
        \foreach \k in {0,...,4}{
            \draw[gray!70,dashed,thick] (\k,0) -- (\k,-0.6);
            \draw[gray!70,dashed,thick] (\k,4) -- (\k,4.6);
            \draw[gray!70,dashed,thick] (0,\k) -- (-0.6,\k);
            \draw[gray!70,dashed,thick] (4,\k) -- (4.6,\k);
        }
        \foreach \k in {1,...,3}{
            \draw[black!80,thick] (\k,0) -- (\k,4);
            \draw[black!80,thick] (0,\k) -- (4,\k);
        }
        \draw[red!80!black,very thick] (0,0) rectangle (4,4);
        \foreach \i in {1,...,3}{
            \foreach \j in {1,...,3}{
                \filldraw[blue!60!black] (\i,\j) circle (0.09);
            }
        }
        \foreach \k in {0,...,4}{
            \filldraw[fill=white,draw=blue!80!black,thick] (\k,0) circle (0.1);
            \filldraw[fill=white,draw=blue!80!black,thick] (\k,4) circle (0.1);
            \filldraw[fill=white,draw=blue!80!black,thick] (0,\k) circle (0.1);
            \filldraw[fill=white,draw=blue!80!black,thick] (4,\k) circle (0.1);
        }
    \end{tikzpicture}
    \caption{Open boundary conditions on a region $A$ of the infinite square lattice, for the toric code $G = \Z_2$. The region is the set of $16$ shaded plaquettes, on every one of which the magnetic constraint $B_p$ is imposed. The magnetic constraint is not imposed at the plaquettes to the exterior of the red boundary legs.
    The electric constraint $A_v$ is imposed at the $9$ interior vertices (blue) and dropped at the $16$ boundary vertices (white), where some of the edges it would act on (dashed gray) are not included as part of the system. 
    The $40$ edges of $A$ carry qubits (because $L^2(\Z_2) = \C^2$), $24$ in the interior (black edges) and $16$ on the boundary (red edges). 
    }
    \label{fig:openbcs}
\end{figure}
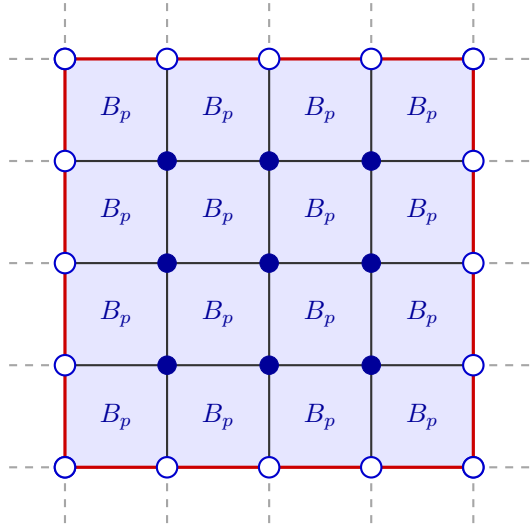

The main text imposes neither the rough nor the smooth boundary condition on $\partial \Sigma$, but the \emph{open} boundary condition of Sec.~\ref{sec:bcs}. A convenient way to think about it is the following. Consider the model on an infinite lattice, and choose a finite region $A$, specified as a set of plaquettes: see Fig.~\ref{fig:openbcs}. The kinematic Hilbert space $\Ha_A$ is the tensor product of $L^2(G)$ over the edges of those plaquettes, and edges outside $A$ are discarded. A vertex of $A$ is \emph{interior} if every edge incident to it lies in $A$, and a \emph{boundary vertex} otherwise. We impose the magnetic constraint on every plaquette of $A$, without exception, and the electric constraint only at the interior vertices: at a boundary vertex some of the edges the constraint would act on are no longer part of the system, so it cannot be imposed there. This is precisely the open boundary condition of Sec.~\ref{sec:bcs}, with the boundary vertices playing the role of the $n$ marked points.\footnote{Although the reduced lattice of Fig.~\ref{fig:reduced} may not appear to have the red lines of Fig.~\ref{fig:openbcs}, they can be included by drawing an arc between neighboring boundary points and adding a plaquette (bounding the resulting triangle) with the magnetic constraint imposed there.}

In the language of Appendix~\ref{app:topbcs}, this is an alternating pattern of rough and smooth segments at the lattice scale. Each boundary vertex is a place where a bulk line can leave $A$ along one of the discarded edges, without leaving an excitation behind, because there is no electric constraint there to detect it. It is therefore a rough segment, shrunk to a single vertex. The boundary edge joining two adjacent boundary vertices bounds a plaquette of $A$, and the flatness constraint is not imposed on the plaquette in the exterior of this boundary edge. It is therefore a smooth segment, one edge long. The boundary edges carry kinematic degrees of freedom, but these are removed by the adjacent plaquette constraints, so the smooth segments carry no physical degrees of freedom of their own, and the physical Hilbert space depends on the region only through the number $n$ of boundary vertices. In the example of Fig.~\ref{fig:openbcs}, the region has $40$ edges, $9$ interior vertices and $16$ plaquettes. The $9$ electric and $16$ magnetic constraints are independent, and each constraint can be used to fix the state at a single leg.
So, the physical Hilbert space has dimension $|G|^{40 - 9 - 16} = |G|^{15} = |G|^{n-1}$, with $n = 16$ the number of boundary points. This agrees with the dimension of $\Ha_{\mathrm{phys}}(\Sigma^{(n)})$ for a finite group computed from the reduced lattice of Sec.~\ref{sec:disk}, which has $n$ boundary legs meeting at a single bulk vertex.

The picture of $A$ as a region carved out of a larger lattice also explains the role of the boundary Hamiltonian in Sec.~\ref{sec:bcs}. The $n$ marked points are the only places where information can enter or leave $A$, so a boundary Hamiltonian coupling the boundary legs is a Hamiltonian for these $n$ degrees of freedom. If we tune that Hamiltonian to a critical point and send $n \to \infty$, so that the marked points become dense along $\partial \Sigma$, the boundary becomes gapless, and we can think of the resulting sequence of open boundary conditions as a renormalization group flow toward a conformal boundary condition. We now review what a conformal boundary condition is.

\subsection{Conformal boundary conditions} \label{app:confbcs}

A conformal boundary condition is the opposite regime from a topological one: the degrees of freedom on the boundary become gapless and are described by a 2d conformal field theory (CFT). This is the familiar situation in three-dimensional gravity, where the Brown--Henneaux (asymptotically AdS) boundary conditions of the gravitational description \cite{Brown1986} induce a current algebra or Virasoro CFT on the edge \cite{Witten1989Jones,Elitzur:1989nr,Coussaert_1995}; for the non-chiral, doubled bulk theories of this paper, the edge theory is correspondingly a full CFT with $c_L = c_R$. Unlike their topological counterparts, conformal boundary conditions are not classified by a rigid set of bulk data, even when the gauge group is finite. Instead, they require additional, non-topological input (typically, a choice of boundary Hamiltonian tuned to criticality), and they come in families rather than finite lists \cite{Kong:2017etd,Kong:2019byq,Kong:2019cuu}.

For example, consider the toric code again. Interpolating between the rough and smooth boundary conditions along an edge, the definition of open boundary conditions, drives the boundary through a phase transition whose critical point is an effective transverse field Ising chain living on the edge; this transition has been realized in an explicit lattice model in \cite{ChenJianKongYouZheng2019}. At the self-dual point the chain is critical, and the edge is described by the $c = 1/2$ Ising CFT. But nothing about the bulk singled out that particular critical point as the unique choice. 
The only thing the bulk demands of its conformal boundary is that the edge theory carry a non-anomalous $\Z_2$ symmetry. The Ising CFT satisfies this with its spin flip symmetry, but there are other 2d CFTs with $\Z_2$ symmetries described by the same SymTFT. For example, the $c=1$ compact boson $X$ of radius $R$ has a $\Z_2$ symmetry inside its momentum $U(1)$ symmetry, corresponding to the half period shift $X \to X + \pi R$. Tuning the edge Hamiltonian to a compact boson critical point rather than to the Ising point therefore produces a second conformal boundary condition for the \emph{same} bulk, now with $c=1$. Indeed, \emph{any} 2d CFT whose global symmetries are parameterized by the TQFT associated with the toric code can be realized as a conformal boundary condition in this model \cite{Kong:2017etd,Ji:2019jhk,Gaiotto:2020iye}. So the determination of ``which'' CFT lives on the boundary requires input which is not constrained by the bulk theory alone. This is the sense in which conformal boundary conditions come in families: the bulk fixes the symmetry that the edge must realize, and nothing more.
In our setting, the additional input is the boundary Hamiltonian of Appendix~\ref{app:openbcs}. Once it is chosen and tuned to criticality, it specifies the complete CFT on the boundary. 

\section{The microcanonical measure on \texorpdfstring{$\widehat{G}$}{Ghat}}
\label{app:microcanonical}

In Sec.~\ref{sec:centertrace} we argued that the trace on the center $\mathcal{Z}(\mathcal{A}_R)$ should be defined using the measure on $\widehat{G}$ dual to the flat measure on the space of conjugacy classes of $G$, rather than the Plancherel measure, which is dual to the Haar measure on $G$. In this appendix we make that prescription more explicit and check it for $G = \SU(2)$ and $G = \SL(2,\R)$.

\subsection{Conjugacy classes, characters, and the Weyl denominator}

Let $G$ be a transformable group and let $T_H$ denote its Cartan subgroups, labeled as in \cite{Balasubramanian:2025rcr} by subgroups $B \subset A$ of the split torus, with $T_{\{e\}} = T_K$ the maximally compact Cartan subgroup and $T_A$ the maximally split one. Every regular element of $G$ is conjugate to an element of some $T_H$, so a class function on $G$ is determined by its restriction to the collection of Cartan subgroups, modulo the action of the Weyl groups $W(T_H)$. The space of conjugacy classes is therefore
\begin{align}
    [G] \cong \bigsqcup_H T_H / W(T_H) \,,
\end{align}
up to a set of measure zero, and there is an obvious flat measure on this set: the Haar measure $dt$ of each $T_H$, restricted to a fundamental chamber of $W(T_H)$ (i.e., a fundamental domain of the Weyl group action on $T_H$).

This flat measure $dt$ is not the same as the one the Haar measure of $G$ induces on $[G]$. The relation between the two is the Weyl integration formula, which for a compactly supported function $f$ reads 
\begin{align}
    \int_G dg \, f(g) &= \sum_H \int_{T_H / W(T_H)} dt \, |\Delta_H(t)|^2 \hat{f}_H(t) \,, \label{eq:weylintegration}\\
    \hat{f}_H(t) &= \int_{G/T_H} dx f(xtx^{-1})\,,
\end{align}
where $\Delta_H(t)$ is the Weyl denominator of $T_H$.\footnote{The relative normalization of $dt$ between different Cartan subgroups is fixed by Harish-Chandra's normalization of the orbital integrals; see \cite{herb2011plancherelformulaplanchereltheorem} for the general statement. The relative normalization between $dt$ and $dx$ follows from using the same normalization for the Lie algebra of each $T_H$ and the tangent space of $G/T$ that is induced from the Lie algebra of $G$.}
Note that $\hat{f}_H(t)$ is a class function of $G$. So the Jacobian relating the Haar measure to the flat measure on conjugacy classes is precisely $|\Delta_H(t)|^2$.

The reason this is useful is that the same Weyl denominator appears in Harish-Chandra's character formula. On each Cartan subgroup, the products
\begin{align}
    \widehat{\chi}_\pi(t) := \Delta_H(t) \chi_\pi(t) \label{eq:normalizedcharacter}
\end{align}
are bounded functions, and in fact are given by a finite sum of exponentials. We will refer to $\widehat{\chi}_\pi$ as the normalized character. Furthermore, we define
\begin{align}
    F^H_f(t) := \Delta_H(t) \hat{f}_H(t)
\end{align}
as the normalized orbital integral of $f$ \cite{HarishChandra1976,herb2011plancherelformulaplanchereltheorem}. Combining \eqref{eq:weylintegration} and \eqref{eq:normalizedcharacter}, the Haar pairing of a character with a compactly supported test function $f$ becomes the flat pairing of the normalized character with the normalized orbital integral:
\begin{align}
    \int_G dg \, \chi^*_\pi(g) f(g) = \sum_H \frac{1}{|W(T_H)|} \int_{T_H} dt \, \widehat{\chi}^*_\pi(t) F^H_f(t) \,,
\end{align}
which is the pairing that defines the flat dual measure on each $T_H$. The measure on $\widehat{G}$ dual to this pairing is what we called the microcanonical measure $d\pi$, and by construction it is the flat Fourier dual of $dt$ on each Cartan subgroup, restricted to a Weyl chamber.

\subsection{Warm-up: \texorpdfstring{$G = \SU(2)$}{G=SU(2)}}

The compact case is a useful warm-up because everything is standard. There is a single Cartan subgroup $T = \mathrm{U}(1)$ (rotations around a fixed axis), its Weyl group is $\Z_2$ (which sends $\theta \to -\theta$), and the Weyl denominator is $\Delta(\theta) = e^{i\theta} - e^{-i\theta} = 2i\sin(\theta)$. 
So if we want to weigh the conjugacy classes of $\SU(2)$ evenly, we should use the flat measure $d\theta$. The Haar measure, in contrast, would assign a measure $4\sin^2(\theta) d\theta$ to each conjugacy class.

The spin-$j$ character is $\chi_j(\theta) = \sin((2j+1)\theta)/\sin(\theta)$, so the normalized character is
\begin{align}
    \widehat{\chi}_j(\theta) = e^{i(2j+1)\theta} - e^{-i(2j+1)\theta} \,,
\end{align}
which is a Fourier mode on the circle labeled by the integer $d_j = 2j+1$, antisymmetrized under the Weyl group. Fourier analysis on $\text{U}(1)$ with the flat measure $d\theta$ therefore assigns the counting measure for $j$ on $\widehat{\SU}(2)$ as the dual measure, one unit per value of $j$. With the normalization $\mathrm{Vol}(\SU(2)) = 1$ for the Haar measure, one can check directly that $\int dg \, \chi_j(g)^* \chi_\ell(g) = \delta_{j\ell}$. The Plancherel measure, by contrast, assigns a weight $\mu(j) = d_j$ to each representation.

\subsection{\texorpdfstring{$G = \SL(2,\R)$}{G=SL(2,R)}}

Now consider the case $G=\SL(2,\R)$. As reviewed in \cite{Balasubramanian:2025rcr}, $\SL(2,\R)$ has two conjugacy classes of Cartan subgroups: the maximally compact one $T_K = \SO(2)$, with elements the rotations $k_\theta$, and the maximally split one $T_A = \pm \R^+$, with elements $\pm a_t = \pm \, \mathrm{diag}(e^t, e^{-t})$. The corresponding Weyl denominators are
\begin{align}
    \Delta_K(k_\theta) = e^{i\theta} - e^{-i\theta} \,, && \Delta_A(a_t) = e^{t} - e^{-t} \,.
\end{align}
The Weyl group of the split Cartan subgroup is $\Z_2$, acting on the representation label by $\lambda \to -\lambda$, while the Weyl group of the compact Cartan subgroup is trivial, since the element that would conjugate $k_\theta$ to $k_{-\theta}$ has determinant $-1$ and so does not lie in $\SL(2,\R)$.

The discrete series $D_n$, labeled by an integer $|n| \geq 2$, is attached to the compact Cartan subgroup, where its character is $\chi_n(k_\theta) = -\mathrm{sign}(n) e^{i(n - \mathrm{sign}(n))\theta}/(e^{i\theta} - e^{-i\theta})$. The normalized character is therefore
\begin{align}
    \widehat{\chi}_n(k_\theta) = -\mathrm{sign}(n)\, e^{i(n-\mathrm{sign}(n))\theta} \,,
\end{align}
a single Fourier mode on the circle. Fourier analysis on $T_K$ with the flat measure $d\theta$ thus assigns the counting measure to the discrete series, and because $W(T_K)$ is trivial the label runs over all $n \neq 0$ rather than over $|n|$ only. Since the Plancherel measure of the discrete series is $\mu(n) = |n|-1$, we find
\begin{align}
    \frac{d\mu(n)}{dn} = |n| - 1\,.
\end{align}

The principal series $P^\pm_\lambda$ is attached to the split Cartan subgroup, where its character is $\chi_{\lambda,\pm}(\pm a_t) = (e^{i \lambda t} + e^{-i\lambda t})/|e^t - e^{-t}|$, so that
\begin{align}
    \widehat{\chi}_{\lambda,\pm}(\pm a_t) = e^{i\lambda t} + e^{-i\lambda t} = 2\cos(\lambda t) \,,
\end{align}
up to the sign that distinguishes the two components. This is the Weyl-symmetric Fourier kernel on $T_A \cong \R$, and Fourier inversion with the flat measure $dt$ assigns to the principal series the flat measure $d\lambda$ on the chamber $\lambda > 0$, with the factor of $|W(T_A)| = 2$ accounting for the identification $\lambda \sim -\lambda$. The Plancherel measure of the principal series is $d\mu(\lambda,\pm) = \tfrac{1}{2}\lambda \tanh^{\pm 1}(\pi \lambda /2) d\lambda$, so
\begin{align}
    \frac{d\mu(\lambda,+)}{d\lambda} = \frac{\lambda}{2} \tanh\left(\frac{\pi \lambda}{2}\right) \,, &&
    \frac{d\mu(\lambda,-)}{d\lambda} = \frac{\lambda}{2} \coth\left(\frac{\pi \lambda}{2}\right) \,.
\end{align}
Collecting these, the area operator \eqref{eq:areaoperator} for $G = \SL(2,\R)$ has eigenvalues $\ln(|n|-1)$ on the discrete series and $\ln\big(\tfrac{\lambda}{2}\tanh^{\pm 1}(\pi\lambda/2)\big)$ on the principal series.

\newpage
\bibliographystyle{JHEP}
\bibliography{biblio}

\end{document}